\documentclass[pdflatex]{sn-jnl}

\usepackage{graphicx}%
\usepackage{multirow}%
\usepackage{amsmath,amssymb,amsfonts}%
\usepackage{amsthm}%
\usepackage{mathrsfs}%
\usepackage[title]{appendix}%
\usepackage{xcolor}%
\usepackage{textcomp}%
\usepackage{manyfoot}%
\usepackage{booktabs}%
\usepackage{algorithm}%
\usepackage{algorithmicx}%
\usepackage{algpseudocode}%
\usepackage{listings}%
\usepackage{float}
\usepackage{caption}
\usepackage{subcaption}
\usepackage{xcolor} 
\usepackage{placeins}
\usepackage{multirow}
\usepackage{booktabs}
\usepackage{adjustbox}
\usepackage{natbib}

\theoremstyle{thmstyleone}%
\theoremstyle{thmstyletwo}%

\theoremstyle{thmstylethree}%

\begin{document}

\title[Evaluating the performance of GCM trajectories ]{Evaluating the performance of GCM trajectories using Weather Type frequencies for persistence and transitions: the Iberian Peninsula and Lamb classification
}




\author*[1]{\fnm{Elsa} \sur{Barrio-Torres}}\email{e.barrio@unizar.es}

\author[2]{\fnm{Swen} \sur{Brands}}\email{brandssf@ifca.unican.es}

\author[1]{\fnm{Jes\'us} \sur{As\'in}}\email{jasin@unizar.es}

\author[1]{\fnm{Jes\'us} \sur{Abaurrea}}\email{abaurrea@unizar.es}

\author[1]{\fnm{Zeus} \sur{Gracia-Tabuenca}}\email{zeus@unizar.es}

\author[1]{\fnm{Jorge} \sur{Castillo-Mateo}}\email{jorgecm@unizar.es}

\affil*[1]{\orgdiv{Department of Statistical Methods}, \orgname{University of Zaragoza}, \orgaddress{\street{Pedro Cerbuna 12}, \city{Zaragoza}, \postcode{50009},\country{Spain}}}

\affil[2]{\orgdiv{Instituto de F\'isica de Cantabria (UC-CSIC)}, \orgname{Organization}, \orgaddress{\street{Street}, \city{Santander}, \postcode{-}, \state{State}, \country{Spain}}}


\abstract{The Iberian Peninsula is highly vulnerable to the impacts of climate change, particularly intense summer heatwaves and droughts driven by persistent large-scale atmospheric circulation patterns. To assess the regional impact of these events, it is important to ensure Global Climate Models (GCMs) accurately reproduce these circulation dynamics. This study evaluates the performance of 36 historical CMIP6 GCM trajectories (1979--2005) in reproducing atmospheric circulation over the Iberian Peninsula in the summer months (June--September) using the Lamb Weather Type (WT) classification scheme. Using ERA5 reanalysis as the observational reference, we introduce a methodological framework---applicable to any region worldwide---to evaluate GCM performance. This approach extends traditional daily frequency analysis by evaluating both the daily frequency distribution of WTs and their 24-hour dynamic evolution (i.e., transition probabilities and persistence). Model performance is quantified using the Overlap coefficient. A filtering process is applied where only trajectories that successfully reproduce both daily and conditional distributions with a minimum Overlap threshold $t_{sim}$ across a set number of grid points are retained. The findings show that while several models can adequately reproduce daily WT frequencies (16 out of 36), some struggle to capture day-to-day atmospheric transitions. This leads to a final selection of 12 trajectories over the Iberian Peninsula. Model performance across the region is then evaluated using integrated metrics assessing daily reproduction, conditional reproduction, and transition dynamics. Overall, models from the \texttt{ec\_earth3} family---specifically the \texttt{ec\_earth3\_aerchem} trajectory---exhibit the best and most consistent performance across the region. Additionally, the results highlight a geographical performance gap: while models generally represent circulation well in the northwest, they face significant challenges in the central and southern Mediterranean regions of the Peninsula. Ultimately, this study establishes that assessing WT persistence and transitions provides a far more discriminative, objective tool for GCM selection than evaluating daily distributions alone.}




\maketitle

\section{Introduction}
 
Global climate models (GCM) are numerical representations of the coupled climate system. They resolve atmosphere, land-surface, ocean, biosphere and cryosphere dynamics, as well as their feedbacks and mutual interactions \citep{Bordoni_et_al_2025}. They are a cornerstone of climate research, contributing to system understanding, the detection of anthropogenic climate change emerging from natural variability, its attribution to human activities and impact on society and ecosystems \citep{Masson-Delmotte_et_al_2021}.

Within the Coupled Model Intercomparision Project (CMIP), GCM experiments to study the past, present and future climate have been designed, executed and compared with each other for the last 30 years \citep{Meehl_et_al_2000, Taylor_et_al_2012, Eyring_et_al_2016, Dunne_et_al_2025}. From one CMIP phase to the next, the number of available models and experiments has been steadily increasing, and so is the demand for model data to be applied in regional-scale impact studies, as anthropogenic climate change is now being recognized as one of the most important challenges to humankind. The need to systematically assess the degree of realism of the models compared to observations, commonly referred to as "performance" \citep{Giorgi_Francisco_2000}, "fidelity" \citep{Shukla_et_al_2006}, or "plausibility" \citep{Sobolowski_et_al_2025}, has long been emphasized by regional climate scientists \citep{Hulme_et_al_1993, Brands_et_al_2013, McSweeney_et_al_2014, Cannon_2020, brands2022circulation, brands2023global}, but only recently has been systematically assessed in large community efforts such as the Coordinated Regional Climate Downscaling Experiment (CORDEX) \citep{Sobolowski_et_al_2025}. Similarly, CMIP itself has been giving a relatively low priority to the assessment of GCM performance for decades, but is now implementing a \emph{Rapid Evaluation Framework} \citep{Hoffmann_et_al_2025} to automatically evaluate model performance as soon as the data become available on the nodes of the Earth System Grid Federation (ESGF). Despite the obvious impossibility to assess model performance for future time periods, some studies have shown that the uncertainty of future outcomes can be reduced by giving more weight to those models that most reliably reproduce key feedback processes seen in observations \citep{Klein_and_Hall_2015,Palmer_et_al_2023}.

Atmospheric circulation is a key process (or diagnostic) for model evaluation since it drives the regional-scale weather and climate conditions, particularly in the extra-tropics \citep
{Shepherd_2014}. 


Research on climate change in Spain is critically important due to the country’s high vulnerability to its impacts. According to the 2024th edition of the `Informe CLIVAR-SPAIN sobre el clima en España' (\url{http://clivar.es/wp-content/uploads/2025/02/Libro-INFORME-CLIVAR-SPAIN-clima-en-Espana-v6.pdf}), large regions of Spain face water stress, hot and dry summers, and low levels of precipitation. The report notes that Spain has been warming at a rate above the global average since the 1980s, with particularly sharp increases during the summer and a growing frequency and intensity of heatwaves and droughts. Indeed, the 21st century has witnessed the highest occurrence of severe droughts in the past 150 years. 

Long-term records indicate that average temperatures have risen by approximately 0.1--0.3 $^\circ$C per decade since the 1960s--1970s, with especially rapid warming during summer months \citep{serrano2017, penaangulo2021, corell2025}. Across the country, hot extremes have become more frequent, including warm days, warm nights, and tropical nights, with the strongest increases reported along the Mediterranean and Cantabrian coasts \citep{kenawy2011, miro2006}. Heatwaves are now more frequent, longer-lasting, and more intense, and their season extends earlier into spring and later into autumn; in addition, their spatial extent over the Iberian Peninsula has increased by around 4\% per decade \citep{paredesfortuny2023, diazposo2023}. 
Local and regional studies (e.g., Barcelona, Extremadura, Asturias, and the southeastern Mediterranean coast) further confirm marked increases in heatwave frequency and duration, together with a substantial loss of summer climate comfort in coastal tourist regions \citep{serra2024, acero2018}. 
Overall, the literature identifies Spain as a climate change hotspot, where sustained warming is associated with increasingly intense and persistent summer heat and growing impacts on health, ecosystems, tourism, and fire risk \citep{serranonotivoli2023, diazposo2023, achebak2018}.

Consistent with these warming trends, increases have also been documented in the occurrence of record-breaking maximum temperatures. \citet{castillo2025a} showed that during the 2012--2021 decade, the number of record-breaking temperature events was nearly twice as large as would be expected under stationary climate conditions, with the strongest increase observed in summer. \citet{castillo2025b} extended the analysis to a bivariate framework for daily maximum and daily minimum temperature records up to 2023, revealing different but correlated patterns of non-stationarity in the two series. In China, \citet{sun2025} showed that anthropogenic forcing has drastically increased the likelihood of record-breaking temperatures since the 1980s, particularly in regions like the Tibetan Plateau. Their projections suggest that current extreme heat events will likely become the annual norm by the 2060s.

Large-scale atmospheric circulation plays a key role in driving extreme heat events over Spain. \citet{ventura2023} showed that historical heatwaves are largely controlled by four recurring synoptic weather patterns associated with persistent Atlantic anticyclonic blocking, which together explain over 50\% of observed events. 
Future projections suggest increasing heatwave intensity linked to higher 500 hPa geopotential heights caused by a northward shift of the subtropical ridge. 
Consistent with this circulation pattern, an analysis of Iberian heatwaves showed that the most intense events are often associated with very warm, dry, and stable Saharan air masses moving northward into the peninsula, guided by a cyclonic circulation over the northeastern Atlantic and a strong subtropical ridge \citep{sousa2019}. 
In line with these findings, \citet{serranonotivoli2023} found that the summer of 2022--the hottest in Spain's recorded history--was driven by a pronounced subtropical ridge at the 500 hPa geopotential height and North Atlantic blocking, illustrating how persistent large-scale anticyclones and warm-air transport are now responsible for the most extreme heat events.
Relationship of heat events and atmospheric situations is analyzed in order to characterize situations that would be predictive. In Iberian Peninsula the occurrence of record-breakings events in maximum daily temperature has been associated with patterns in 700 hPa fields \citep{barrio2026}. 


The specific relationship between Lamb Weather Types (LWT) and extreme heat events have also been explored. 
\citet{perez2023} used an extensive 74-year database (1948–2021) to analyze climate evolution in Spain through Lamb weather types (LWT), finding that anticyclonic conditions are the most frequent while easterly flows generally correlate with higher atmospheric pressure. They also found that North Africa's intense thermal gradients significantly influence regional heat waves depending on the specific air flow.  \citet{fernandez2024signature} found the Pure Anticyclonic LWT to be associated with atmospheric blocking episodes, which drive extreme heat events and are themselves sustained by them.






In this study, we examine the distribution of LWT over the Iberian Peninsula during the warm season (June--September). To generalize to other ways of defining Weather Types, we will refer to LWT as WT throughout the paper. Due to large-scale circulation patterns strongly influencing extreme summer temperatures in this region, identifying models that accurately reproduce WT dynamics is a critical first step for statistical downscaling. To this end, we use ERA5 reanalysis data as an observational reference to evaluate 36 CMIP6 historical trajectories from GCMs. Our primary goal is to establish objective criteria for selecting or discarding GCM runs based on two complementary metrics: 

\begin{enumerate} 

	\item \textbf{Daily Distributions:} Assessing the models' ability to reproduce the overall frequency of WT occurrences. 

	\item \textbf{WT Transitions or Persistence:} Evaluating day-to-day WT evolution using transition probability matrices, based on conditional relative frequencies associated with preceding weather types. Note that this includes the persistence.

\end{enumerate} 

These comparisons are performed both locally (node-by-node) and regionally. The methodological framework can be applied to any region in the world. Its implementation presents two main challenges. First, the distributions being compared are categorical with up to 27 classes--some of which are rare--complicating the selection of appropriate distance measures. Second, because model performance varies spatially, we must use summary metrics that integrate data across all grid points while accounting for spatial heterogeneity. 

The rest of this paper is organized as follows: Section 2 describes the datasets, presents exploratory analyses, and details the proposed methodology. Section 3 outlines the evaluation results, followed by the selection of GCM runs in Section 4. Finally, Section 5 discusses the main findings and suggests directions for future research.

\section{Datasets and Exploratory Data Analysis}

This section outlines the data and distribution comparison metrics used to evaluate WT over the Iberian Peninsula. First, we describe the extraction of WT data from the ERA5 observational reference and the ensemble of historical GCM trajectories, and the mathematical framework used to calculate their relative frequencies. Subsequently, we present an exploratory analysis of the ERA5 observations to understand the daily distributions and WT persistence of these weather types across the region.

\subsection{Data Sources}
WT come from the Jenkinson--Collison weather typing scheme (JC-WT), which is an automated classification system that transforms continuous regional sea-level pressure (SLP) data into a set of 27 distinct categories, each representing a recurrent (i.e., typical) circulation pattern. Originally created for the British Isles in the early 1970s \citet{Jenkinson_and_Collison_1977}, this method has since been widely applied to classify near-surface atmospheric circulation \citet{Trigo_DaCamara_2000,Jones_et_al_2013}. 

The scheme organizes instantaneous SLP patterns centered on a specific location based on the direction of the geostrophic flow (or lack thereof) and the sign and magnitude of the vorticity. To achieve this, SLP differences are calculated between grid cells arranged in a cross pattern around the target region. The resulting weather types (WTs) consist of a purely cyclonic (C) type, a purely anticyclonic (A) type, eight directional types (N, NE, E, \dots, NW), and 16 hybrid types that combine A or C circulation with the directional types. A 27th type represents unclassified (U) cases, corresponding to situations with chaotic, weak flow, or incompatible hybrid types. Unlike other SLP-based classification methods, JC-WT exclusively uses SLP gradients, offering a straightforward and interpretable approach that avoids the need for statistical transformations of input data. For a complete description of the method, the interested reader is referred to \citet{fernandez2023exploring}.

This study considers the ESCENA spatial domain, covering 30$^\circ$--60$^\circ$N and 41.25$^\circ$W--33.75$^\circ$E with a 2.5$^\circ \times$ 2.5$^\circ$ grid resolution. However, the analysis is strictly focused on the grid box over the Iberian Peninsula, spanning 35$^\circ$--45$^\circ$N and 8.75$^\circ$W--3.75$^\circ$E (30 grid points). 

For each grid point $s$ within the ESCENA domain, the focus is on the probability $P(\text{WT}_{d}=i \cap \text{WT}_{d-1}=j; s)$. This is the probability of a day $d$ being characterized by $\text{WT}_d=i$ and the previous day $d-1$ being characterized as $\text{WT}_{d-1}=j$. This probability is estimated using the \textbf{joint relative frequency (joint rf)} derived from the corresponding trajectory for each grid box. 

At a given grid point, the \textbf{joint rf} is calculated using the complete daily series of WT data at 12:00 UTC, restricted to the summer months (June, July, August, and September) covering the period 1979--2005. This calculation relies on two simplifying hypotheses: that $P(\text{WT}_{d}=i \cap \text{WT}_{d-1}=j; s)$ does not undergo relevant seasonal changes across the summer, and that it maintains a stationary behavior across the 1979--2005 period. It should be noted that if a much longer time period was considered, then evident non-stationarity would make these hypotheses invalid. Additionally, if the entire year was considered, seasonality should be addressed. Furthermore, spatial integration across the Iberian Peninsula is not possible due to the region's high climate variability. 

ERA5 reanalysis data \citet{hersbach2020era5} serves as the observational reference, representing the observed evolution of WT. Throughout this analysis, parameters derived from this database are denoted using the subscript ERA. Table \ref{tab:cmip6_models} details the 36 GCM historical trajectories evaluated in this study; parameters obtained from these models use the subscript $n$ to denote the $n$-th trajectory.

\begin{table}[h!]
\centering
\captionsetup{justification=centering}
\caption{CMIP6 models considered \citep{brands2022circulation}.}
\vspace{0.3cm}
\label{tab:cmip6_models}

\begin{minipage}{0.48\textwidth}
\centering
\begin{tabular}{lll}
\textbf{Institute} & \textbf{Model} & \textbf{Ensemble} \\
\multirow{2}{*}{ACCESS} 
  & CM2        & r1i1p1f1 \\
  & ESM1.5     & r1i1p1f1 \\
\addlinespace

AWI 
  & ESM-1-1-LR & r1i1p1f1 \\ 
\addlinespace

BCC 
  & CSM2-MR    & r1i1p1f1 \\
\addlinespace

\multirow{3}{*}{CMCC}
  & CM2-HR4    & r1i1p1f1 \\
  & CM2-SR5    & r1i1p1f1 \\
  & ESM2       & r1i1p1f1 \\
\addlinespace

\multirow{3}{*}{CNRM}
  & CM6-1-HR   & r1i1p1f2 \\
  & CM6-1      & r1i1p1f2 \\
  & ESM2-1     & r1i1p1f2 \\
\addlinespace

\multirow{5}{*}{EC-Earth}
  & EC-Earth3-AerChem & r1i1p1f1 \\
  & EC-Earth3-CC      & r1i1p1f1 \\
  & EC-Earth3         & r1i1p1f1 \\
  & EC-Earth3-Veg-LR  & r1i1p1f1 \\
  & EC-Earth3-Veg     & r1i1p1f1 \\
\addlinespace

FGOALS
  & G3          & r3i1p1f1 \\

\multirow{2}{*}{GFDL}
  & CM4         & r1i1p1f1 \\
  & ESM4        & r1i1p1f1 \\
\addlinespace

GISS
  & E2.1-G      & r1i1p1f1 \\
\addlinespace
\end{tabular}
\end{minipage}%
\hfill
\begin{minipage}{0.48\textwidth}
\vspace{1.35cm}
\centering
\begin{tabular}{lll}
\textbf{Institute} & \textbf{Model} & \textbf{Ensemble} \\

HadGEM
  & GC31-MM     & r1i1p1f3 \\
\addlinespace

IITM
  & ESM         & r1i1p1f1 \\
\addlinespace

INM
  & CM5         & r2i1p1f1 \\
\addlinespace

IPSL
  & CM6A-LR     & r1i1p1f1 \\
\addlinespace

KACE
  & 1.0-G       & r1i1p1f1 \\
\addlinespace

KIOST
  & ESM         & r1i1p1f1 \\
\addlinespace

\multirow{2}{*}{MIROC}
  & ES2L        & r5i1p1f2 \\
  & MIROC6      & r3i1p1f1 \\
\addlinespace

\multirow{3}{*}{MPI}
  & ESM1.2-HAM  & r1i1p1f1 \\
  & ESM1.2-HR   & r1i1p1f1 \\
  & ESM1.2-LR   & r1i1p1f1 \\
\addlinespace

MRI
  & ESM2.0      & r1i1p1f1 \\
\addlinespace

NESM
  & NESM3       & r1i1p1f1 \\
\addlinespace

\multirow{2}{*}{NorESM}
  & NorESM2-LM  & r1i1p1f1 \\
  & NorESM2-MM  & r1i1p1f1 \\
\addlinespace

SAM
  & SAM0-UNICON & r1i1p1f1 \\
\addlinespace

TAI
  & TAIESM1     & r1i1p1f1 \\
\addlinespace
\end{tabular}
\end{minipage}

\end{table}

\subsubsection{Calculation of Relative Frequencies}

To assess how well the models reproduce WT dynamics, we analyze two key aspects of the WT distribution: the frequency of their occurrence, and the frequency of transitions (or persistence) between two consecutive days.

The working data consists of the relative frequencies of WTs on two consecutive days, $d$ and $d-1$, in grid point $s$. This \textbf{joint relative frequency (joint rf)} estimates the corresponding probabilities $P(\text{WT}_{d}=i \cap \text{WT}_{d-1}=j;s)$. For ERA5 data, this is denoted as $\mathbf{rf}_{ERA}(i,j; s)$. It is assumed that this frequency is constant for any given day $d$ within the analyzed period, meaning it is not tied to one specific date.

The \textbf{daily relative frequency (rf)} for a specific $\text{WT}=i$ at a given grid point is used to estimate its overall probability of occurrence:
\begin{equation*}\label{eq:prob}
    P(\text{WT}_{d}=i; s) = \sum_{j=1}^{27} P(\text{WT}_{d}=i \cap \text{WT}_{d-1}=j; s)
\end{equation*}
We define $rf_{ERA}(i; s)$ as the relative frequency of $\text{WT}=i$ at grid point $s$ in the ERA dataset, obtained by summing the joint frequencies across the $\text{WT}=j$ dimension:
\begin{equation}\label{eq:rf}
    rf_{ERA}(i; s) = \sum_{j=1}^{27} rf_{ERA}(i,j; s)
\end{equation}
To estimate the transition dynamics, we compute the \textbf{conditional relative frequency (cond rf)} of $\text{WT}=i$ occurring, given that $\text{WT} = j$ occurred on the previous day at grid point $s$. This cond rf, denoted as $rf_{ERA}(i|j; s)$, is calculated as follows:
\begin{equation}
    rf_{ERA}(i|j; s) = \frac{rf_{ERA}(i,j; s)}{rf_{ERA}(j; s)}
\end{equation}
This serves as the estimation for the corresponding conditional probability:
\begin{equation*}
    P(\text{WT}_{d}=i | \text{WT}_{d-1}=j; s) = \frac{P(\text{WT}_{d}=i \cap \text{WT}_{d-1}=j; s)}{P(\text{WT}_{d-1}=j; s)}
\end{equation*}
Consequently, the persistence probability---the likelihood of $\text{WT}_{i}$ remaining the same for two consecutive days---is defined as:
\begin{equation*}
    P(\text{WT}_{d}=i | \text{WT}_{d-1}=i; s) = \frac{P(\text{WT}_{d}=i \cap \text{WT}_{d-1}=i; s)}{P(\text{WT}_{d-1}=i; s)}
\end{equation*}

To estimate persistence in a specific $\text{WT = i}$, we compute the \textbf{persistence relative frequency (per rf)}, which is a specific case of the \textbf{conditional relative frequency (cond rf)}. It corresponds to the probability that a weather type $\text{WT} = i$ occurs, given that the same weather type $\text{WT} = i$ was observed on the previous day at grid point $s$. This \textbf{per rf}, denoted $rf_{ERA}(i|i; s)$, is calculated as follows:

\begin{equation}
    rf_{ERA}(i|i; s) = \frac{rf_{ERA}(i,i; s)}{rf_{ERA}(i; s)}
\end{equation}

For the GCM data, we apply the same statistical framework. We use the notation $rf_{n}(i; s)$ to represent the daily relative frequency of $\text{WT}_{i}$ at grid point $s$ for GCM trajectory $n$, and $rf_{n}(i|j; s)$ for the conditional relative frequency.

\vspace{0.5cm}
\textbf{Data structure}

The working databases for both ERA5 and each GCM run share an identical file structure. Each multidimensional array contains dimensions for longitude (31), latitude (13), $\text{WT}_{i}$ (27), and $\text{WT}_{j}$ (27). A single variable, $rf(i,j; s)$, is defined over these dimensions to estimate $P(\text{WT}_{d}=i \cap \text{WT}_{d-1}=j; s)$. This structure allows the array to store the matrices for every grid box within the ESCENA spatial domain. While calculations are performed for the entire ESCENA region, the results presented in this study are restricted to the grid over the Iberian Peninsula (35$^\circ$--45$^\circ$N and 8.75$^\circ$W--3.75$^\circ$E), which comprises a $5 \times 6$ grid, i.e., 30 grid points.

The 27-year period (1979--2005) is enough to capture the atmospheric dynamics of the Iberian Peninsula, including interannual variability, a representative range of circulation patterns, and enough occurrences of each weather type to reliably estimate transition probabilities. We focus on the summer months (June--September) to reduce seasonal variability and concentrate on a climatically homogeneous period.

\subsection{Exploratory Data Analysis}

\subsubsection{Observed Daily Frequencies of WTs}

\begin{figure}[h!]
    \centering
    \includegraphics[width=1\linewidth]{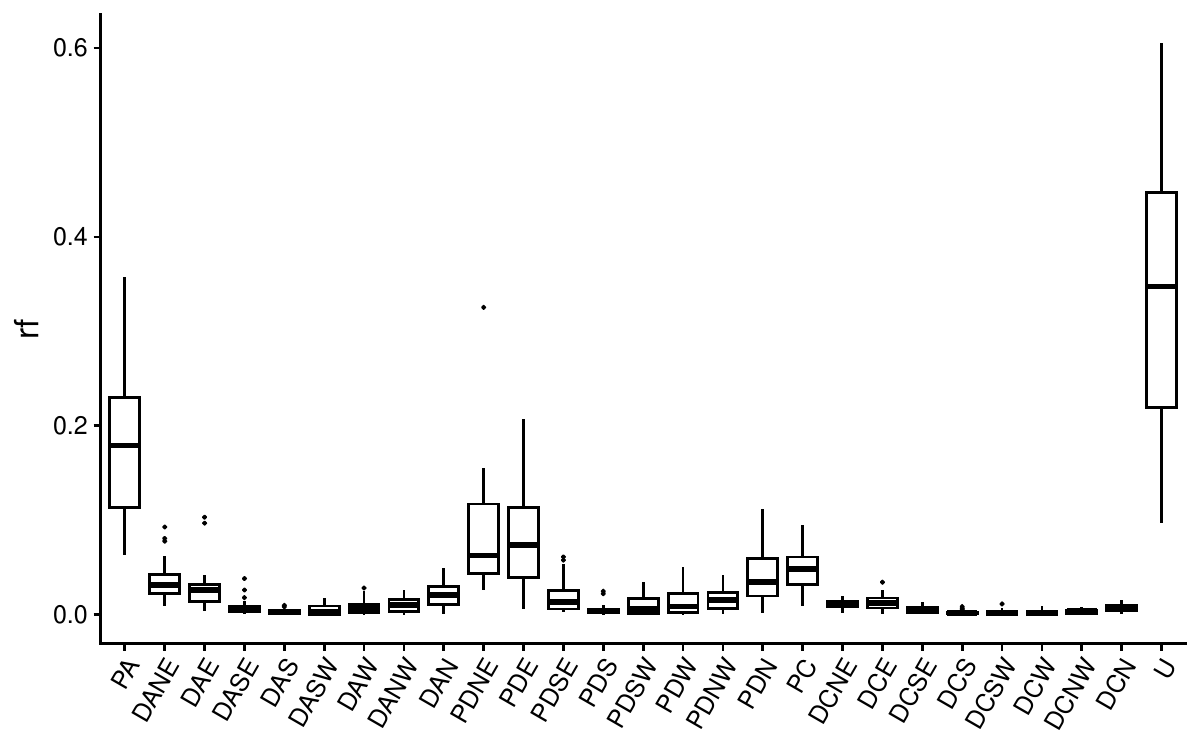}
    \caption{\textbf{rf} of ERA5 data for each WT across 30 points in the Iberian Peninsula grid box.}
    \label{fig:box.marg}
\end{figure}

Figure \ref{fig:box.marg} presents boxplots of the daily relative frequency (\textbf{rf}) for each WT derived from the ERA5 dataset. These boxplots are based on \textbf{rf} values computed across all 30 grid points over the Iberian Peninsula. The figure shows that some weather types (WTs) occur substantially more often than others. The unclassified type (U) is the most frequent, followed by the purely anticyclonic (PA) type. Among the purely directional types, those from the northeast (PDNE) and east (PDE) are also relatively common.

There is notable spatial variability in the occurrence of several weather types, as reflected in the wide spread of their boxplots. For example, type U exhibits considerable variability, with frequencies ranging approximately from 0.1 to 0.6 across the grid. Other frequent types, such as PA, PDNE, and PDE, also display marked regional differences. In contrast, most directional cyclonic types, as well as several directional anticyclonic types, are rare and show limited spatial variability.

Overall, purely directional types tend to be relatively frequent, particularly PDNE, PDE, and PDN. Hybrid types that combine anticyclonic circulation with directional components are generally infrequent, although those associated with northeast (DANE), east (DAE), and north (DAN) flows occur more often within this group. Hybrid cyclonic-directional types are very rare; among them, the eastward type (DCE) is the most frequent, though its occurrence remains low.

Table \ref{tab:WT.reduced} details a reduced set of four primary WTs selected for subsequent analysis. These represent some of the most frequent weather types (PA, PDNE, PC, and U). The table includes their index numbers, abbreviated names, and summary statistics. For a comprehensive overview, Table \ref{app:marg:frec:ERA5} in Appendix \ref{app:daily:frec:ERA5} provides information on the full set of all 27 weather types.

\begin{table}[h!]
    \centering
    \caption{Reduced set of the most frequent weather types (WTs) selected for subsequent analysis. For each WT, the full name is provided along with summary statistics of its relative frequency ($rf$) in ERA5, including the 25th percentile (Q25), median, 75th percentile (Q75), and maximum. Additionally, the products of both the minimum and Q25 with the sample size are shown, to estimate the number of days associated with each WT.}
    \begin{tabular}{cllllllll}
    \hline
         & Full Name & Min & Q.25 & Median & Q.75 & Max & \#Min & \#Q.25 \\
        \hline
        1 & PA & 0.064 & 0.113 & 0.179 & 0.230 & 0.357 & 210 & 373 \\ 
        10 & PDNE & 0.026 & 0.044 & 0.063 & 0.117 & 0.325 & 85 & 144 \\ 
        18 & PC & 0.009 & 0.032 & 0.048 & 0.061 & 0.095 & 30 & 104 \\ 
        27 & U & 0.097 & 0.219 & 0.348 & 0.447 & 0.605 & 320 & 722 \\ 
        \hline
    \end{tabular}
    \label{tab:WT.reduced}
\end{table}

Appendix \ref{cond.rf.box} provides analogous boxplots illustrating the conditional relative frequencies (\textbf{cond rf}). These plots are generated by conditioning on each WT from the reduced set defined in Table \ref{tab:WT.reduced}.

\subsubsection{WT Persistence}

\begin{figure}[h!]
    \centering
    \includegraphics[width=0.49\linewidth]{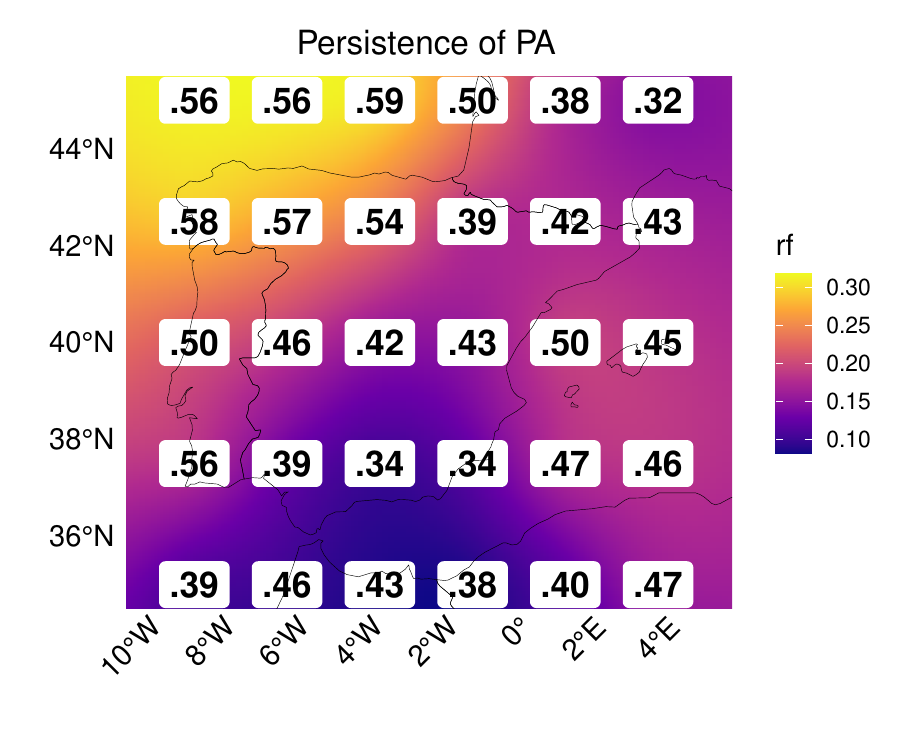}
    \includegraphics[width=0.49\linewidth]{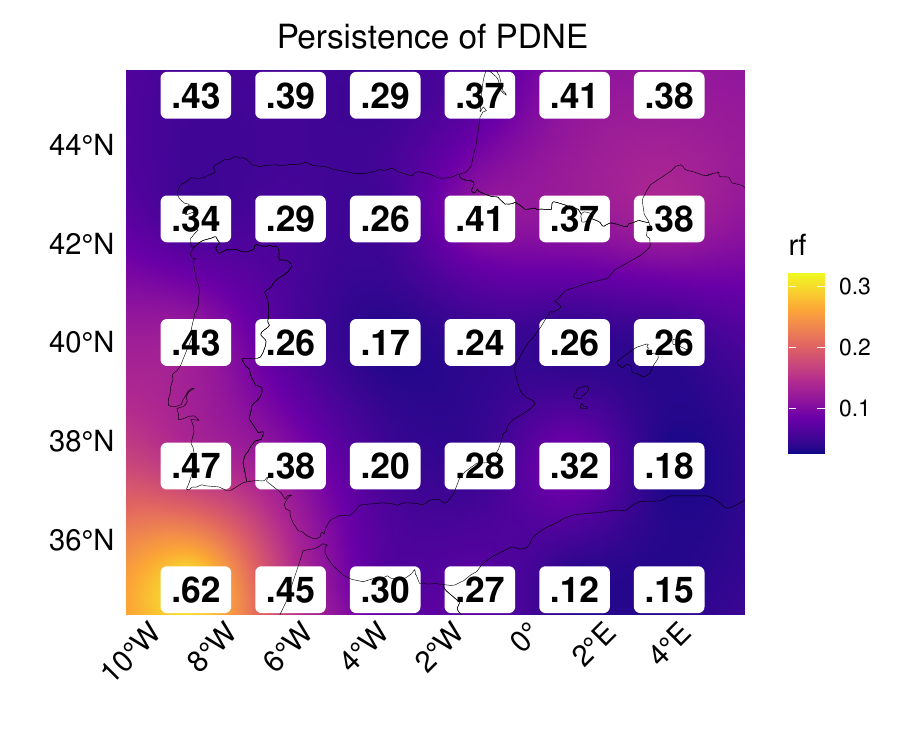}
    
    \includegraphics[width=0.49\linewidth]{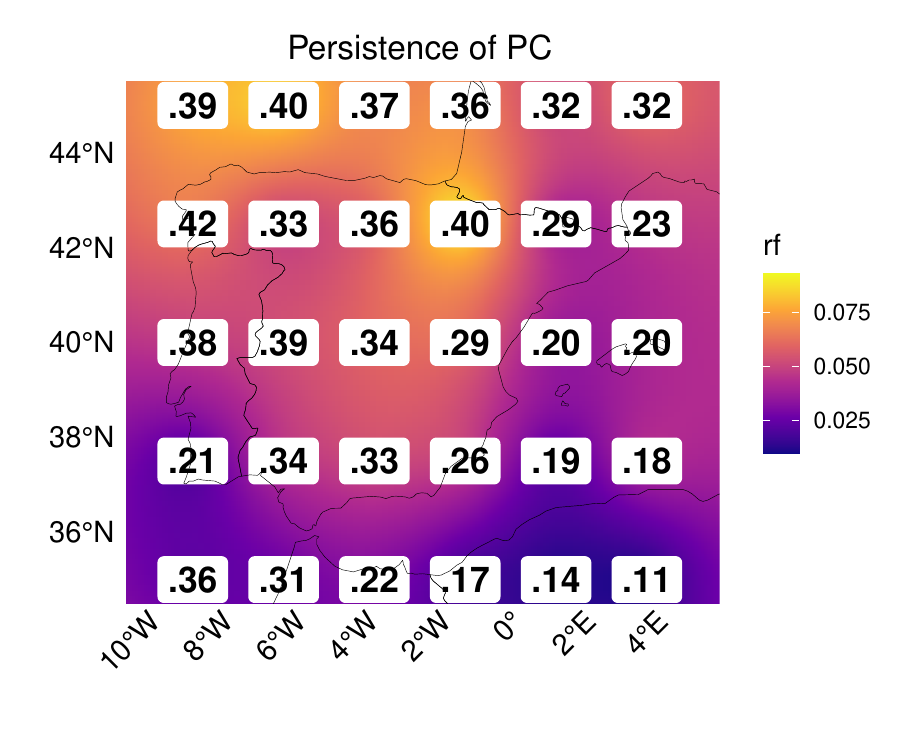}
    \includegraphics[width=0.49\linewidth]{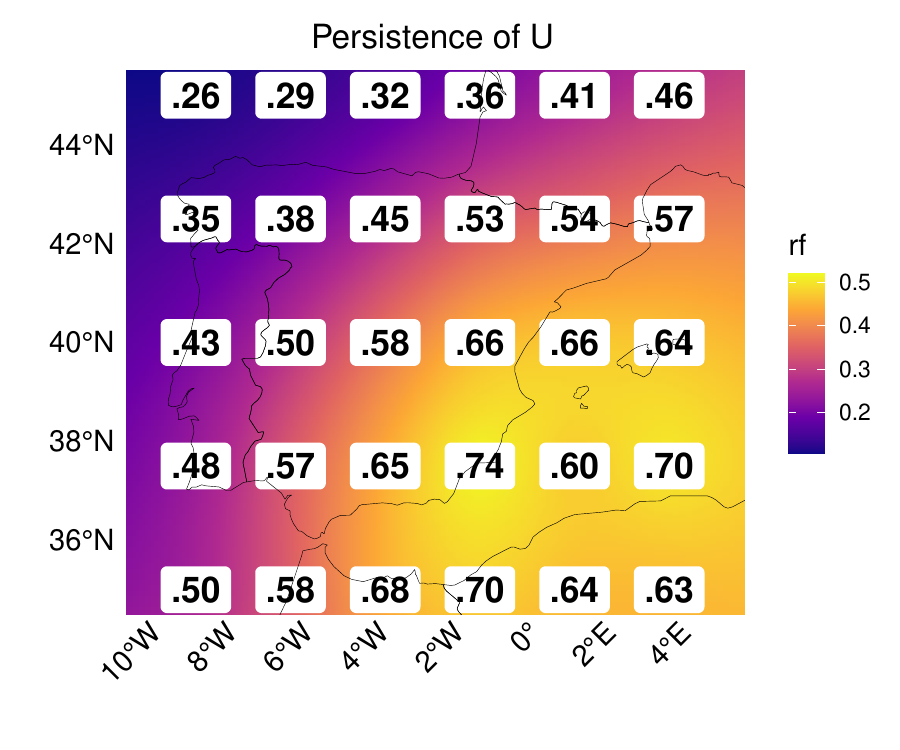}

    \caption{Persistence of WT from Table~\ref{tab:WT.reduced} across the Iberian Peninsula grid points, estimated using the \textbf{cond rf} for the current day. Panels are arranged as in Table \ref{tab:WT.reduced}, with the background showing interpolated daily rf values of the corresponding WT. Persistence across the Iberian Peninsula grid box points, analyzed using the \textbf{cond rf} at each grid point by conditioning on the same WT for the current day. Each plot illustrates the persistence of a weather type from the summary set defined in Table \ref{tab:WT.reduced}, and the background shows interpolated daily \textbf{rf} values of the corresponding WT.}
    \label{fig:persistence}
\end{figure}

An important characteristic of WT dynamics is their persistence, defined as the occurrence of the same weather type on two consecutive days. The numbers in the maps in Figure \ref{fig:persistence} illustrate the \textbf{cond rf} of a specific WT occurring on a given day, conditioned on the exact same WT having occurred the previous day. Using the ERA5 trajectory data, this provides a direct measure of persistence likelihood across each grid point in the Iberian Peninsula. The background color represents the daily \textbf{rf} of the corresponding each WT, to provide a measure of how frequent each WT is at each point.

The plots display persistence maps for the four most frequent WTs, and Table \ref{table:ERA5:persistence} summarizes the relative frequencies of these persistence events.

\begin{table}[h!]
    \centering
        \caption{Summary statistics of persistence relative frequencies across grid points for the most frequent WTs in ERA5. For each WT, the table reports the minimum, 25th percentile (Q25), mean, 75th percentile (Q75), maximum, and standard deviation (sd) of the relative frequencies.}
    \begin{tabular}{lrrrrrr}
        \hline
        WT & Min & Q.25 & Mean & Q.75 & Max & sd \\ 
        \hline
        PA & 0.32 & 0.39 & 0.46 & 0.50 & 0.59 & 0.08 \\ 
        PDNE & 0.12 & 0.26 & 0.32 & 0.38 & 0.62 & 0.11 \\ 
        PC & 0.11 & 0.21 & 0.29 & 0.36 & 0.42 & 0.09 \\ 
        U & 0.26 & 0.43 & 0.53 & 0.64 & 0.74 & 0.13 \\ 
        \hline
    \end{tabular}
    \label{table:ERA5:persistence}
\end{table}

Distinct spatial patterns emerge in the persistence maps shown in Figure \ref{fig:persistence}. Types U and PA exhibit the highest relative frequencies of persistence. Type PA displays a relatively homogeneous spatial distribution across the region, with values ranging from 0.32 to 0.59. In contrast, Type U shows greater spatial variability, with persistence values reaching 0.74 in the southeast and dropping to around 0.26 in the northwest. 

Similarly, Types PDNE (10) and PC (18) both exhibit a northwest-to-southeast decreasing gradient. For these types, persistence values are highest in the northwest (near 0.40) and lowest in the southeast (around 0.15). These spatial gradients reflect the differing atmospheric dynamics and topographic influences between the Mediterranean and Atlantic regions of the peninsula.


\section{Methods}\label{sec:methods}

In this section, we present a new algorithm designed to select GCM trajectories. To do so, we introduce several measures of similarity and dissimilarity between the distributions of two categorical variables. These elements include categorical dissimilarity metrics and summary measures that allow local results to be aggregated and interpreted across the entire region of interest.

The proposed methodology is general and can be applied to any region of the world using reanalysis data and a collection of trajectories from GCMs. The associated code is publicly available to researchers through a GitHub repository.

\subsection{Proposed Algorithm for GCM Trajectory Selection}

\begin{figure}[h!]
    \centering
    \includegraphics[width=1\linewidth]{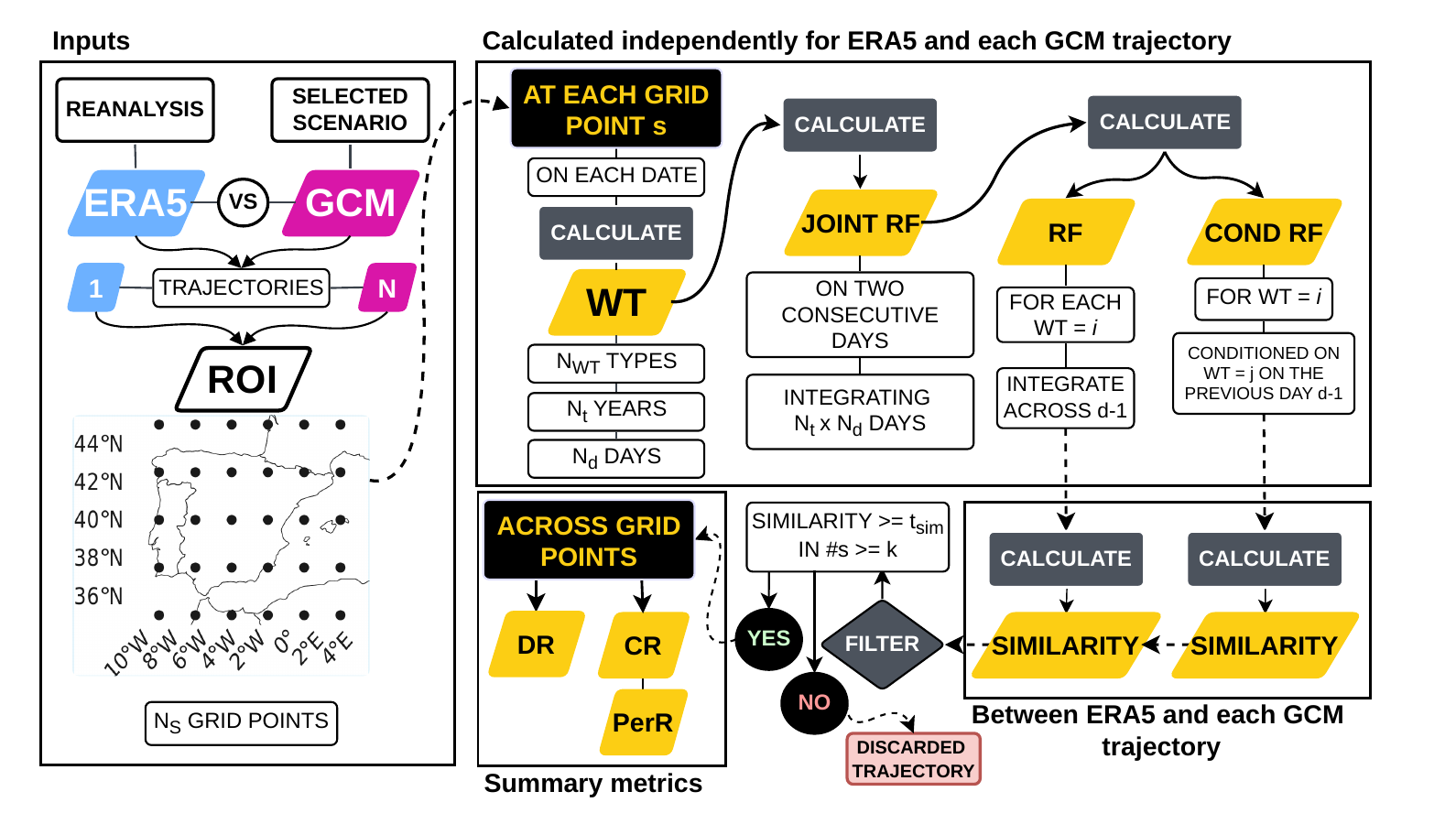}
    \caption{Flow chart for the proposal algorithm.}
    \label{fig:algorithm}
\end{figure}

The procedure for selecting GCM trajectories consists of the following steps:

\begin{itemize}

\item \textbf{Definition of the Region of Interest (ROI).}  
The region of interest is first defined. This typically includes the $N_S$ grid points surrounding the study area in order to adequately capture its spatial variability.

\item \textbf{WT extraction.}  
A reanalysis dataset is used as the reference representing the observed climate. A classification into $N_{\text{WT}}$ weather types (WTs) is defined. WTs are extracted from daily SLP data using the methodology implemented in the Brands classification tools \citet{brands2022b}. This produces daily WT time series at each grid point $s$ for the selected period. 

The same procedure is applied to extract WTs from the GCM trajectories.

\item \textbf{Joint rf calculation.}  
At each spatial point $s$, the time series of weather types (WTs) is used to compute the joint relative frequency \textbf{joint rf} of two consecutive days, $d$ and $d-1$, thereby removing the explicit time dimension. This yields a dataset of dimensions $N_{\text{WT}} \times N_{\text{WT}} \times N_{S}$, representing the  WT on day $d$, the WT on day $d-1$, and the spatial grid, respectively. This is computed for both GCM trajectories and ERA5.

\item \textbf{rf calculation.}  
The daily relative frequency (\textbf{rf}) for a specific $\text{WT}=i$ is computed at each grid point as the sum of the \textbf{joint rf} over all previous-day states $\text{WT}=j$. This represents the estimated probability of occurrence of $\text{WT}=i$.

\item \textbf{Conditional RF calculation.}  
The conditional relative frequency (\textbf{cond rf}) of $\text{WT}=i$ given that $\text{WT}=j$ occurred on the previous day at grid point $s$ is then computed. This provides a measure of transition dynamics and persistence.

\item \textbf{Similarity calculation.}  
Using either the \textbf{rf} or the \textbf{cond rf} for a selected set of WTs, similarity measures are computed between each GCM trajectory and ERA5 at each grid point $s$.

\item \textbf{Filter.}  
A filter is applied to the similarity fields derived from \textbf{rf} and \textbf{cond rf}. Only GCM trajectories that achieve a similarity $\geq t_{sim}$ in at least $k$ grid points are retained. $t_{sim}$ and $k$ are predefined thresholds.

\item \textbf{Summary metrics.}  
A ranking of the surviving trajectories is performed using three metrics that integrate results across the $N_S$ grid points. 

The first, Daily Reproduction (DR), quantifies agreement in daily frequencies through the sum of overlap values. The second, Conditional Reproduction (CR), evaluates agreement in transition dynamics using conditional frequencies with local and regional weighting. The third, Persistence Reproduction (PerR), measures how well trajectories reproduce the persistence of WT over consecutive days.

\end{itemize}

The steps of the procedure are summarized in Figure \ref{fig:algorithm}, where the elements that must be selected by the researcher are: $N_S = 30$, $N_{WT} = 27$, $t_{sim} = 0.8$, and $k= 10$. It should be noted that the criteria for ranking trajectories require the researcher to choose measures that combine similarity across all grid points.

Several methodological challenges arise in this framework. One challenge is the selection of the most appropriate distance measure to compare the daily and conditional distributions of ERA5 and a given GCM. In addition, thresholds for what constitutes an acceptable reproduction of these distributions must be defined.

Another challenge concerns the sample size available for estimating conditional distributions. In the Spanish database, the daily distribution is estimated from a sample of 3,294 days (122 summer days per year $\times$ 27 years). However, when conditioning on a specific WT on the previous day, the effective sample size is reduced. For instance, if a selected WT occurs at a grid point with a relative frequency of 0.01, the corresponding conditional distribution across 27 categories would be estimated using only about 32 data points. 

Therefore, the influence of infrequent WTs on the comparison between distributions must be carefully considered. In particular, a minimum relative frequency for both daily and conditional distributions should be established to ensure robust and meaningful comparisons.

\subsection{Measures for Comparing Distributions of Categorical Variables}

The comparison of categorical distributions is a well-documented challenge in the literature. \citet{alamuri2014} provide a comprehensive taxonomy of similarity measures designed specifically for this purpose, while \citet{sekhon2011multivariate} demonstrates the practical necessity of comparing such distributions to ensure balance and comparability between different datasets.

In this study, we consider three commonly used measures for comparing categorical distributions. Two of these are similarity measures: the Overlap coefficient and the Bhattacharyya coefficient. The third measure is a distance metric, the Hellinger distance.

Let a categorical variable have $J$ categories, with probability distributions $\mathbf{p}_1 = (p_{1,1},\dots,p_{1,J})$ and $\mathbf{p}_2 = (p_{2,1},\dots,p_{2,J})$, such that $\sum_{j=1}^J p_{1,j} = \sum_{j=1}^J p_{2,j} = 1$.

\textbf{Overlap coefficient.} The overlap between two distributions is defined as:
$$
O_{12} = \sum_{j=1}^J \min(p_{1,j}, p_{2,j}).
$$
The closer $O_{12}$ is to 1, the more similar the distributions are.

\textbf{Dissimilarity index.} The Dissimilarity Index is defined as
$$
\Delta_{12} = \frac{1}{2} \sum_{j=1}^J |p_{1,j} - p_{2,j}|.
$$
Since it is linearly related to the overlap, $\Delta_{12} = 1 - O_{12}$, it is redundant in practice.

\textbf{Bhattacharyya coefficient.} This coefficient is given by
$$
B_{12} = \sum_{j=1}^J \sqrt{p_{1,j} \, p_{2,j}},
$$
with values closer to 1 indicating more similar distributions.

\textbf{Hellinger distance.} The Hellinger distance is defined as
\begin{align*}
d_{H,12} &= \frac{1}{\sqrt{2}} \sqrt{\sum_{j=1}^{J} \left( \sqrt{p_{1,j}} - \sqrt{p_{2,j}} \right)^2 } \\
         &= \sqrt{1 - \sum_{j=1}^{J} \sqrt{p_{1,j} \, p_{2,j}}} \\
         &= \sqrt{1 - B_{12}},
\end{align*}
where values closer to 0 indicate more similar distributions.
Note that the conditional probabilities can be reasonably estimated (provided the sample size is sufficiently large) when the condition is defined for some of the most frequent WT. However, conditioning on less frequent WT results in a smaller available sample size, and therefore, the conditional distribution relies on more limited information.

\subsection{Regional Summary Metrics for GCM WT Representation}

GCM trajectories are compared across the grid points of the ROI (in our case Iberian Peninsula) using the selected distance measure, and a final ranking of the best-performing trajectories is provided. The spatial variability of the region is explicitly considered in these comparisons.

To summarize the reproduction of daily distributions across the Iberian Peninsula, we define the daily reproduction score ($DR$), which is based on a selected distance measure. For instance, if the chosen measure is the overlap, then for a given grid point $s$, the overlap between the ERA5 daily distribution and the distribution associated with trajectory $n$ is denoted by $O_{ERA,n}(s)$. The daily reproduction score for GCM trajectory $n$, $DR(n,S)$, is then defined as the sum of these overlap values over all grid points $s$ within the region:

\begin{equation}
DR(n,S) = \sum_{s\in S} O_{\text{ERA},n}(s),
\end{equation} 

If the overlap values are always summed over all grid points $s$ in the region, then the dependence on $S$ becomes implicit, and the notation $DR(n, S)$ can be simplified to $DR(n, \cdot)$. This applies to all equations. However, one may wish to use a reduced set of grid points; therefore, we will retain the notation $S$ in the following equations.

To summarize the ability of a GCM trajectory to reproduce persistence and transitions between days (i.e., conditional distributions), we define a set of scores. These scores differ according to the weighting assigned to the distance measure at each grid point: one score applies local weights at each grid point, while another applies a regional weighting across all grid points. In addition, the scores are computed using either the full set of weather types (WTs) or a subset of selected relevant WTs.

For example, using the overlap measure, the conditional reproduction score with local weights and considering all WTs, denoted by $CR_{\text{loc}}(n,S)$, is defined based on $O_{\text{ERA},n}(s \mid j)$, which represents the overlap at grid point $s$ conditioned on the weather type on the previous day being $\text{WT} = j$. Let $rf_{\text{ERA}}(j; s)$ denote the relative frequency of $\text{WT} = j$ at grid point $s$. The expression is given by:

\begin{equation}
    CR_{loc}(n,S) = \sum_{j \in WT} \sum_{s \in S} O_{\text{ERA},n}(s \mid j) \cdot rf_{\text{ERA}}(j; s)
\end{equation}

The score $CR_{\text{loc}}(n,S)$ is computed using conditional distributions for all weather types (WTs). A variant, denoted by $CR_{\text{loc}}^{*}(n,S)$, is instead calculated using a subset of the most relevant WTs. The selection of these relevant WTs is left to the researcher and depends on the chosen criterion; for instance, one may focus on the most frequent WTs within the region of interest.

In this work, given the objective of studying the dynamics, the selection criterion is based on ensuring a sufficient sample size to estimate conditional transition frequencies, i.e., imposing a minimum relative frequency for inclusion.  


Other options have been considered to identify the strengths and weaknesses of GCM trajectories. The score $CR_{reg}(n,S)$ is defined considering a regional weight of each WT is defined avoiding situations where \textbf{rf} of a WT displays high variability in the region, to adress this, for a given WT, we calculate the median of its \textbf{rf} across all grid points as a summary of the distributions across the region, in order to quantify its impact. For example, using the overlap measure, we define the following expression for a specific WT: 
$$
CR(n,S|j) = \sum_{s \in S} O_{\text{ERA},n}(s \mid j),
$$

where the sum is over all grid points $s$ in the region. Based on these indices, a regional summary for trajectory $n$ is computed as:

\begin{equation}
    CR_{reg}(n,S) = \sum_{j \in WT} CR(n,S|j) 
    \times \frac{\mathrm{median}\big(rf_{\text{ERA}}(j,s),\, s \in S\big)}{\sum_{i\in WT} \mathrm{median}\big(rf_{\text{ERA}}(i,s),\, s \in S\big)}.
\end{equation}

A specific index is also defined to evaluate the reproduction of persistence, i.e., the conditional relative frequency that the WT remains the same from one day to the next:

\begin{equation}
    PerR(n,S) = \sum_{i \in WT} \sum_{s \in S} \big\| rf_{n}(i;s \mid i) - rf_{\text{ERA}}(i;s \mid i) \big\|
\end{equation}

where $rf_n(i\mid i;s )$ is the GCM conditional relative frequency for persistence, and $rf_{\text{ERA}}(i \mid i;s)$ is the corresponding ERA5 value.

Corresponding versions $CR_{reg}^*(n,S)$ and $PerR^*(n,S)$ are computed using a selected set of relevant WTs, denoted by $WT^*$. Trajectories can also be compared using the number of grid points where the overlap of the conditional distribution exceeds a threshold (e.g., 0.8).  

Note that $DR$ and the various $CR$ indices are defined as sums over $N$, the total number of grid points. Dividing these indices by $N$ produces normalized versions in the interval $[0,1]$, allowing comparison between regions of different spatial extent.

To compare $rf$ and conditional $rf$ of WTs between ERA5 and a selected GCM, the \textit{comp.prop()} function from the R package \textit{MatchIn} is used. This function compares two distributions of the same categorical variable and returns four different measurements, each taking values between 0 and 1.

\section{Results}

\subsection{Evaluation of Similarity Measures and WT Selection Strategies}

This section evaluates the measures introduced in Section~\ref{sec:methods} for comparing distributions. First, the performance of several similarity measures is evaluated in order to determine which ones are most informative for comparing GCM trajectories. In addition, the effect of infrequent WT on the behavior of these similarity measures is examined.

\subsubsection{Measures of Similarity Between Distributions}

\begin{figure}[h!]
    \centering
    \includegraphics[width=0.8\linewidth]{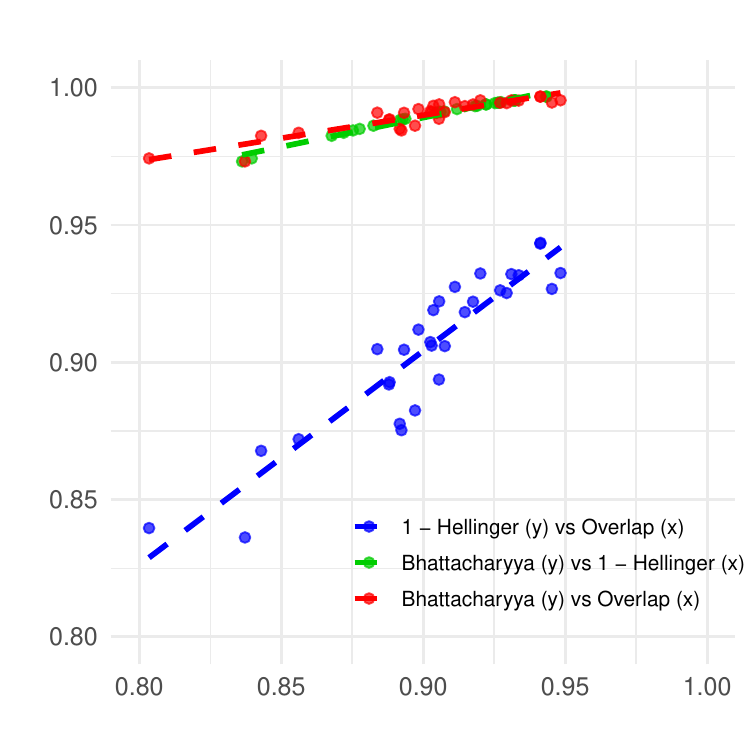}
    \caption{Overlap, Bhattacharyya coefficient, and \(1-\)Hellinger distance between the \textbf{rf} of the ERA5 trajectory and the GCM trajectory \texttt{ec\_earth3\_aerchem} across the Iberian Peninsula grid points. The measures are plotted against one another.}
    \label{fig:comp.distances}
\end{figure}

We compare the behavior of three similarity measures---Overlap, the Bhattacharyya coefficient, and \(1-\)Hellinger distance---in order to determine the most appropriate measure for comparing GCM trajectories.  

    For this purpose, each measure is calculated between the \textbf{rf} of the ERA5 trajectory and three different GCM trajectories. Results using the trajectory \texttt{ec\_earth3\_aerchem} are presented here for illustration, while results for the remaining trajectories are provided in Appendix \ref{sec:dist:comp:gcm}.

Figure \ref{fig:comp.distances} compares the three measures by plotting them against each other across 30 grid points. The Bhattacharyya coefficient exhibits very limited variability, taking values only between approximately 0.95 and 1. 

When selecting a similarity measure for model comparison, it is desirable to use a metric that can clearly discriminate between similar and dissimilar distributions. This requires sufficient variability in the values of the measure. Because the Bhattacharyya coefficient shows very little variation, it provides limited discriminatory power in this context. For this reason, it is discarded and the analysis focuses on comparing the Overlap measure and \(1-\)Hellinger distance.

\subsubsection{Selection of WTs for Distance Calculations}

Some WT categories occur very infrequently, which may affect the reliability of similarity measures. Therefore, it is important to assess whether all WT categories should be included when computing distances between distributions.

\begin{figure}
    \centering
    \includegraphics[width=0.49\linewidth]{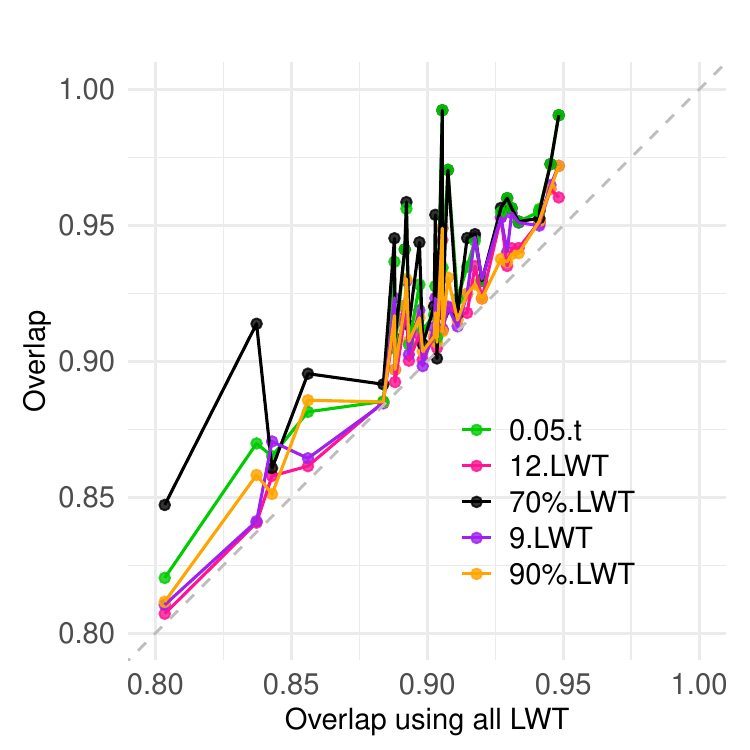}
    \includegraphics[width=0.49\linewidth]{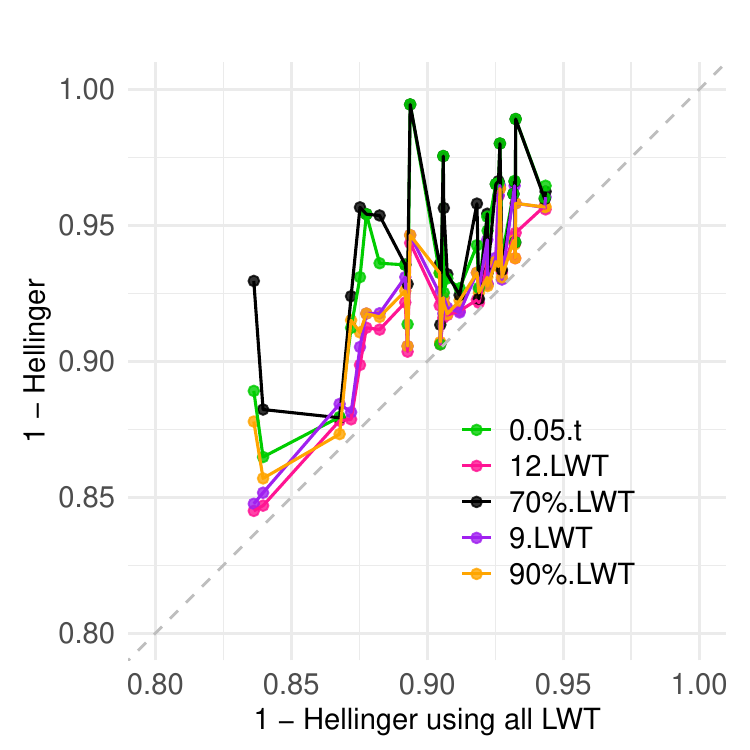}
    \caption{Comparison of different strategies for selecting WT categories in distance calculations. Overlap (top) and \(1-\)Hellinger (bottom) values obtained with each strategy are plotted against the default method, which includes all WT categories.}
    \label{fig:diff.calc.methods}
\end{figure}

To investigate this, the Overlap and \(1-\)Hellinger measures are calculated between the \textbf{rf} of the ERA5 trajectory and the GCM trajectory \texttt{ec\_earth3\_aerchem} using several alternative strategies for selecting the WT categories included in the calculations. The following approaches are considered:

\begin{itemize}
\item Including all WT categories (default approach);
\item Including WT categories that cumulatively account for at least 70\% or 90\% of the ERA5 distribution;
\item Including only the most frequent 9, or 12 WT in ERA5;
\item Including WT with \textbf{rf} greater than 0.05 in ERA5 (labeled as 0.05.t in Figure~\ref{fig:diff.calc.methods}).
\end{itemize}

Note that the selection of WT categories is evaluated at each grid point and therefore varies, as the most frequent categories differ across points.

Figure \ref{fig:diff.calc.methods} compares the different approaches for selecting WT categories by plotting the resulting similarity values relative to those obtained when all WT categories are included. The objective is to identify subsets of WT that produce results comparable to the full set while reducing the number of categories, to simplify the analysis and focus on the most frequent types. Additionally, because some WT categories are very infrequent and have \textbf{rf} close to zero, we examine how sensitive the distance measures are to the presence of such near-zero values.

When only WT with \textbf{rf} greater than 0.05, or those cumulatively accounting for 70\% of the ERA5 distribution, are included, some grid points contain very few categories. This limited representation produces substantial differences in Overlap values compared to the full set, indicating that these approaches are unreliable and that the measure may be sensitive to the absence of certain categories.

By contrast, similar Overlap values are obtained when using all 27 WT categories, the most frequent 12 or 9 categories, or the set accounting for 90\% of the ERA5 distribution. This indicates that the Overlap measure is largely insensitive to the inclusion of infrequent categories, supporting the use of all WT categories in the calculations.

The behavior of \(1-\)Hellinger differs somewhat. This measure displays lower variability across grid points, limiting its ability to distinguish between well-performing and poorly performing models. Furthermore, when all WT categories are included, the presence of many near-zero \textbf{rf} values makes the Hellinger-based measure highly sensitive, leading to increased variability across grid points.

Overall, the Overlap measure is more suitable for comparing daily distributions, as it exhibits greater variability and is less affected by infrequent categories. Consequently, the Overlap coefficient is selected as the similarity measure for comparing GCM trajectories, and all WT categories are included in the calculations.

Analogous figures obtained using other GCM trajectories are presented in Appendix \ref{sec:dist:comp:gcm}. Similar conclusions are reached: approaches based on a small subset of WT categories (e.g., cumulative 70\% or \textbf{rf} greater than 0.05) show high volatility, whereas approaches including most WT categories produce results very close to those obtained when all categories are used.

Figures comparing conditional distributions given the WT of the previous day are presented in Appendix \ref{sec:dist:comp:cond}. Similar patterns are observed when conditioning on frequent types such as U or PA. However, when conditioning on less frequent types (e.g., PC or PDNE), the results are less stable due to the smaller sample size available for estimating the conditional distributions.

\subsection{Comparative Performance in Daily and Transition Distribution}

This section details the selection procedure used to identify GCM trajectories capable of reproducing climate patterns over the Iberian Peninsula. We evaluated 36 trajectories based on their ability to replicate the daily and conditional distribution of WT. 

\subsubsection{Daily Distributions}

Daily WT distributions are obtained from the full set of 3,294 days for both ERA5 and each GCM. Similarity between the ERA5 and GCM distributions is quantified using the Overlap coefficient, $O_{ERA,n}(s)$, which ranges from 0 (completely different) to 1 (identical distributions).

\begin{figure}[h!]
    \centering
    \includegraphics[width=1\linewidth]{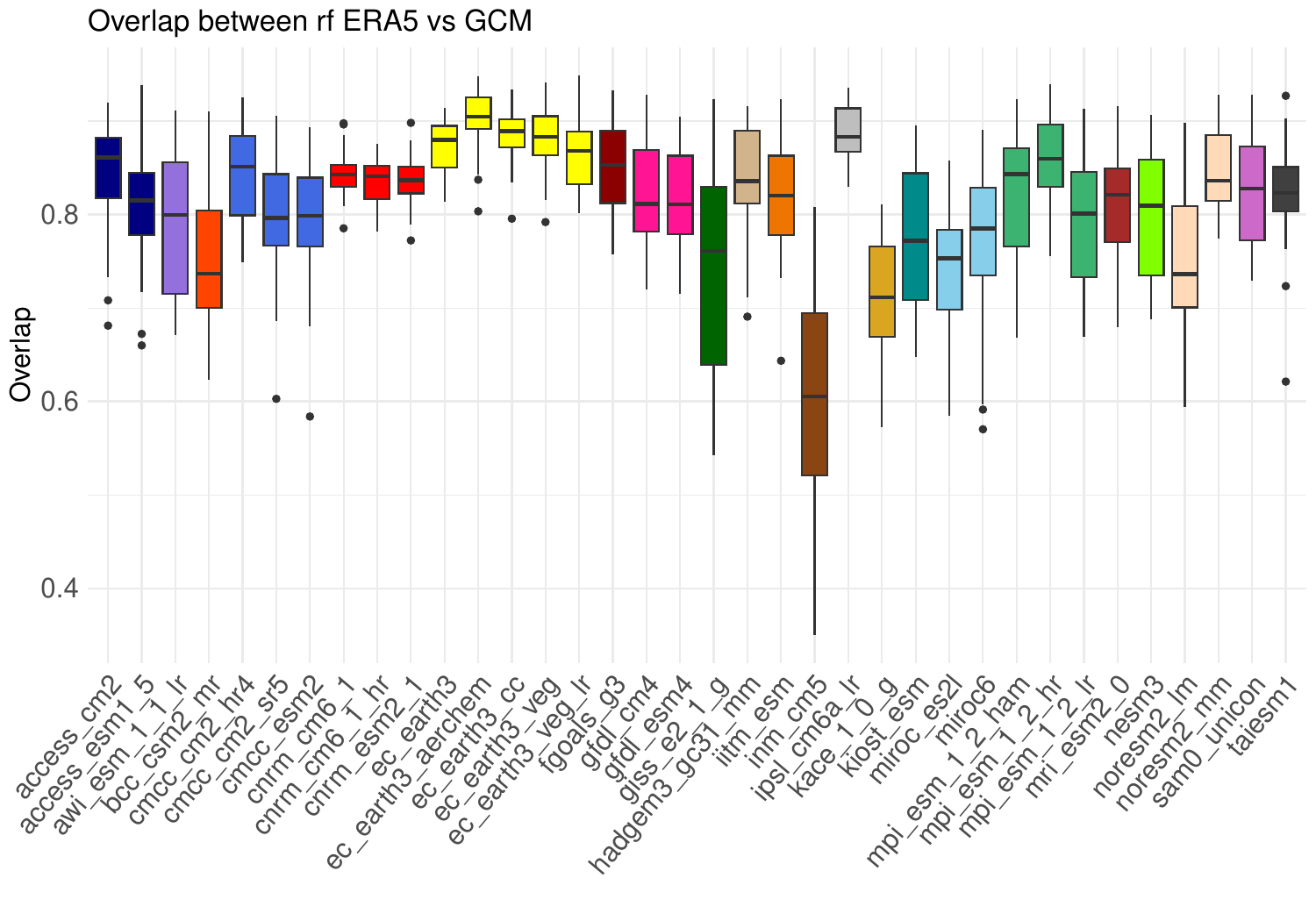}
    \includegraphics[width=0.7\linewidth]{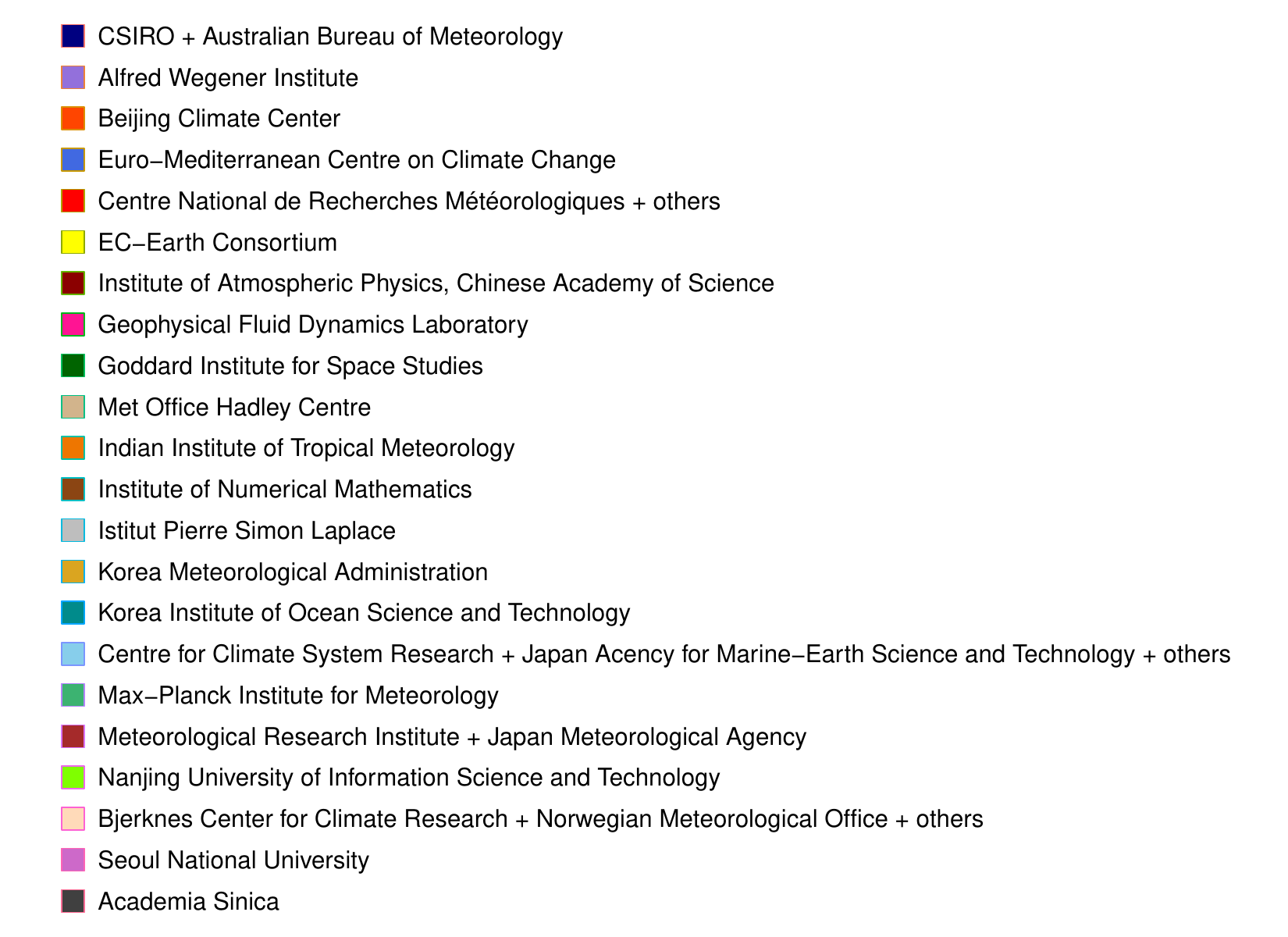}
    \caption{Boxplots of Overlap values between ERA5 \textbf{rf} and 36 GCM trajectories across 30 Iberian Peninsula grid points. Boxplots are colored by institute. The x-axis indicates the GCM trajectory names.}
    \label{fig:box.per.gcm}
\end{figure}

GCM performance is assessed across the 30 grid points of the Iberian Peninsula. 
Figure \ref{fig:box.per.gcm} shows boxplots of Overlap values for each GCM, colored by institute. Models from the \texttt{ec\_earth3} family generally perform best, while the trajectory \texttt{inm\_cm5} consistently ranks lower across grid points.

To consider a GCM trajectory as adequately reproducing the daily distribution, fewer than one-third of the grid points (i.e., fewer than 10) should have $O_{ERA,n}(s) \leq 0.80$. Applying this criterion eliminates 20 of the 36 GCM trajectories. The remaining 16 trajectories reproduce the daily distribution to varying degrees and will be the focus of the subsequent analysis.

Table \ref{tab:best_performers} highlights the eight best-performing GCM trajectories based on the 25\%, 10\%, and 5\% quantiles of the overlap, as well as the highest minimum overlap value. A score of 1 indicates that a trajectory ranks among the top eight for a given metric, and 0 otherwise. Additional metrics reported include the daily reproduction score (DR) and the number of grid points with $O_{ERA,n}(s) \le 0.8$.

\begin{table}[h!]
\centering
\caption{Summary of the 16 surviving GCM trajectories (those with fewer than one-third of grid points having Overlap below 0.8). The table indicates the eight top-performing trajectories for the 25\%, 10\%, and 5\% quantiles, as well as the minimum Overlap values. A value of 1 denotes that the trajectory ranks among the top eight for that metric, and 0 otherwise. DR denotes the daily reproduction score, and $\#O \leq 0.8$ indicates the number of grid points with Overlap $\leq 0.8$.}
\begin{tabular}{rcccccc}
\hline
Trajectory & Q25 & Q10 & min & $\#O \leq 0.8$ & DR \\ 
  \hline
  access\_cm2 & 0 & 0 & 0 & 7 & 25.27 \\ 
  cmcc\_cm2\_hr4 & 0 & 0 & 0 & 8 & 25.37 \\ 
  cnrm\_cm6\_1 & 1 & 1 & 1 & 1 & 25.34 \\ 
  cnrm\_cm6\_1\_hr & 0 & 0 & 1 & 4 & 25.01 \\ 
  cnrm\_esm2\_1 & 0 & 0 & 0 & 4 & 25.09 \\ 
  ec\_earth3 & 1 & 1 & 1 & 0 & 26.22 \\ 
  ec\_earth3\_aerchem & 1 & 1 & 1 & 0 & 27.06 \\ 
  ec\_earth3\_cc & 1 & 1 & 1 & 1 & 26.58 \\ 
  ec\_earth3\_veg & 1 & 1 & 1 & 1 & 26.47 \\ 
  ec\_earth3\_veg\_lr & 1 & 1 & 1 & 0 & 25.92 \\ 
  fgoals\_g3 & 0 & 0 & 0 & 6 & 25.48 \\ 
  hadgem3\_gc31\_mm & 0 & 0 & 0 & 6 & 25.06 \\ 
  ipsl\_cm6a\_lr & 1 & 1 & 1 & 0 & 26.61 \\ 
  mpi\_esm\_1\_2\_hr & 1 & 1 & 0 & 4 & 25.77 \\ 
  noresm2\_mm & 0 & 0 & 0 & 5 & 25.36 \\ 
  taiesm1 & 0 & 0 & 0 & 7 & 24.61 \\ 
\hline
\end{tabular}
\label{tab:best_performers}
\end{table}

From Table \ref{tab:best_performers}, the trajectories \texttt{cnrm\_cm6\_1}, \texttt{cnrm\_cm6\_1\_hr}, \texttt{cnrm\_esm2\_1}, \texttt{ec\_earth3},\texttt{ec\_earth3\_aerchem}, \texttt{ec\_earth3\_cc}, \texttt{ec\_earth3\_veg},\texttt{ec\_earth3\_veg\_lr}, \texttt{ipsl\_cm6a\_lr}, and \texttt{mpi\_esm\_1\_2\_hr} score 1 in at least one of the summary values. 


The best daily reproduction--defined as having Overlap values $\le 0.8$ in four or fewer grid points--occurs in three \texttt{cnrm} trajectories (1, 4, and 4), five \texttt{ec\_earth3} trajectories (0, 0, 1, 1, 0), and in \texttt{ipsl\_cm6a\_lr} (0) and \texttt{mpi\_esm\_1\_2\_hr} (4). Trajectories with Overlap values exceeding 0.8 in 5–9 grid points include \texttt{acces\_cm2} (7), \texttt{cmcc\_cm2\_hr4} (8), \texttt{fgoals\_g3} (6), \texttt{hadgem3\_gc31\_mm} (6), \texttt{noresm2\_mm} (5), and \texttt{taiesm1} (7). 


\subsubsection{Transition Distributions}

The ability of the GCM trajectories to reproduce conditional distributions is also analyzed. 
In this case, we consider the distribution of WT given that the observed WT on the previous day was a specific type $j$, where $j$ corresponds to a WT of interest.

For the Iberian Peninsula in summer, the conditioning WT were selected based on two criteria: the WT must be climatologically relevant, and frequent enough to allow reliable estimation of the conditional distribution. Based on these criteria, we consider anticyclonic (PA), cyclonic (PC), pure North East directional (PDNE), and undefined (U) types. From Figure \ref{fig:box.marg}, these categories have sufficient data: for at least half of the grid points, the sample size exceeds 155 days, and for 80\% of the grid points, it exceeds 82 days. The minimum sample size per grid point is 210 for PA, 86 for PDNE, 31 for PC, and 320 for U.

Table~\ref{tab:cond:overlap:n} shows for our conditioning WT, the number of grid points with $O_{ERA,n}(s) \le 0.8$.

\begin{table}[h!]
\centering
\caption{$\#O \leq 0.8$ considering different conditional distributions. The first column it the trajectory being used, and the rest represent the conditioning WT.}
\label{tab:cond:overlap:n}
\begin{tabular}{lrrrr}
  \hline
Trajectory & $|PA$ & $|PC$ & $|PDNE$ & $|U$ \\ 
  \hline
access\_cm2 &   0 &  12 &   3 &   1 \\ 
  cmcc\_cm2\_hr4 &   2 &  14 &   3 &   1 \\ 
  cnrm\_cm6\_1 &   5 &   7 &  11 &   0 \\ 
  cnrm\_cm6\_1\_hr &   0 &   5 &   8 &   0 \\ 
  cnrm\_esm2\_1 &   6 &   6 &   7 &   0 \\ 
  ec\_earth3 &   1 &   7 &   0 &   0 \\ 
  ec\_earth3\_aerchem &   0 &   4 &   0 &   0 \\ 
  ec\_earth3\_cc &   1 &   6 &   1 &   0 \\ 
  ec\_earth3\_veg &   1 &   7 &   0 &   0 \\ 
  ec\_earth3\_veg\_lr &   1 &   8 &   1 &   0 \\ 
  fgoals\_g3 &   8 &   9 &   8 &   1 \\ 
  hadgem3\_gc31\_mm &   2 &   7 &   5 &   0 \\ 
  ipsl\_cm6a\_lr &   1 &  13 &   7 &   0 \\ 
  mpi\_esm\_1\_2\_hr &   0 &   6 &   2 &   0 \\ 
  noresm2\_mm &   0 &   6 &   2 &   0 \\ 
  taiesm1 &   8 &   9 &   2 &   3 \\ 
   \hline
\end{tabular}
\end{table}
\textbf{Conditioning on WT U}.  
The selected trajectories generally reproduce this conditional distribution well. 
Fourteen trajectories have no grid points with Overlap values below or equal to 0.80, while the trajectories \texttt{taiesm1} and \texttt{cmcc\_cm2\_hr4} have only one grid point below this threshold.

\textbf{Conditioning on WT PA}.  
Trajectories \texttt{taiesm1}, \texttt{cnrm\_esm2\_1} and \texttt{cnrm\_cm6\_1} show poorer reproduction, with 14, 11, and 10 grid points below or equal to 0.8, respectively.
Similarly, \texttt{cnrm\_cm6\_1\_hr} and \texttt{fgoals\_g3} do not perform satisfactorily, with eight and seven grid points below the threshold, respectively.

\textbf{Conditioning on WT PC}.  
The atmospheric situation defined by PC on the previous day is difficult to reproduce overall.
The trajectories with more than eight grid points having Overlap values below or equal to 0.80 are 
\texttt{access\_cm2}, \texttt{cmcc\_cm2\_hr4}, \texttt{fgoals\_g3},  \texttt{ipsl\_cm6a\_lr}, and \texttt{taiesm1}.

\textbf{Conditioning on WT PDNE}.  
Conditional distributions associated with PDNE are reproduced less accurately in some models. 
In particular, the \texttt{cnrm} trajectories show 8, 6, and 6 grid points with Overlap values below or equal to 0.80. 
\texttt{fgoals\_g3} struggles with this WT, and has 10 points with an Overlap below or equal to 0.8.

\subsubsection{GCM Filtering and Selection\label{sec:gcm:filter}}

The first filtering step evaluates whether a GCM trajectory adequately reproduces both the daily and conditional distributions. A trajectory is considered acceptable if, for a given set of conditioning weather types (WT) defined by the researcher (i.e., PA, PC, PDNE, and U), fewer than one-third of the grid points (i.e., fewer than 10) exhibit values of $O_{ERA,n}(s) \leq 0.80$. 

Applying this criterion to the daily distribution results in the rejection of 20 out of 36 GCM trajectories. Subsequently, when the same criterion is applied to the conditional WT distributions, 4 of the remaining 16 trajectories are further discarded. 

After this filtering process, a total of 12 trajectories remain for analysis in the following sections.

\subsection{Spatial and Regional Performance Patterns}




\subsubsection{Spatial Performance}

\begin{figure}[h!]

    \includegraphics[width=0.33\linewidth]{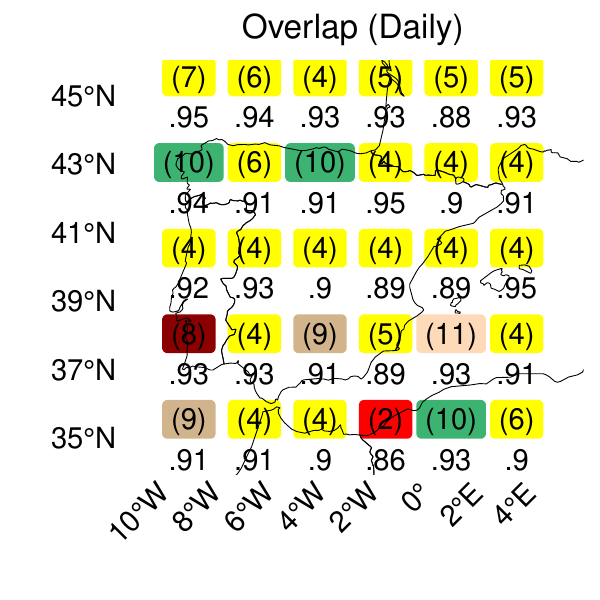}%
    \includegraphics[width=0.23\linewidth]{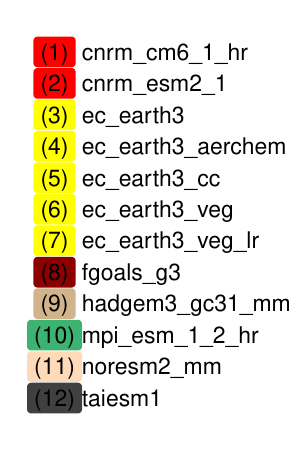}
    \includegraphics[width=0.33\linewidth]{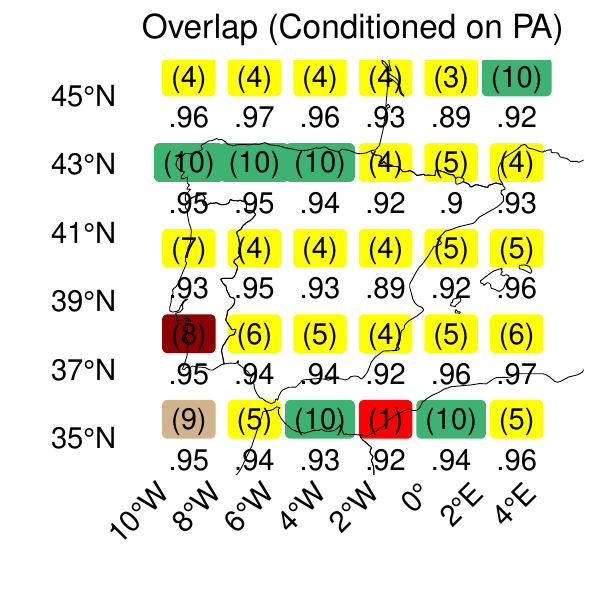}%
    
    \includegraphics[width=0.33\linewidth]{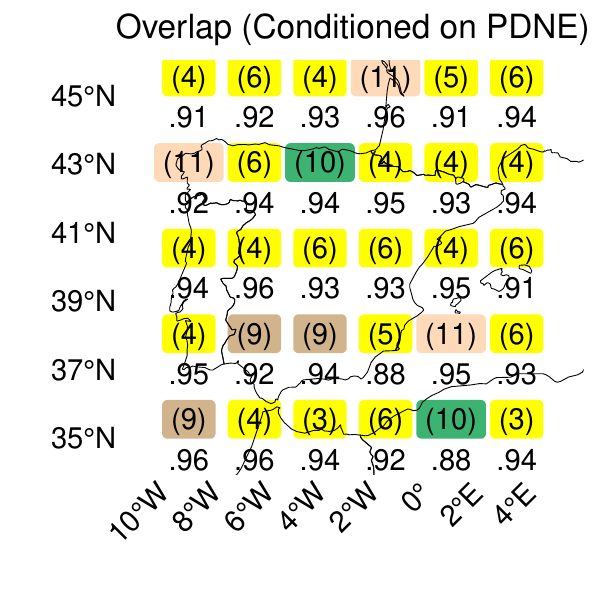}%
    \includegraphics[width=0.33\linewidth]{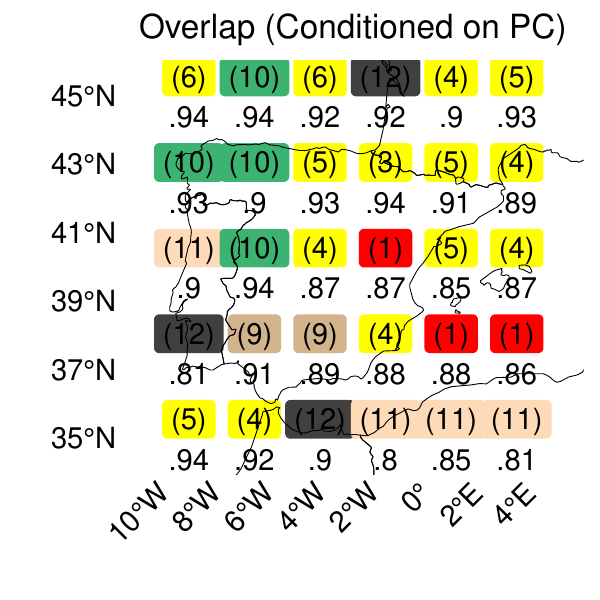}%
    \includegraphics[width=0.33\linewidth]{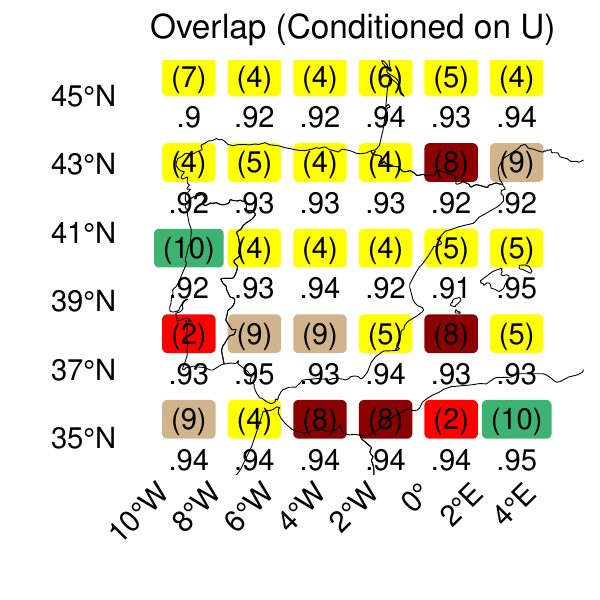}
    \caption{Comparison of the twelve best-performing GCM trajectories as identified in Section~\ref{sec:gcm:filter}. Each point represents the highest Overlap value between ERA5 and the twelve GCM trajectories, with the corresponding model indicated in brackets, and colored according to the institute the model belongs to. The first plot shows results for \textbf{rf}, while the remaining plots display \textbf{cond rf} values conditioned to WT from the summary set defined in Table \ref{tab:WT.reduced}.}
    \label{fig:comp}
\end{figure}

Figure~\ref{fig:comp} presents a comparison of the twelve best-performing GCM models (Section~\ref{sec:gcm:filter}). Each point shows the highest Overlap between ERA5 and the GCM trajectories, labeled by model (in brackets) and colored by institute. The first panel corresponds to rf, whereas the remaining panels show cond rf conditioned on WT from the reduced summary set (Table \ref{tab:WT.reduced}).

From Figure \ref{fig:comp}, it is evident that the models dominating the largest number of grid points--across different conditioning WT and daily distributions--consistently belong to the \texttt{ec\_earth3} family. However, not all trajectories within the \texttt{ec\_earth3} family perform equally well. The trajectories \texttt{ec\_earth3\_aerchem} and \texttt{ec\_earth3\_cc} dominate most points across both \textbf{rf} and \textbf{cond rf}. Specifically, \texttt{ec\_earth3\_aerchem} appears in 14 grid points for \textbf{rf}, and in 9, 10, 6, and 10 points when conditioning on PA, PDNE, PC, and U, respectively. For \texttt{ec\_earth3\_cc}, the corresponding counts are 4, 7, 2, 5, and 6. The trajectory \texttt{ec\_earth\_veg} is also relatively frequent. In contrast, \texttt{ec\_earth\_veg\_lr} (where ``lr'' denotes low resolution) and \texttt{ec\_earth3} are infrequent, each dominating only three grid points in total across all relative frequency types.

The conditioning WT showing the greatest variability in winning models is \textbf{PC} followed by \textbf{U}, where models from different families frequently dominate.

\texttt{mpi\_esm\_1\_2\_hr} primarily dominates north-western points, it is particularly relevant in this area when conditioning to PC and PA, where it dominates four and three north-western points, respectively. 

Models from the \texttt{cnrm} family dominate three south-eastern grid points when conditioning on PC, two southern points when conditioning on U, and one southern point when conditioning on PA using daily frequency. 

\texttt{fgoals\_g3} dominates one south-western point in the daily frequency under PA conditioning, and four south-eastern points under U conditioning. 

\texttt{hadgem3\_gc31\_mm} generally dominates some south-western grid points.

\texttt{noresm2\_mm} rarely dominates, but shows localized dominance in the south-east under PC conditioning and in three dispersed grid points under PDNE conditioning. 

Finally, \texttt{taiesm1} only dominates two south-western and one northern grid points under PC conditioning.

Overall, \texttt{ec\_earth3} dominates across most conditions, while the other models show more localized wins depending on the conditioning WT.



\subsubsection{Regional Performance}

\begin{figure}[h!]
    \centering
    \includegraphics[width=0.4\linewidth]{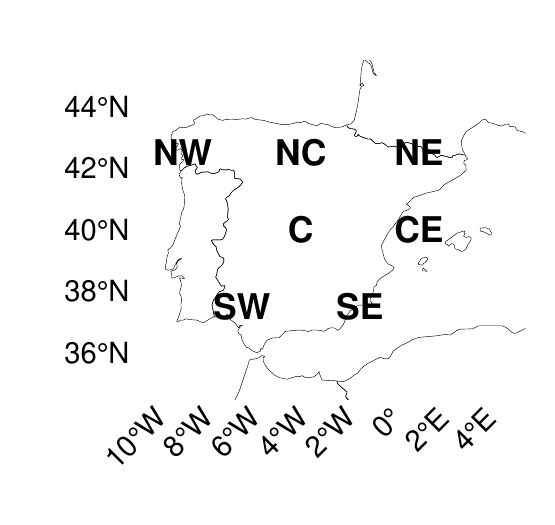}
    \caption{7 key points in Iberian Peninsula, and their label.}
    \label{fig:key.points.map}
\end{figure}

\begin{figure}[h!]
\centering

\begin{subfigure}{0.32\linewidth}
    \includegraphics[width=\linewidth]{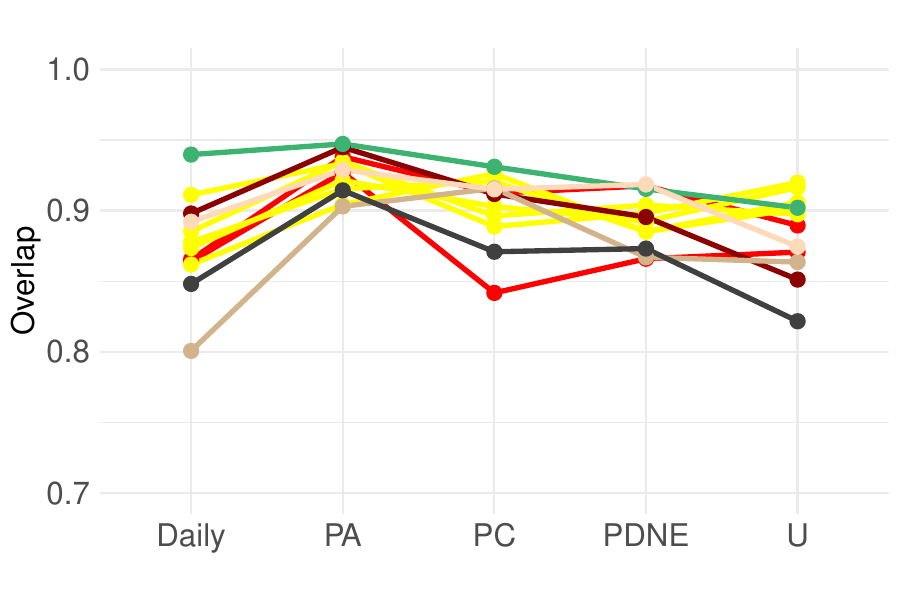}
\end{subfigure}
\begin{subfigure}{0.32\linewidth}
    \includegraphics[width=\linewidth]{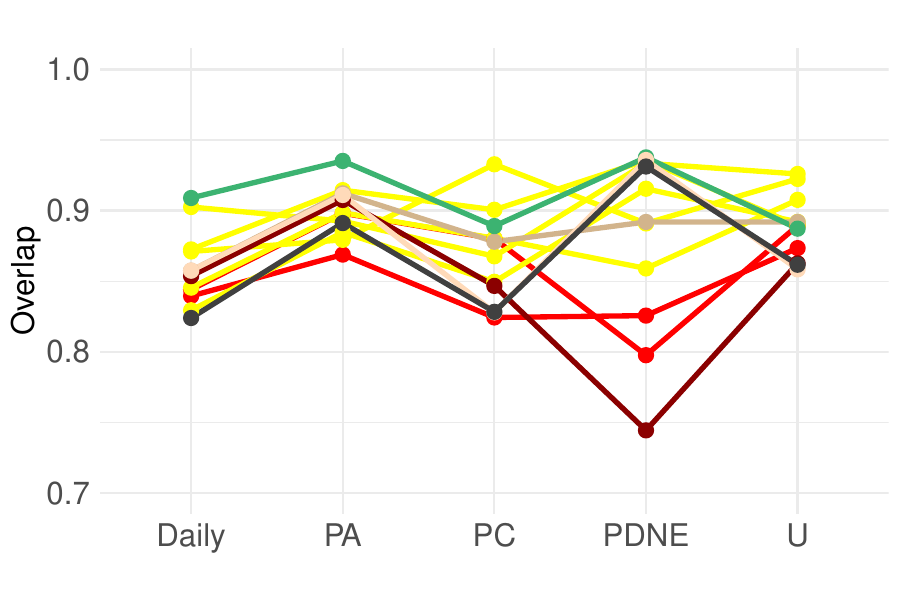}
\end{subfigure}
\begin{subfigure}{0.32\linewidth}
    \includegraphics[width=\linewidth]{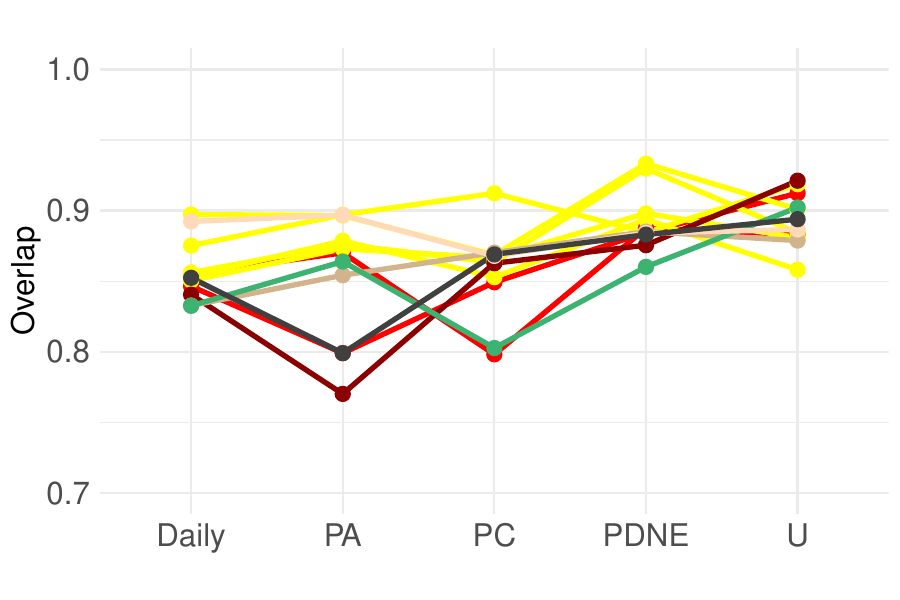}
\end{subfigure}

\vspace{0.3cm}

\begin{subfigure}{0.32\linewidth}
    \includegraphics[width=\linewidth]{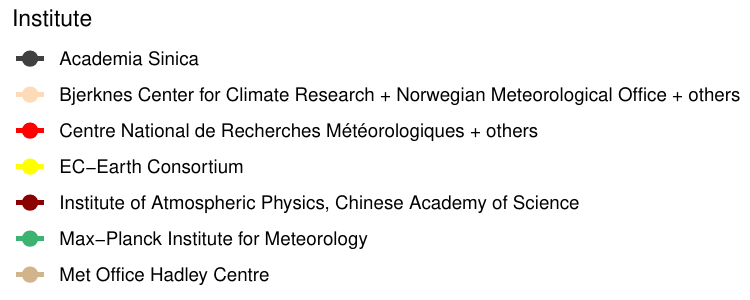}
\end{subfigure}
\begin{subfigure}{0.32\linewidth}
    \includegraphics[width=\linewidth]{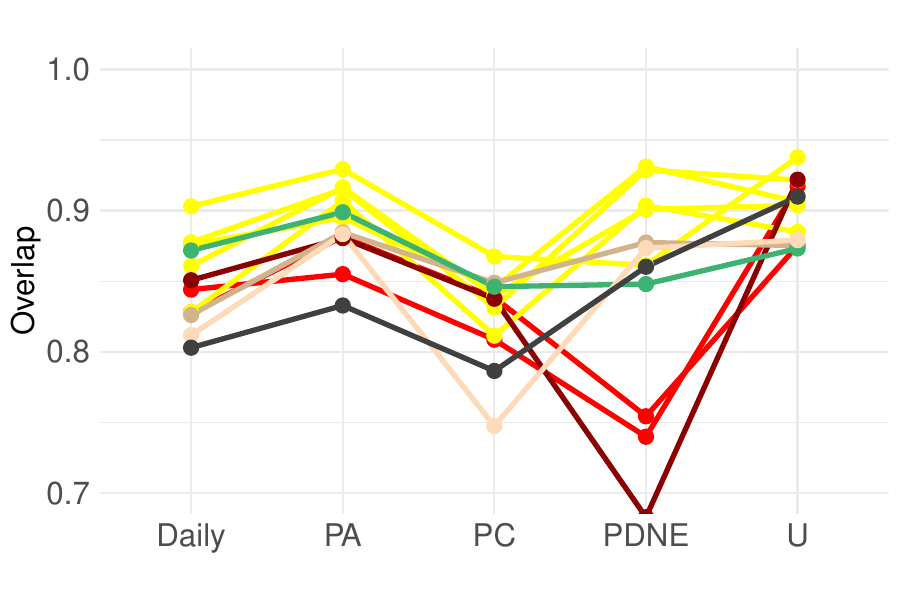}
\end{subfigure}
\begin{subfigure}{0.32\linewidth}
    \includegraphics[width=\linewidth]{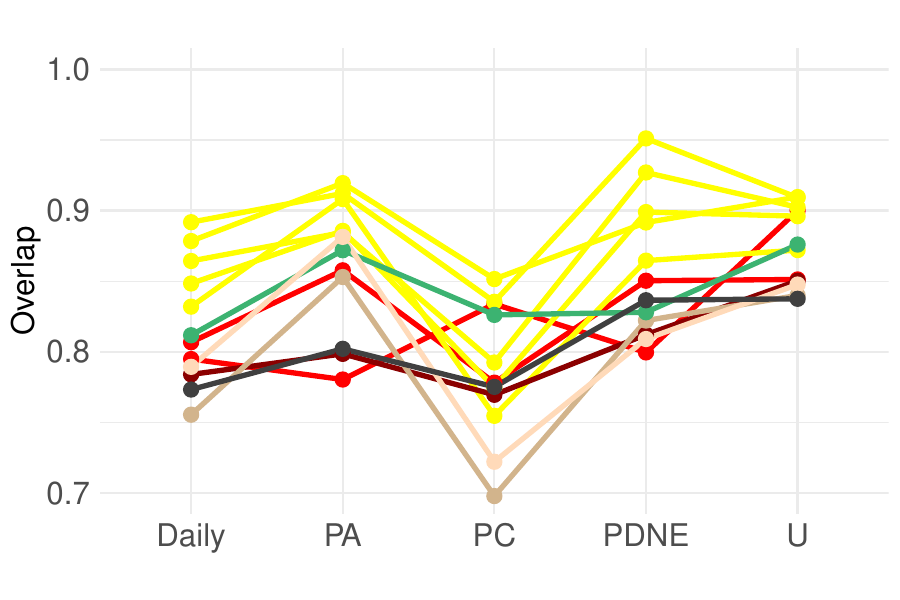}
\end{subfigure}

\vspace{0.3cm}

\begin{subfigure}{0.32\linewidth}
    \includegraphics[width=\linewidth]{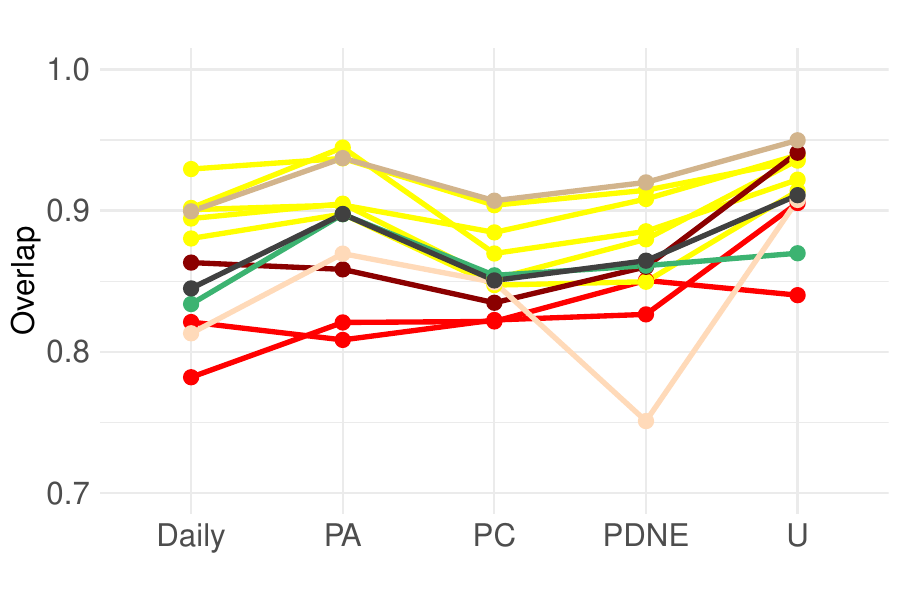}
\end{subfigure}
\begin{subfigure}{0.32\linewidth}
    \includegraphics[width=\linewidth]{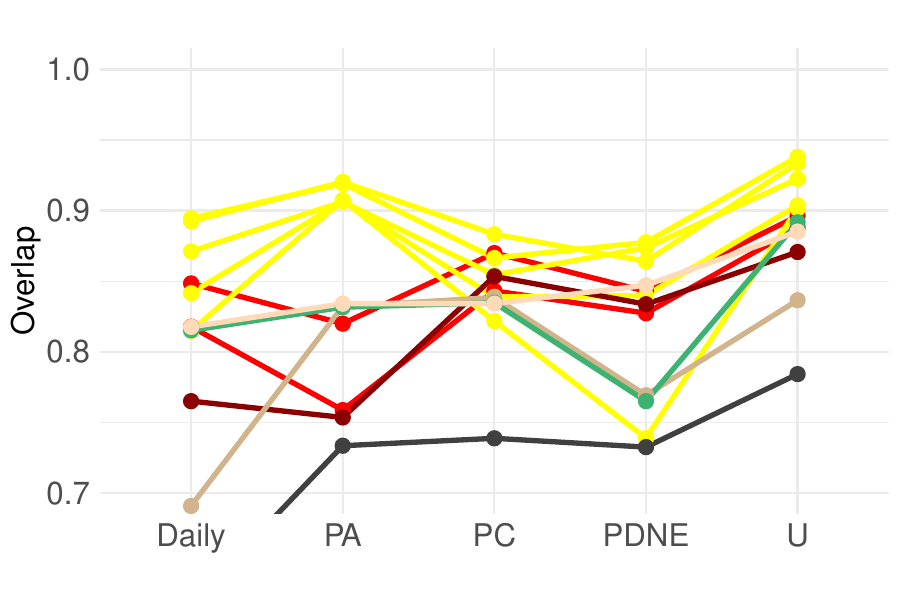}
\end{subfigure}

\caption{Overlap values between \textbf{rf} and \textbf{cond rf} conditioned to WT from Table \ref{tab:WT.reduced}, comparing the ERA5 trajectory and twelve winning GCM trajectories, at specific points across the Iberian Peninsula. The points are shown in Figure \ref{fig:key.points.map} and are in the same positions across the plots. Each line in each plot corresponds to one of the GCM trajectories, color-coded according to the institute they belong to. A legend with the institute names is included.}
\label{fig:seven.key.points}
\end{figure}

The selected twelve GCM trajectories    
are revised in order to evaluate different patterns across the Iberian Peninsula.
We selected seven points, shown in Figure \ref{fig:key.points.map}, six located in Peninsular Spain and one in the Mediterranean sea. We analyzed the behavior of these GCMs at these points using \textbf{rf} and \textbf{cond rf}, conditioned on WT from the summary set defined in Table \ref{tab:WT.reduced}. This is shown in Figure \ref{fig:seven.key.points}.

\begin{figure}
    \centering
    \includegraphics[width=0.32\linewidth]{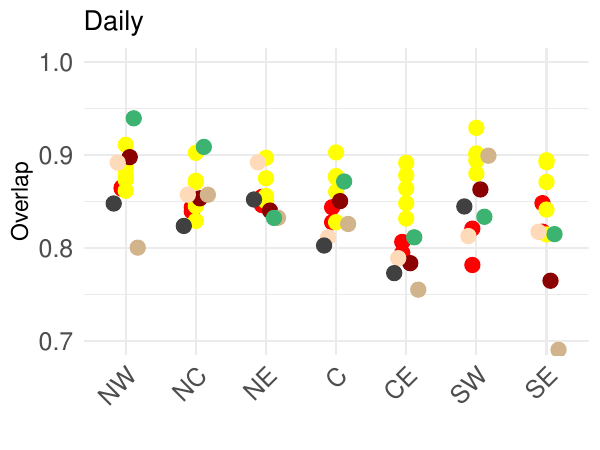}
    \includegraphics[width=0.32\linewidth]{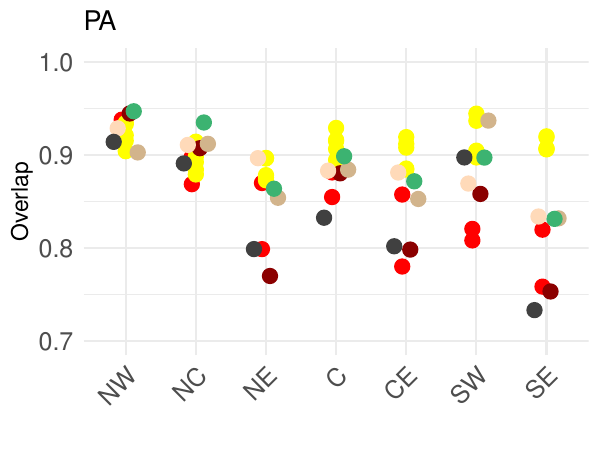}
    \includegraphics[width=0.32\linewidth]{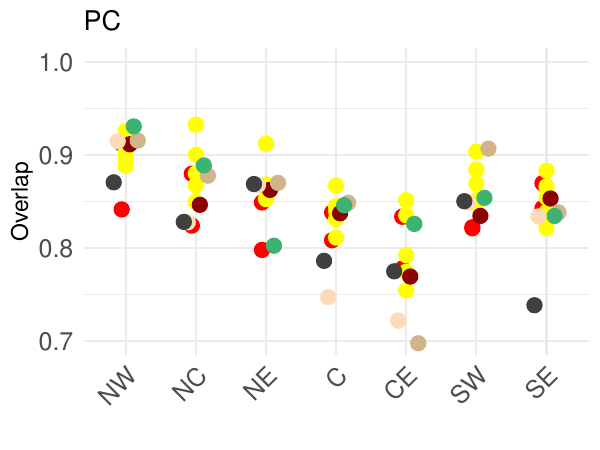}
    \includegraphics[width=0.32\linewidth]{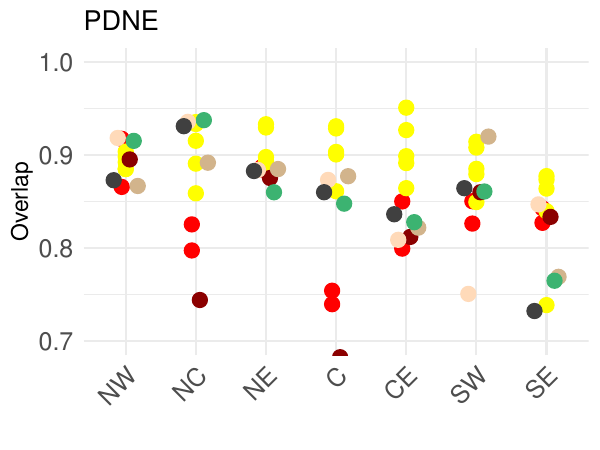}
    \includegraphics[width=0.32\linewidth]{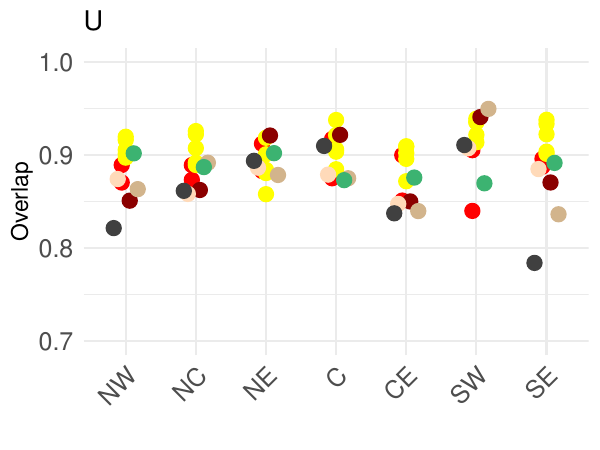}

    \caption{Overlap values between \textbf{rf} and \textbf{cond rf}, conditioned to WT from Table \ref{tab:WT.reduced}, comparing the ERA5 trajectory and twelve winning GCM trajectories, at specific points across the Iberian Peninsula. Each plot shows the overlap across the twelve winning GCM trajectories at the seven key points using \textbf{rf} or \textbf{cond rf}.}
    \label{fig:placeholder}
\end{figure}
From Figure \ref{fig:seven.key.points}, two main patterns emerge. First, a given GCM trajectory can perform very well at the north-western point while performing substantially worse at central, eastern, or southern locations. Second, the \textbf{rf} daily distribution generally shows good agreement across all points, whereas some \textbf{cond rf} distributions exhibit poorer reproduction in specific points.

The daily distributions (\textbf{rf}) are generally well represented, with overlap values $> 0.8$ for most trajectories across the northern points, the central point, and the south-western point (with the exception of one trajectory).  At the center-east and south-eastern points, overlap values remain $> 0.8$ for the \texttt{ec\_earth3}, \texttt{mpi\_esm\_1\_2\_hr}, and \texttt{cnrm} trajectories. However, at the south-eastern point, the \texttt{hadgem3\_gc31\_mm} and \texttt{taiesm1} trajectories show poorer performance, with values $< 0.7$.

For the conditional distribution given $\text{WT} =$ PA, most trajectories perform well overall; however, models outside the \texttt{ec\_earth3} family show weaker performance at some central and southern points.

For \textbf{cond rf} given $\text{WT} =$ PC, most points are well reproduced, except for the central-eastern point, where several trajectories fall below an overlap of 0.8.

For \textbf{cond rf} given $\text{WT} =$ PDNE, good agreement is observed at the northern, central, and south-western points for most trajectories. In contrast, the two \texttt{cnrm} trajectories and \texttt{fgoals\_g3} show the weakest performance at the north-central and central points. Some trajectories also struggle at the central-eastern and the south-eastern points.

Finally, for \textbf{cond rf} given $\text{WT} =$ U, the distributions are generally well represented across all seven points and trajectories.

The trajectory \texttt{taiesm1} performs poorly across both \textbf{rf} and \textbf{cond rf} at the south-western point. The trajectory \texttt{hadgem3\_gc31\_mm} also shows poor performance at this location, although primarily for \textbf{rf}. For this reason, these trajectories may be excluded from further consideration, as they are not well suited for local analyses. In this study, the focus is on trajectories that perform well both globally and locally.


\begin{figure}[h!]
    \centering
    \includegraphics[width=0.48\linewidth]{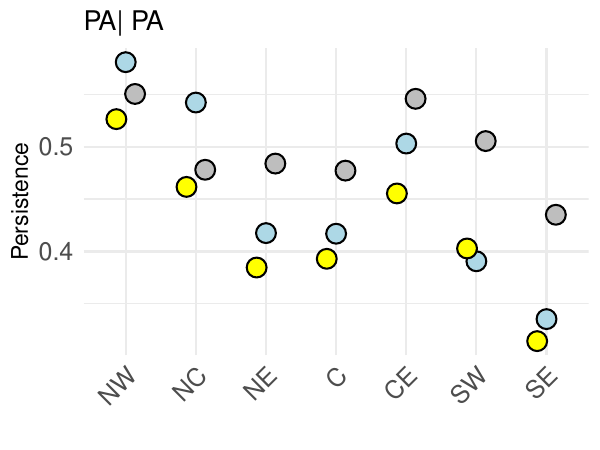}
    \includegraphics[width=0.48\linewidth]{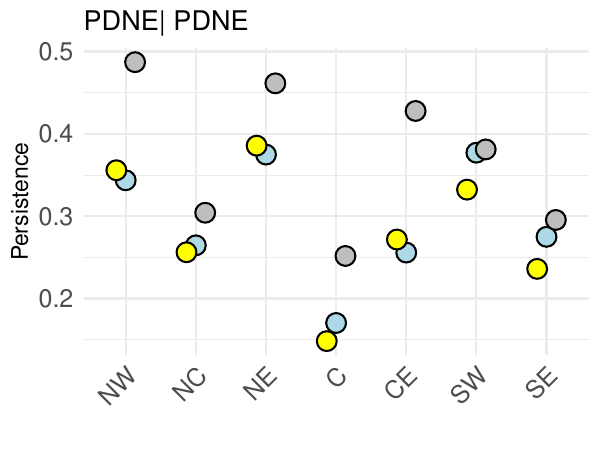}
    \includegraphics[width=0.48\linewidth]{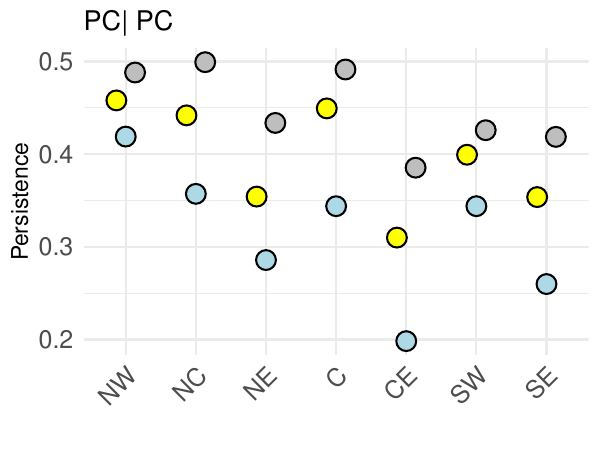}
    \includegraphics[width=0.48\linewidth]{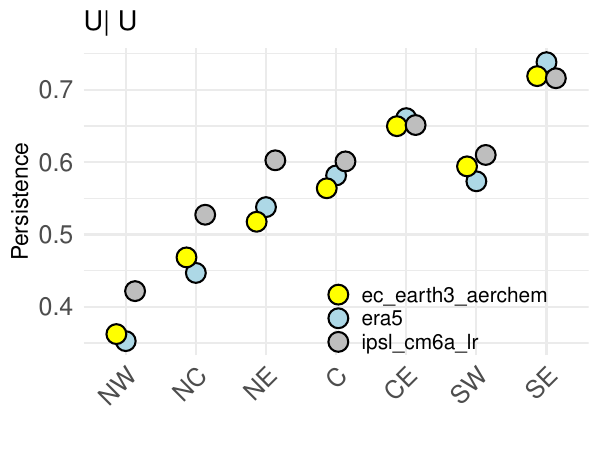}
    \caption{\textbf{per rf} in 7 key points in the Iberian Peninsula (x axis) from the ERA5 and two GCM trajetories. Each plot corresponds to a different WT.}
    \label{fig:placeholder}
\end{figure}

\FloatBarrier

\subsubsection{Evaluation of Performance Scores}

Table \ref{tab:final.ranking} ranks the  GCM trajectories using the similarity scores $DR$, $CR_{loc}$, $CR_{loc}^*$, $CR_{reg}$, $CR_{reg}^*$, and the discrepancy scores $PerR$ and $PerR^*$. 
Values with a superscript $^*$ correspond to a version computed using a selected subset of WT (PA, PDNE, PC, and U), while the unstarred values use the full WT set.


\begin{table}[ht]
\centering
\caption{Scores for 16 selected trajectories for $DR$, $CR_{loc}$, $CR_{loc}^*$, $CR_{reg}$, $CR_{reg}^*$, and $PerR$. Values with a superscript $^*$ correspond to a version computed using a selected subset of WT (PA, PDNE, PC, and U), while the unstarred values use the full WT set. The values in bold correspond to the highest performance in each score, while the ones in italics to the lowest.}
\begin{tabular}{lccccccc}
  \hline
Trajectory & $DR$ & $CR_{loc}$ & $CR_{loc}^*$ & $CR_{reg}$ & $CR_{reg}^*$ & $PerR$ & $PerR^*$ \\ 
  \hline
access\_cm2 & 25.27 & 25.88 & 17.23 & 25.43 & 17.88 & 45.86 & 7.33 \\ 
  cmcc\_cm2\_hr4 & 25.37 & 25.40 & 16.92 & 25.08 & 17.58 & 44.83 & 7.52 \\ 
  cnrm\_cm6\_1 & 25.34 & 25.68 & 17.17 & 25.21 & 17.74 & 51.67 & 8.07 \\ 
  cnrm\_cm6\_1\_hr & 25.01 & 25.57 & 16.92 & 25.19 & 17.62 & 47.02 & 7.81 \\ 
  cnrm\_esm2\_1 & 25.09 & 25.47 & 16.98 & 25.02 & 17.56 & 49.66 & 8.57 \\ 
  ec\_earth3 & 26.22 & 26.38 & 17.51 & 26.03 & 18.20 & 40.03 & 6.11 \\ 
  ec\_earth3\_aerchem & \textbf{27.06} & \textbf{26.81} & \textbf{17.78} & \textbf{26.44} & \textbf{18.49} & 38.84 & \textbf{5.09} \\ 
  ec\_earth3\_cc & 26.58 & 26.69 & 17.68 & 26.30 & 18.37 & 41.29 & 6.14 \\ 
  ec\_earth3\_veg & 26.47 & 26.51 & 17.57 & 26.12 & 18.24 & \textbf{35.59} & 5.77 \\ 
  ec\_earth3\_veg\_lr & 25.92 & 26.07 & 17.30 & 25.79 & 18.01 & 43.27 & 6.42 \\ 
  fgoals\_g3 & 25.48 & 25.43 & 17.14 & 24.97 & 17.71 & 52.13 & 6.80 \\ 
  hadgem3\_gc31\_mm & 25.06 & 25.93 & 17.19 & 25.58 & 17.91 & 41.66 & 6.53 \\ 
  ipsl\_cm6a\_lr & 26.61 & 25.84 & 17.24 & 25.35 & 17.79 & \textit{57.43} & \textit{9.58} \\ 
  mpi\_esm\_1\_2\_hr & 25.77 & 25.97 & 17.37 & 25.57 & 18.05 & 41.70 & 6.02 \\ 
  noresm2\_mm & 25.36 & 25.74 & 17.09 & 25.43 & 17.79 & 43.10 & 6.62 \\ 
  taiesm1 & \textit{24.61} & \textit{25.22} & \textit{16.86} & \textit{24.74} & \textit{17.38} & 45.17 & 7.54 \\ 
   \hline
\end{tabular}
\label{tab:final.ranking}
\end{table}

From Table \ref{tab:final.ranking}, the trajectory \texttt{ec\_earth3\_aerchem} performs best across all scores except $PerR$, where \texttt{ec\_earth3\_veg} wins. In general, trajectories from the \texttt{ec\_earth3} family consistently outperform others, securing the top ranks in the evaluation measures. The lowest values in the table are associated with the \texttt{taiesm1} trajectory, except for the persistence-related scores, where the worst performance is observed for \texttt{ipsl\_cm6a\_lr}.
The $DR$ score for \texttt{hadgem3\_gc31\_mm} is the lowest among the 16 trajectories, but it ranks higher in $CR_{loc}$ and $CR_{loc}^*$ scores. This suggests that this trajectory performs better when evaluated using \textbf{cond rf} compared to \textbf{rf}.
Regarding the range, the most discriminative measures are the persistence scores ($PerR$ and $PerR^*$), which span 35.59--57.43 and 5.09--9.58, respectively. Therefore, the persistence measure may be particularly useful for classifying trajectories.
The different versions of $CR$ vary over smaller intervals and produce similar classifications. $CR_{loc}$ and $CR_{reg}$ are less useful because they are harder to compare directly with $DR$. On the other hand, versions based only on the most frequent WT benefit from sufficient data and produce classification results similar to those using all WT.

\begin{figure}[h!]
    \centering
    \includegraphics[width=0.7\linewidth]{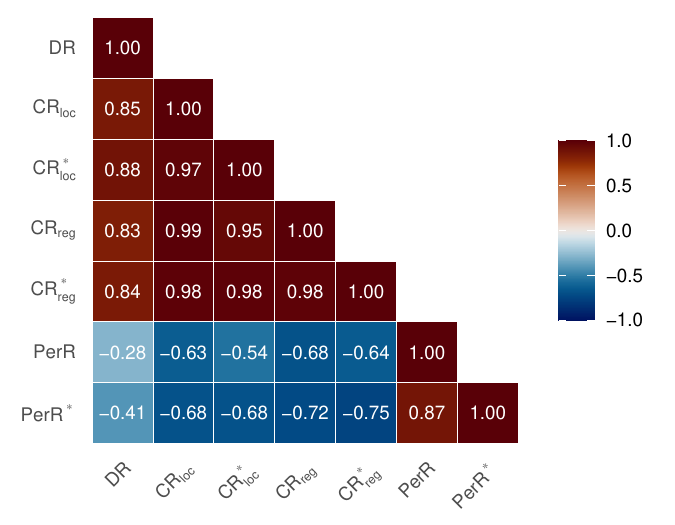}
    \caption{Correlations between measures in Table \ref{tab:final.ranking}.}
    \label{fig:cor:ranking:measures}
\end{figure}

The correlation plot in Figure \ref{fig:cor:ranking:measures} shows the relationships between the measures in Table \ref{tab:final.ranking}. The four $CR$ scores are highly correlated, so considering one of them would be sufficient. These $CR$ measures are also strongly correlated with $DR$, although slightly less so. In contrast, the persistence scores are not highly correlated with $DR$ or the $CR$ scores, providing an additional independent way to rank trajectories.

\FloatBarrier
\section{Discussion and conclusions }

The specific aim of this work is to establish objective criteria for selecting climate models based on their ability to replicate both the daily frequency and the conditional distributions (persistence and transitions) of WT in a region. This study evaluates the performance of 36 CMIP6 GCM trajectories against ERA5 reanalysis data across the Iberian Peninsula, which is a critical climate change `hotspot' due to its vulnerability to heatwaves and droughts \citep{serranonotivoli2023, diazposo2023, achebak2018}. This is essential because these large-scale circulation patterns are the primary drivers of extreme summer temperatures in the region \citep{ventura2023, sousa2019}. This is done using the Lamb Weather Type classification scheme based on the methodology proposed by \citet{brands2022b}.
While previous studies \citep{brands2022circulation} have assessed the daily distribution of weather patterns, this research introduces a significant methodological advancement. The contribution of this work is the systematic evaluation of 24-hour atmospheric dynamics. Rather than looking at daily frequencies alone, this paper proposes evaluating the transition probability matrices--specifically, the probability of a WT occurring conditioned on the WT of the previous day.

To provide a robust comparison between GCM trajectories, this study establishes a methodological framework that evaluates them against a reference observational dataset, typically a reanalysis product.  First, at each grid point $s$ within the ROI, reproduction is assessed using both daily and conditional frequencies. The Overlap coefficient is selected as the primary similarity measure due to its strong discriminative power and robust performance in the presence of rare weather types, compared to alternatives such as the Hellinger distance or the Bhattacharyya coefficient. To ensure reliability, a filtering process is applied where only trajectories that reproduce both daily and conditional distributions with an Overlap of at least a threshold $t_{sim}$ across a minimum number of grid points $k$ are retained. Performance across the $N_S$-point ROI is then ranked using three types of metrics that integrate the results across the $N_S$ points. The first type addresses the daily frequencies, and it is denoted as the Daily Reproduction (DR), defined as the sum of overlap values for daily frequencies. The second type addresses the dynamical evolution, and it is denoted as the Conditional Reproduction (CR), incorporating transition dynamics through local and regional weighting. The third type addresses the persistence of WT, and it is denoted as Persistence Reproduction (PerR), an index designed to evaluate how accurately trajectories capture the persistence of a weather type over two consecutive days. CR and PerR are newly introduced metrics. The proposed methodology allows for the identification of trajectories that fail to adequately capture regional dynamics, specifically those that perform well only in a subregion within the ROI or under particular conditioning WT. 


The evaluation of 36 CMIP6 trajectories over the Iberian Peninsula ($N_s = 30$, ERA5 reanalysis, $t_{sim} = 0.8$, $k = 1/3\,N_s$) shows that reproduction is not uniform across the ROI. Additionally, some WT categories were found to be very infrequent, resulting in small transition probabilities. This was taken into account, and their impact on the distance measures and scores was evaluated. The results show that, under the overlap measure, these small transitions have no effect, and the integrated scores do not exhibit differences when WT categories with low transition probabilities are considered. The initial filtering process indicates that only 16 out of the 36 trajectories are capable of adequately reproducing daily WT distributions. When both daily and conditional distributions are considered, only 12 trajectories from 7 distinct families remain acceptable. Within this subset, trajectories from the \texttt{ec\_earth3} family consistently outperform all others across all metrics. In particular, \texttt{ec\_earth3\_aerchem} emerges as the top-performing trajectory, demonstrating the highest reliability across the entire Peninsula. Some trajectories outperform others only in specific regions or under particular conditioning WT. For example, \texttt{mpi\_esm\_1\_2\_hr} dominates the north-west in daily frequencies, as well as under PA and PC. The trajectory \texttt{cnrm\_esm2\_1} dominates the south-east when conditioning on PC. Finally, \texttt{hadgem3\_gc31\_mm} dominates the south-west under PDNE, PC, and U.

High variability in model performance was observed depending on the geographical region. While the twelve surviving trajectories (after the conditional filter) generally perform well in the North-West, particularly along the Atlantic Coast, they face substantial challenges in accurately representing circulation patterns in the Central and Southern parts of the Iberian Peninsula, near the Mediterranean sea. A key strength of this study was the ability to establish specific behaviors for local climates by analyzing individual grid points. For example, the framework allows for a precise determination of how many and which models can accurately represent a specific location, such as the North-West (NW) point compared to the South-West (SW) point, both in terms of daily frequency and the dynamic behavior conditioned on specific patterns like Anticyclonic (A) or North-East (NE) flows.

This study over the Iberian Peninsula faces several limitations. First, the performance of the GCMs was assessed over a limited number of grid points due to the relatively coarse grid resolution of $2.5^\circ \times 2.5^\circ$. A higher-resolution grid would provide more points for a more detailed analysis and potentially a better understanding of regional variations in model performance. 

To conclude, a primary insight of this research is the high value of using conditional distributions to assess GCM performance. The results indicate that while many trajectories can adequately reproduce the daily frequency of WTs (which does not consider the dynamics), their ability to capture the actual atmospheric behavior--specifically transitions and persistence--is far more variable. The study identifies that the reproduction of transitions for less frequent weather types is where model reliability decreases most significantly, making conditional metrics a more discriminative tool for model selection than daily distributions. To address this, the Overlap measure is used as it is not highly influenced by infrequent WT. In addition, similarity measures based exclusively on relevant WT within the ROI are proposed and denoted with a $^*$. Overall, this study provides valuable insights into the performance of various GCM trajectories in simulating both daily and conditional climate patterns over the Iberian Peninsula. The findings suggest that while the \texttt{ec\_earth3} trajectories are the best performers, there are still areas for improvement, particularly in capturing regional climate variations and conditional distributions.

\subsection{Future work}

While the Overlap measure was found to be robust for comparing the GCMs, future research could benefit from exploring alternative metrics or hybrid approaches. These could offer additional insights into GCM behavior and help address some of the issues encountered with the current measures, especially in cases of low-frequency events or transitions.

The method could be used to assess changes in daily and conditional frequencies under future scenarios, by comparing the current ERA5 circulation with that projected by a GCM under a climate change scenario (i.e., a non-present climate). However, this application is only valid for GCMs that have passed the initial filtering based on their ability to reproduce present-day distributions. Otherwise, it would not be possible to distinguish between changes driven by global warming and those arising from inherent GCM bias.


The method could also be applied using WT classifications other than the Lamb classification, particularly those more appropriate for specific regions such as the Niño region. Further extensions could include pseudo-WTs derived from atmospheric variables at 700, 500, and 300 hPa.  A fully integrated procedure could be developed based on gridded hourly or daily GCM output, implemented using the software developed in \citet{brands2022b}.

\backmatter





\bmhead{Acknowledgements}
Authors thank for supporting of Projects PID2023-150234NB-I00, PID2024-155426OB-I00, RED2024-153680-T Biostatnet,  Gobierno de Arag\'on under PROY\_T21\_24-HIDROGIF Grant and Research Group E46\_23R: Modelos Estoc\'asticos.




\noindent






\bibliographystyle{plainnat}  
\bibliography{Climate_Dynamics/draft}

\begin{appendices}


\newpage

\section{Summary of WT frequencies from ERA5}
\label{app:marg:frec:ERA5}
\subsection{Summary of daily frequencies from ERA5}
\label{app:daily:frec:ERA5}

\begin{table}[h!]
\centering
\caption{Summary statistics of the \textbf{rf} for each WT in ERA5, including selected percentiles (5th, 10th, 20th, 30th, 40th, 60th, 70th, 90th, and 95th), as well as the median, minimum, and maximum values.}
\begin{tabular}{lcccccccccccc}
  \hline
  WT & Min & Q.5 & Q.10 & Q.20 & Q.30 & Q.40 & Median & Q.60 & Q.80 & Q.90 & Q.95 & Max \\ 
  \hline
  PA & 0.064 & 0.080 & 0.088 & 0.109 & 0.123 & 0.140 & 0.179 & 0.192 & 0.260 & 0.317 & 0.317 & 0.357 \\ 
  DANE & 0.009 & 0.012 & 0.016 & 0.020 & 0.024 & 0.027 & 0.031 & 0.035 & 0.045 & 0.064 & 0.079 & 0.093 \\ 
  DAE & 0.004 & 0.006 & 0.007 & 0.013 & 0.015 & 0.025 & 0.026 & 0.029 & 0.034 & 0.040 & 0.072 & 0.103 \\ 
  DASE & 0.001 & 0.002 & 0.002 & 0.003 & 0.004 & 0.005 & 0.006 & 0.007 & 0.009 & 0.015 & 0.023 & 0.038 \\ 
  DAS & 0.001 & 0.001 & 0.001 & 0.001 & 0.002 & 0.002 & 0.003 & 0.003 & 0.005 & 0.006 & 0.007 & 0.010 \\ 
  DASW & 0.000 & 0.000 & 0.000 & 0.001 & 0.001 & 0.002 & 0.003 & 0.005 & 0.011 & 0.014 & 0.015 & 0.018 \\ 
  DAW & 0.000 & 0.000 & 0.001 & 0.001 & 0.003 & 0.005 & 0.007 & 0.008 & 0.016 & 0.020 & 0.023 & 0.028 \\ 
  DANW & 0.000 & 0.001 & 0.002 & 0.003 & 0.004 & 0.007 & 0.010 & 0.011 & 0.018 & 0.022 & 0.025 & 0.026 \\ 
  DAN & 0.001 & 0.004 & 0.007 & 0.009 & 0.013 & 0.015 & 0.021 & 0.025 & 0.031 & 0.036 & 0.045 & 0.050 \\ 
  PDNE & 0.026 & 0.031 & 0.032 & 0.042 & 0.047 & 0.054 & 0.063 & 0.070 & 0.121 & 0.140 & 0.149 & 0.325 \\ 
  PDE & 0.007 & 0.013 & 0.021 & 0.034 & 0.042 & 0.063 & 0.074 & 0.077 & 0.121 & 0.137 & 0.171 & 0.207 \\ 
  PDSE & 0.003 & 0.003 & 0.003 & 0.005 & 0.007 & 0.011 & 0.013 & 0.019 & 0.028 & 0.052 & 0.056 & 0.061 \\ 
  PDS & 0.000 & 0.000 & 0.000 & 0.001 & 0.002 & 0.003 & 0.004 & 0.005 & 0.007 & 0.011 & 0.017 & 0.025 \\ 
  PDSW & 0.000 & 0.000 & 0.001 & 0.001 & 0.002 & 0.004 & 0.006 & 0.009 & 0.022 & 0.028 & 0.030 & 0.035 \\ 
  PDW & 0.000 & 0.000 & 0.001 & 0.001 & 0.003 & 0.007 & 0.009 & 0.013 & 0.024 & 0.034 & 0.040 & 0.050 \\ 
  PDNW & 0.000 & 0.001 & 0.002 & 0.005 & 0.008 & 0.013 & 0.016 & 0.017 & 0.028 & 0.035 & 0.039 & 0.042 \\ 
  PDN & 0.002 & 0.005 & 0.008 & 0.016 & 0.021 & 0.027 & 0.035 & 0.040 & 0.071 & 0.085 & 0.092 & 0.112 \\ 
  PC & 0.009 & 0.013 & 0.019 & 0.025 & 0.036 & 0.045 & 0.048 & 0.051 & 0.064 & 0.072 & 0.084 & 0.095 \\ 
  DCNE & 0.002 & 0.004 & 0.005 & 0.007 & 0.009 & 0.009 & 0.011 & 0.012 & 0.015 & 0.016 & 0.019 & 0.020 \\ 
  DCE & 0.001 & 0.002 & 0.003 & 0.007 & 0.008 & 0.009 & 0.012 & 0.015 & 0.019 & 0.026 & 0.026 & 0.034 \\ 
  DCSE & 0.001 & 0.002 & 0.002 & 0.002 & 0.003 & 0.003 & 0.004 & 0.005 & 0.009 & 0.010 & 0.011 & 0.013 \\ 
  DCS & 0.000 & 0.000 & 0.000 & 0.001 & 0.001 & 0.002 & 0.002 & 0.002 & 0.003 & 0.005 & 0.006 & 0.009 \\ 
  DCSW & 0.000 & 0.000 & 0.000 & 0.000 & 0.001 & 0.001 & 0.002 & 0.003 & 0.006 & 0.007 & 0.007 & 0.011 \\ 
  DCW & 0.000 & 0.000 & 0.000 & 0.000 & 0.001 & 0.001 & 0.002 & 0.003 & 0.004 & 0.006 & 0.008 & 0.009 \\ 
  DCNW & 0.000 & 0.000 & 0.000 & 0.001 & 0.001 & 0.003 & 0.004 & 0.004 & 0.005 & 0.006 & 0.007 & 0.009 \\ 
  DCN & 0.001 & 0.001 & 0.002 & 0.004 & 0.004 & 0.005 & 0.006 & 0.007 & 0.010 & 0.011 & 0.012 & 0.016 \\ 
  U & 0.097 & 0.128 & 0.150 & 0.195 & 0.233 & 0.276 & 0.348 & 0.386 & 0.479 & 0.496 & 0.532 & 0.605 \\ 
   \hline
\end{tabular}
\end{table}

\FloatBarrier

\subsection{Boxplots of the \textbf{cond rf} from the ERA5 trajectory conditioned on a set of representative WT.}

\label{cond.rf.box}

\begin{figure}[h!]
    \centering
    
    \begin{subfigure}{0.49\linewidth}
        \centering
        \includegraphics[width=\linewidth]{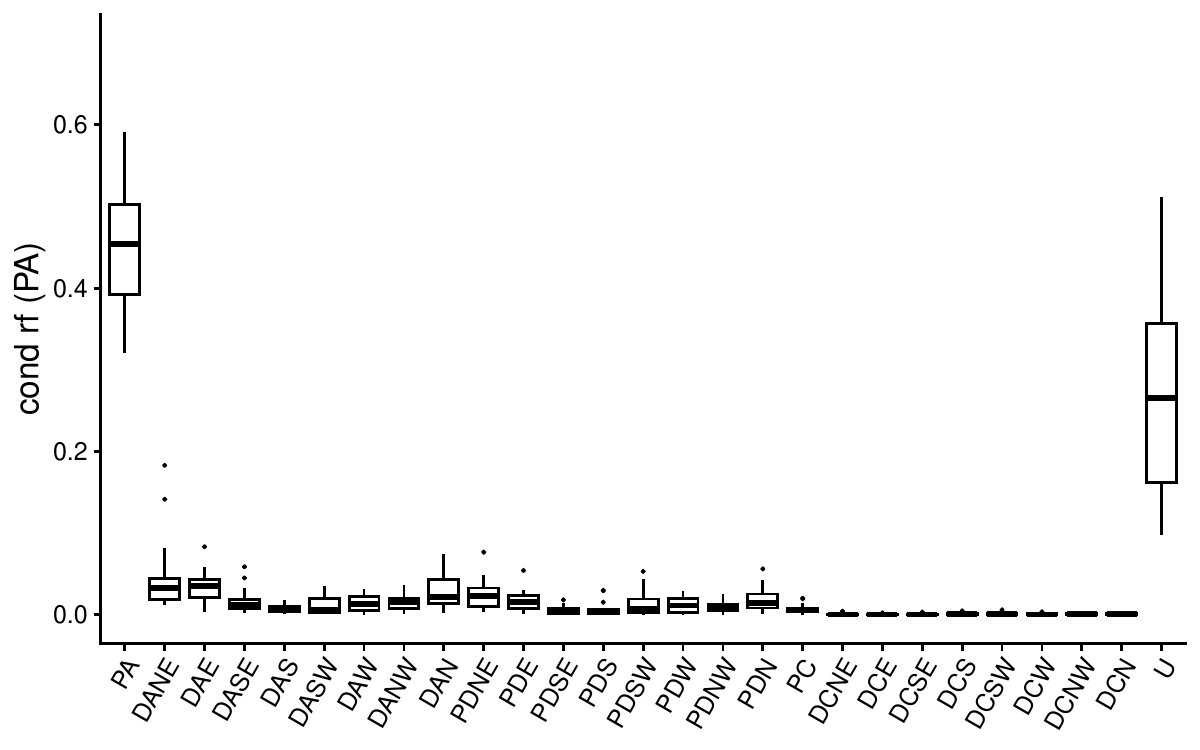}
        \caption{PA}
    \end{subfigure}
    \begin{subfigure}{0.49\linewidth}
        \centering
        \includegraphics[width=\linewidth]{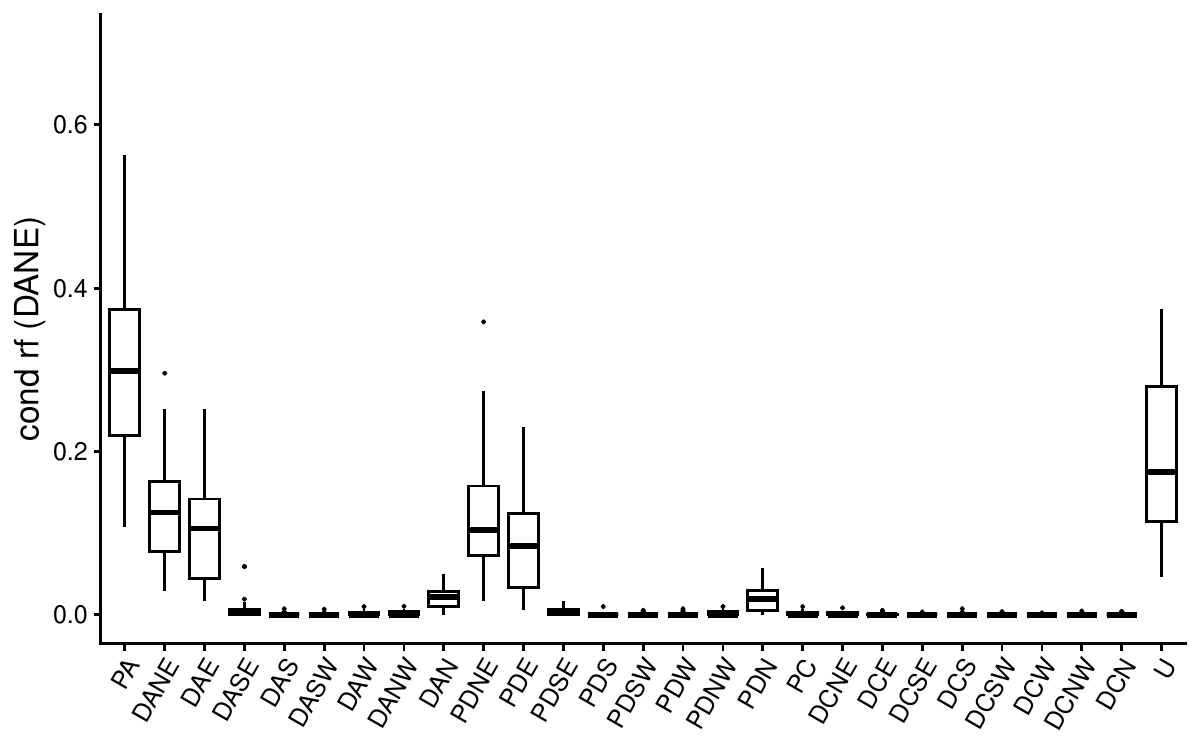}
        \caption{DANE}
    \end{subfigure}
    \begin{subfigure}{0.49\linewidth}
        \centering
        \includegraphics[width=\linewidth]{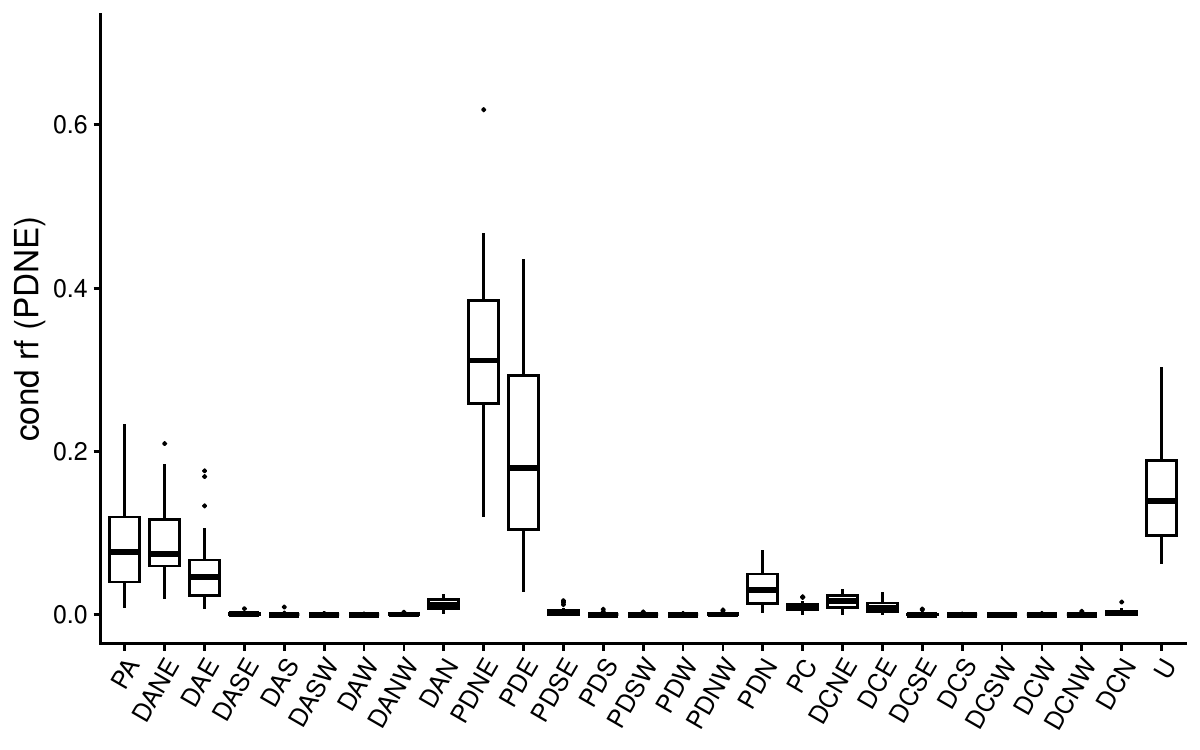}
        \caption{PDNE}
    \end{subfigure}
    \begin{subfigure}{0.49\linewidth}
        \centering
        \includegraphics[width=\linewidth]{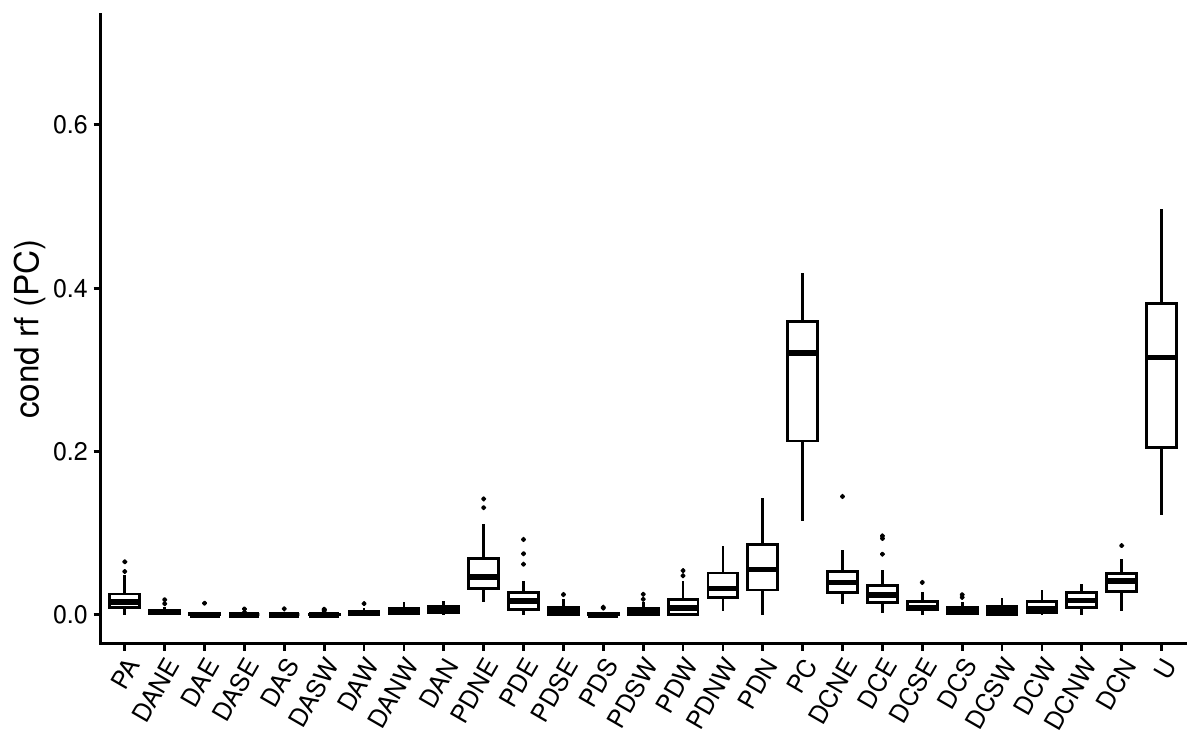}
        \caption{PC}
    \end{subfigure}
    \begin{subfigure}{0.49\linewidth}
        \centering
        \includegraphics[width=\linewidth]{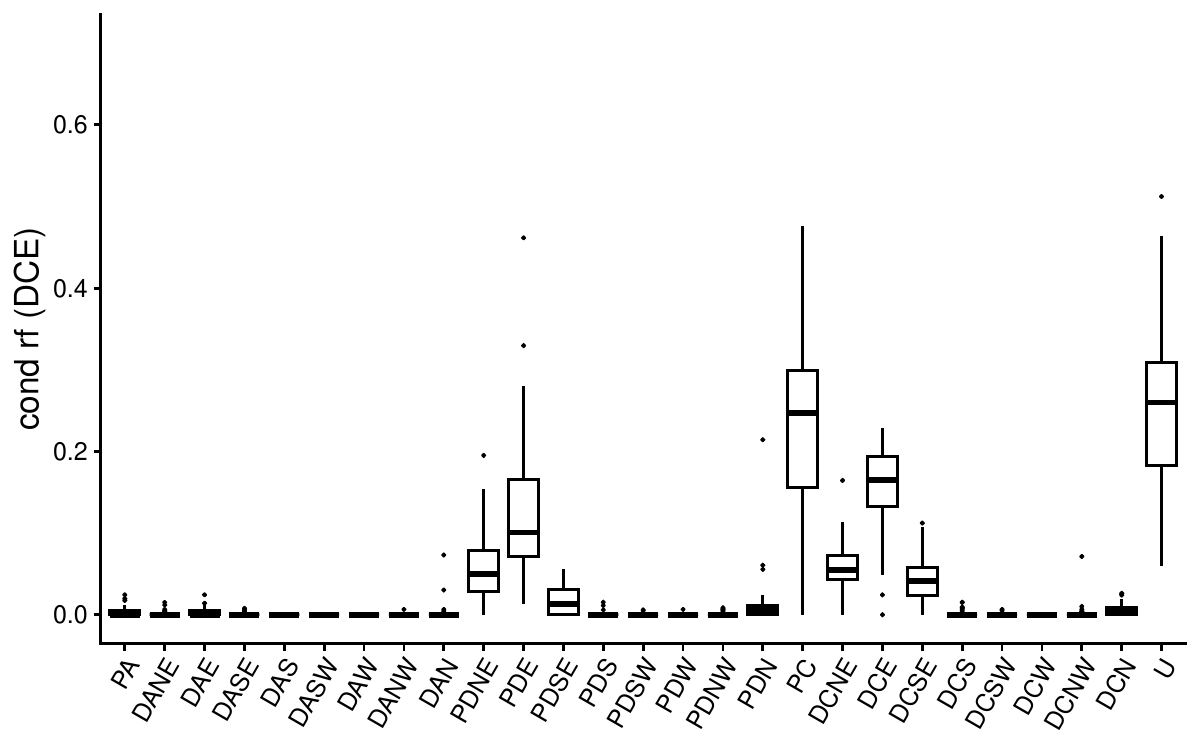}
        \caption{DCE}
    \end{subfigure}
    \begin{subfigure}{0.49\linewidth}
        \centering
        \includegraphics[width=\linewidth]{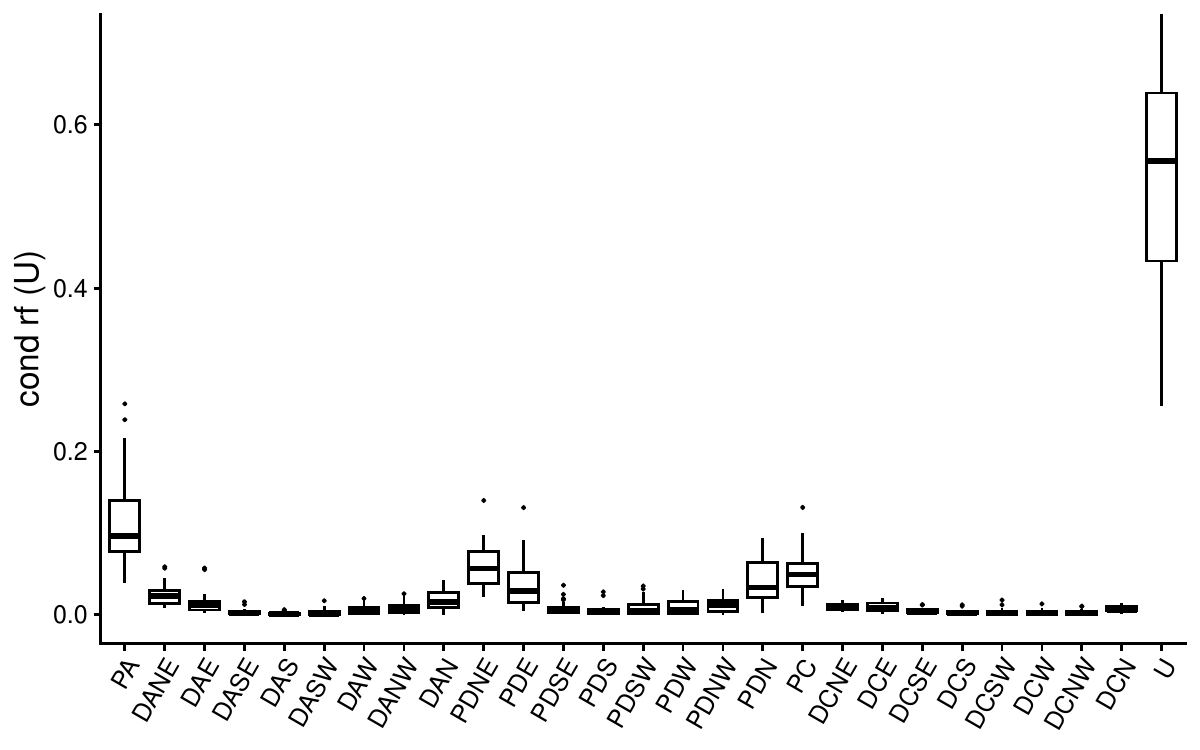}
        \caption{U}
    \end{subfigure}

    \caption{\textbf{cond rf} of the ERA5 trajectory conditioned on $\text{WT}=j$ in the Iberian Peninsula. Panels correspond to $\text{WT} = \{\text{PA, DANE, PDNE, PC, DCE, U}\}$.}
    \label{fig:box.cond}
\end{figure}
\FloatBarrier

\section{Evaluation of Similarity Measures and WT selection strategy using conditional frequencies and using other GCM trajectories with daily frequencies.}

\label{app:comp:tools}

\subsection{Using daily frequency with other GCM trajectories.}

\label{sec:dist:comp:gcm}

\begin{figure}[h!]
    \centering
    
    \begin{subfigure}{1\linewidth}
        \centering
        \includegraphics[width=0.32\linewidth]{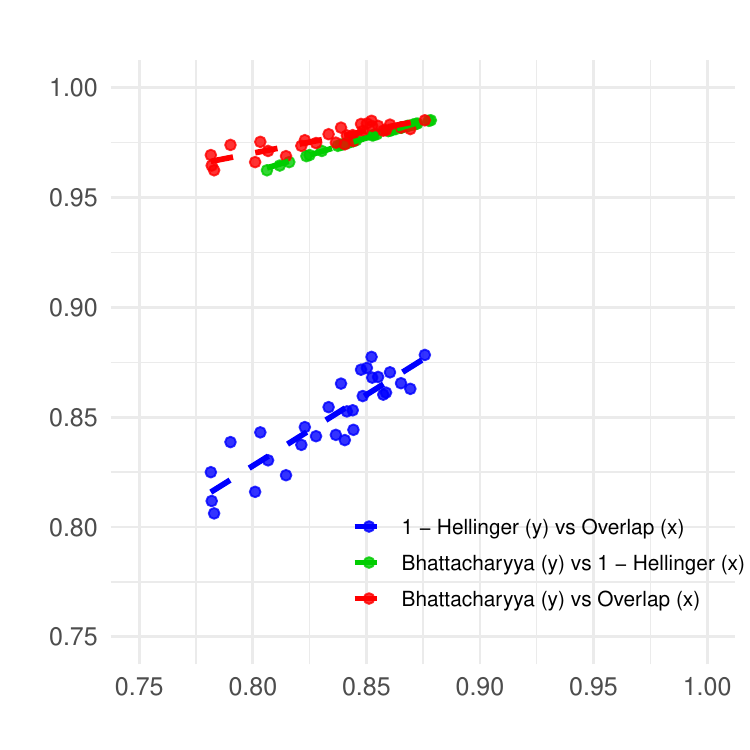}
        \includegraphics[width=0.32\linewidth]{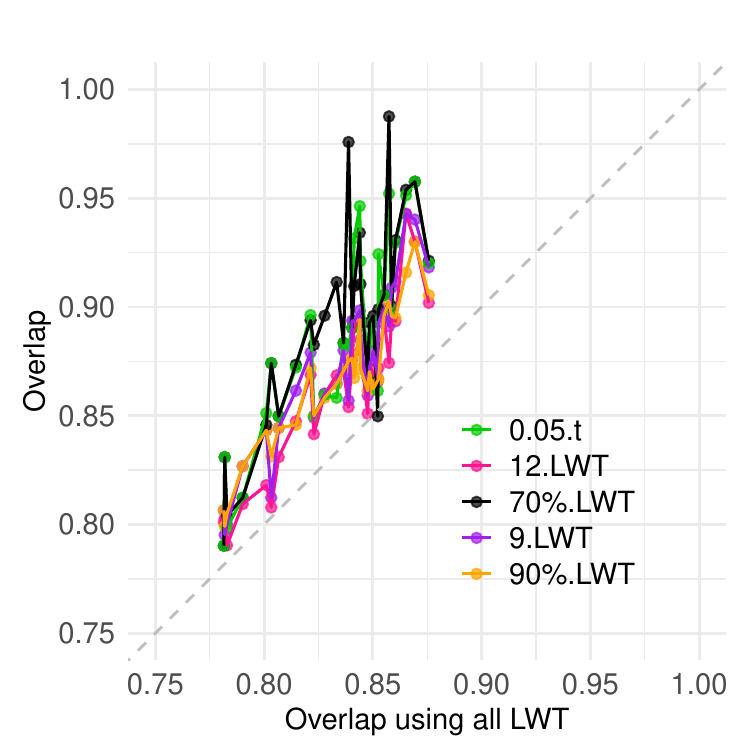}
        \includegraphics[width=0.32\linewidth]{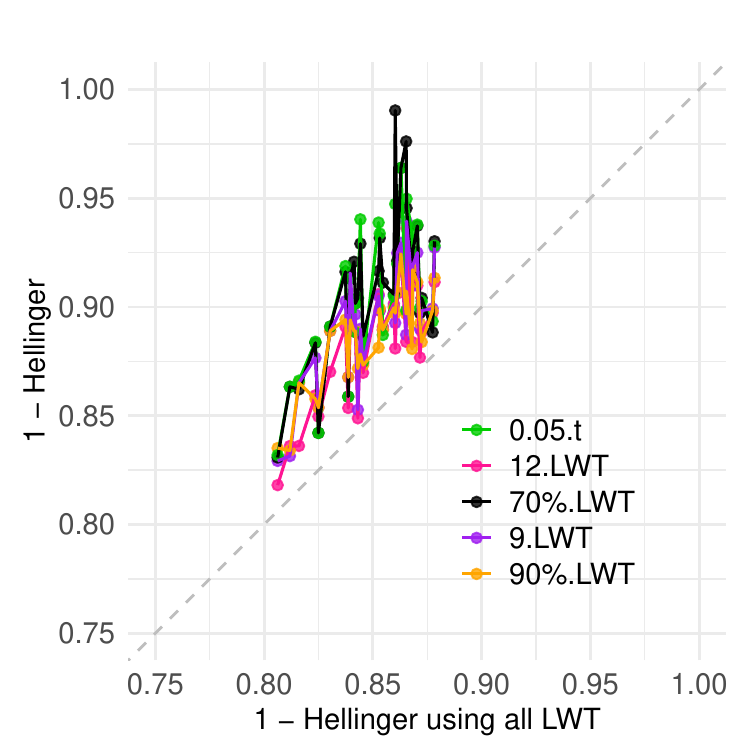}
        \caption{Using GCM trajectory \texttt{cnrm\_cm6\_1\_hr}.}
    \end{subfigure}
    
        \begin{subfigure}{1\linewidth}
        \centering
        \includegraphics[width=0.32\linewidth]{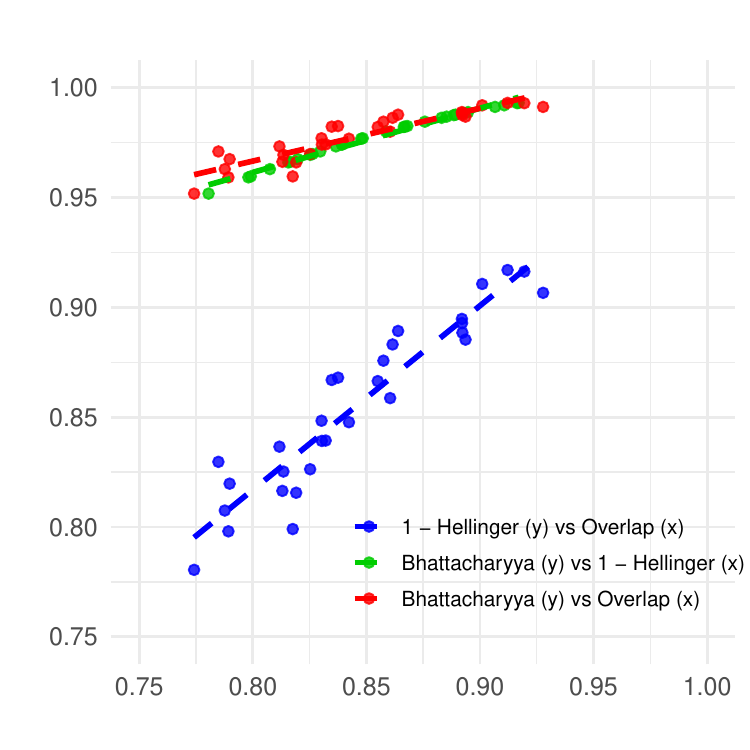}
        \includegraphics[width=0.32\linewidth]{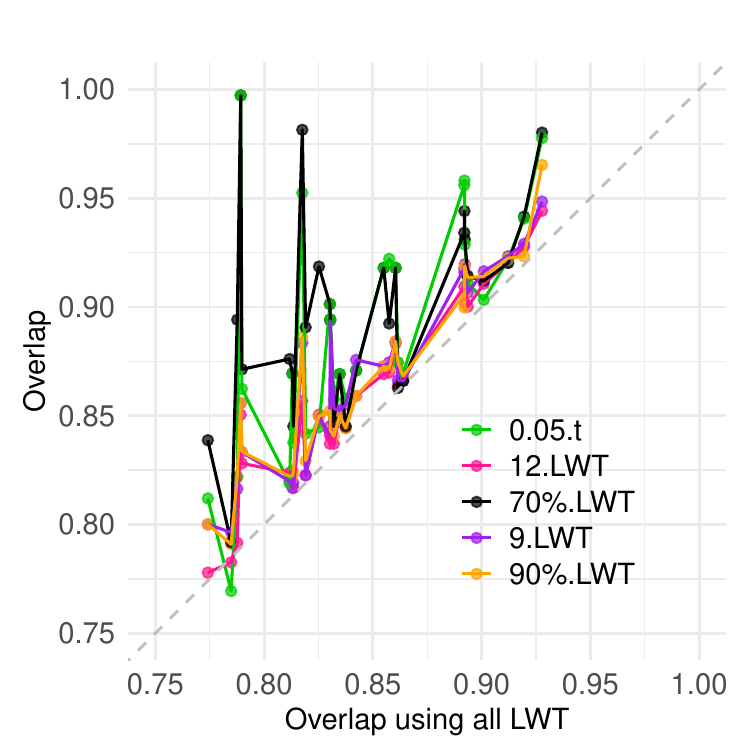}
        \includegraphics[width=0.32\linewidth]{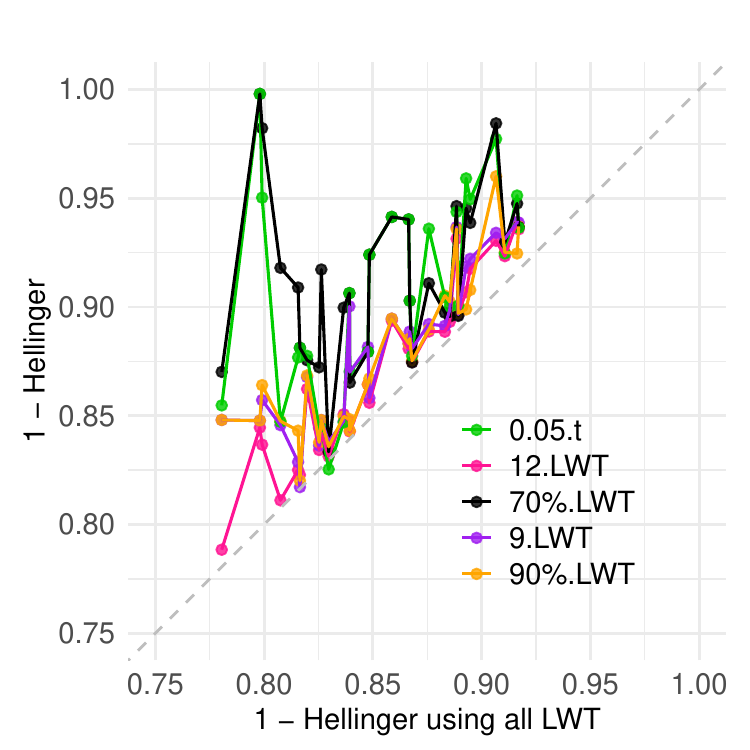}
        \caption{Using GCM trajectory \texttt{noresm2\_mm}.}
    \end{subfigure}
     
    \caption{\textbf{Left column:} Pairwise relationships between the overlap, Bhattacharyya coefficient, and $1-\text{Hellinger}$ distance computed from the \textbf{rf} of ERA5 and two GCM trajectories across grid points in the Iberian Peninsula. \textbf{Middle column:} Overlap values obtained under different WT-selection strategies, plotted against the default approach including all WT categories. \textbf{Right column:} Corresponding $1-\text{Hellinger}$ distances for the same WT-selection strategies, also compared to the default method.}
    \label{fig:box.cond}
\end{figure}
\FloatBarrier

\subsection{Conditioning on relevant WT using \texttt{ec\_earth3\_aerchem}.} \label{sec:dist:comp:cond}

\begin{figure}[h!]
    \centering
    
    \begin{subfigure}{0.8\linewidth}
        \centering
        \includegraphics[width=0.32\linewidth]{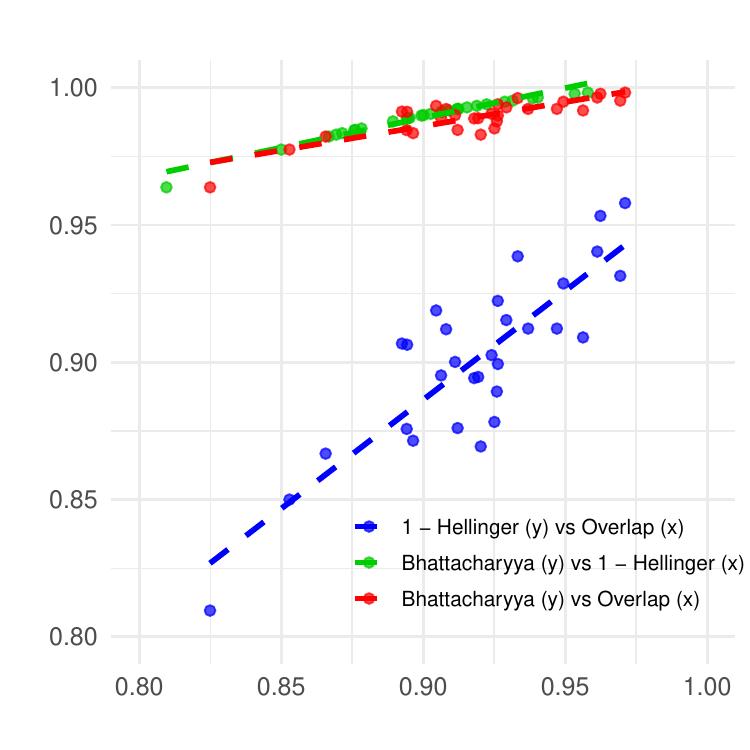}
        \includegraphics[width=0.32\linewidth]{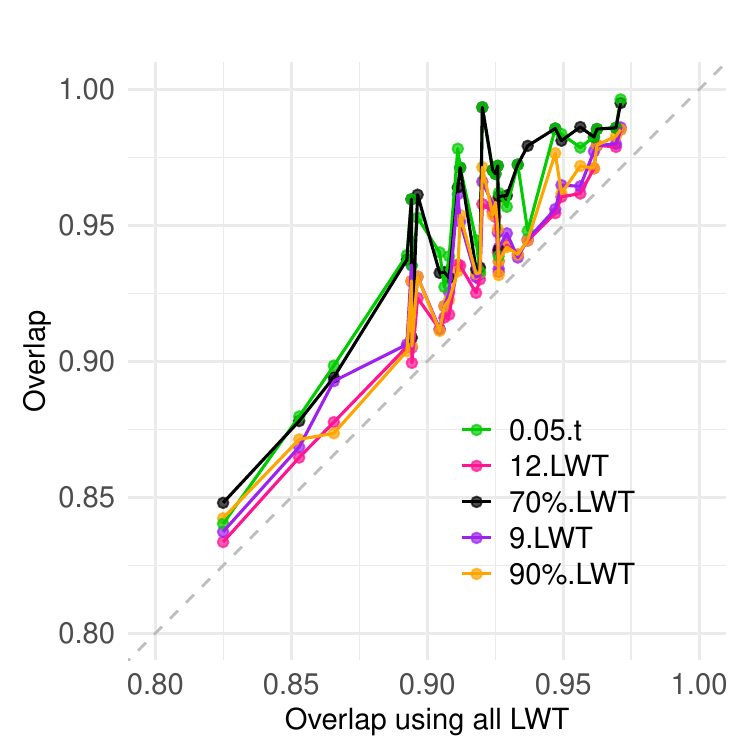}
        \includegraphics[width=0.32\linewidth]{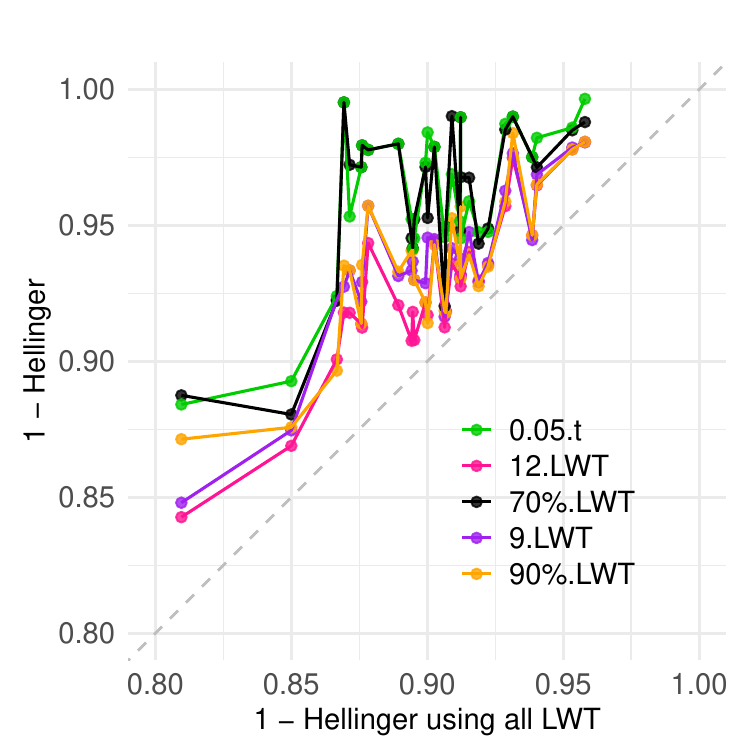}
        \caption{Conditioned on PA}
    \end{subfigure}
    
    \begin{subfigure}{0.8\linewidth}
        \centering
        \includegraphics[width=0.32\linewidth]{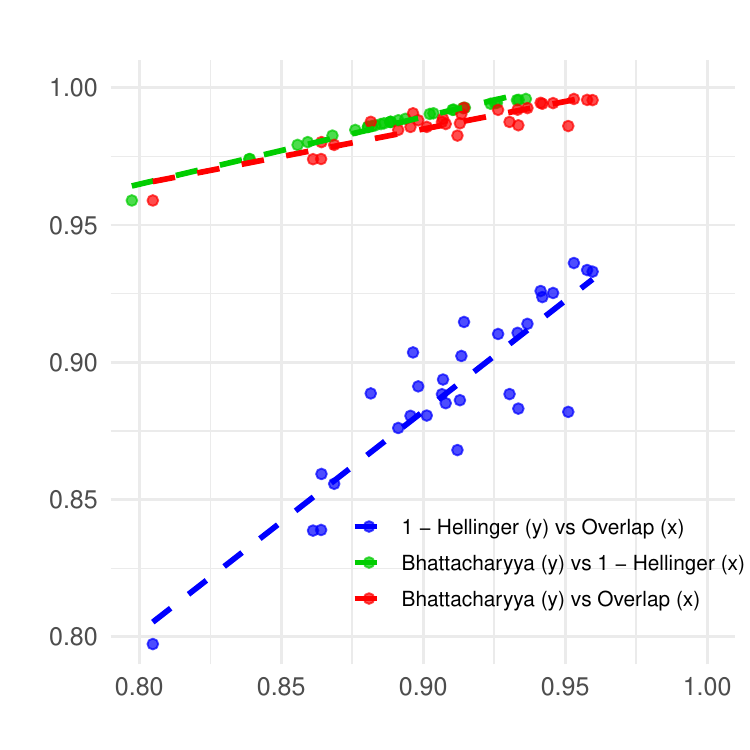}
        \includegraphics[width=0.32\linewidth]{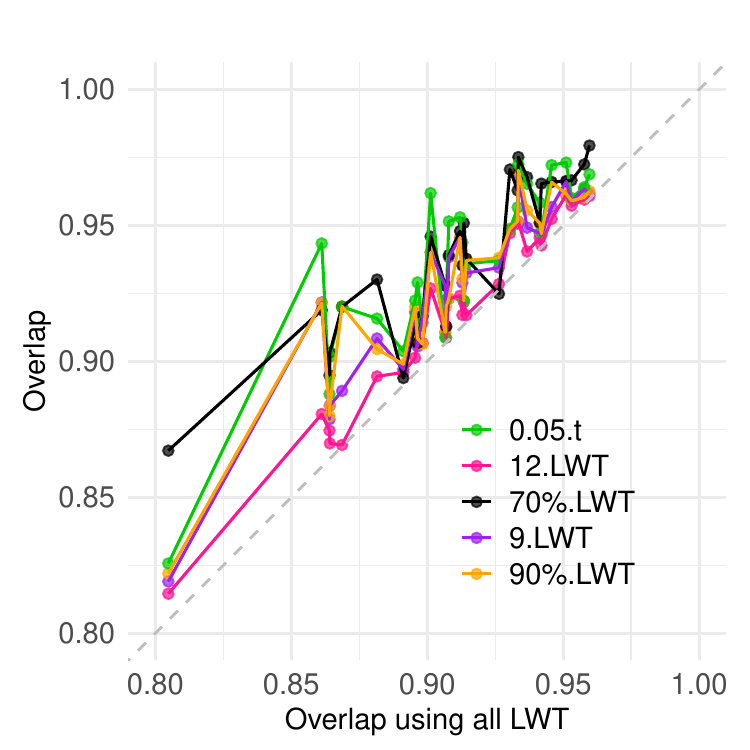}
        \includegraphics[width=0.32\linewidth]{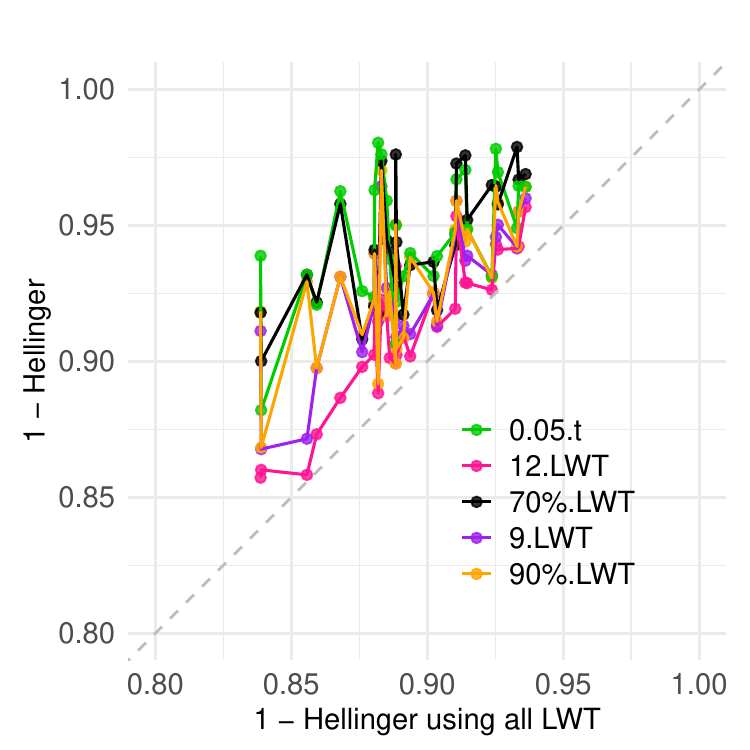}
        \caption{Conditioned on PDNE}
    \end{subfigure}
     
    \begin{subfigure}{0.8\linewidth}
        \centering
        \includegraphics[width=0.32\linewidth]{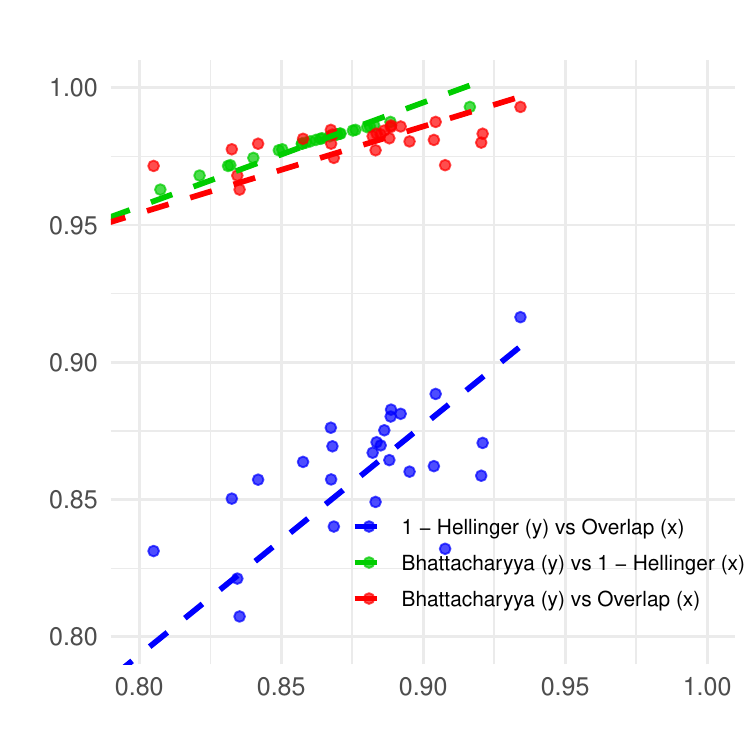}
        \includegraphics[width=0.32\linewidth]{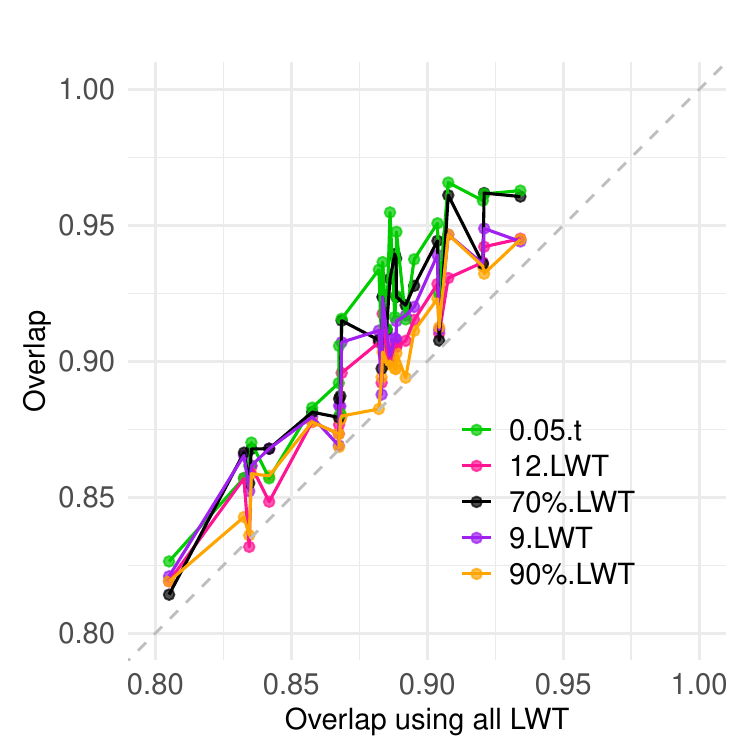}
        \includegraphics[width=0.32\linewidth]{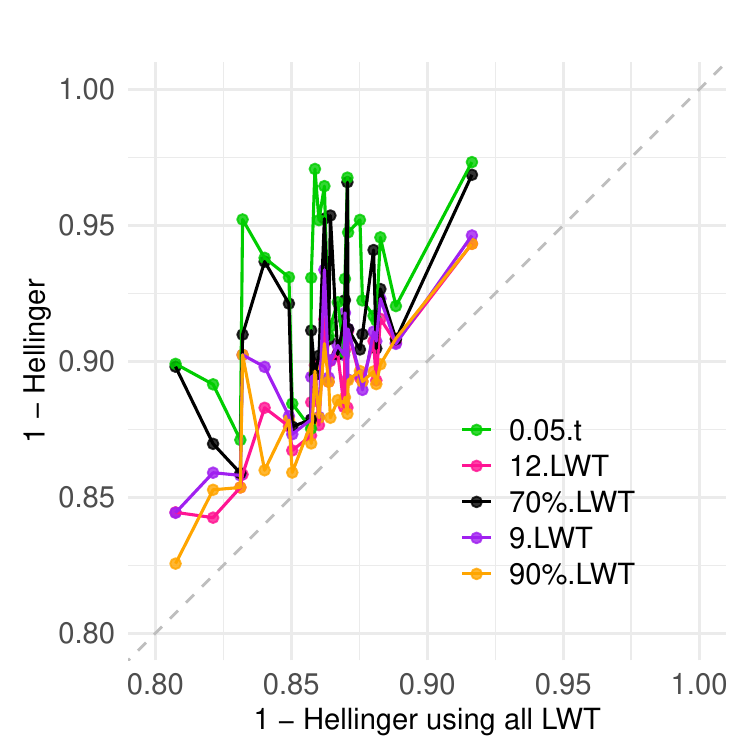}
        \caption{Conditioned on PC}
    \end{subfigure}
 
    \begin{subfigure}{0.8\linewidth}
        \centering
        \includegraphics[width=0.32\linewidth]{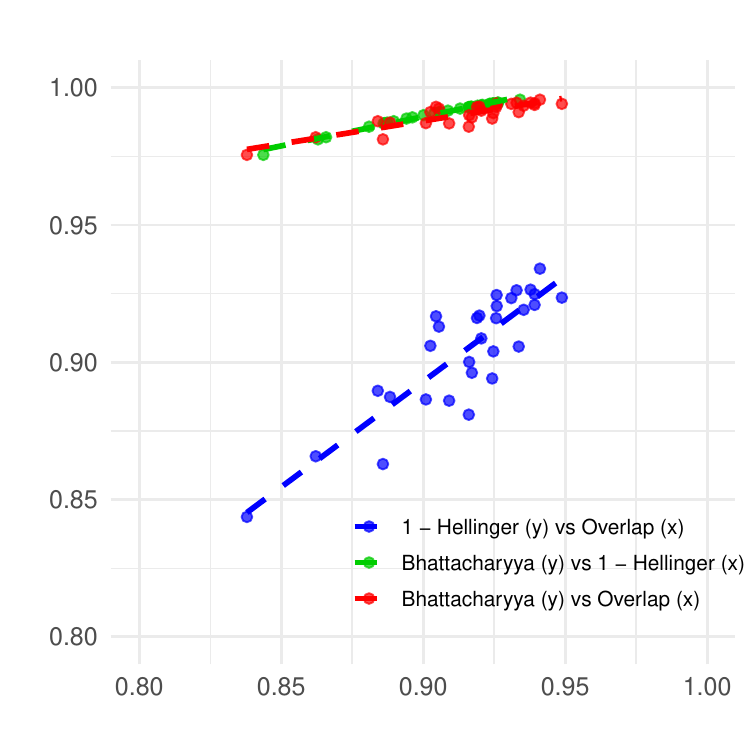}
        \includegraphics[width=0.32\linewidth]{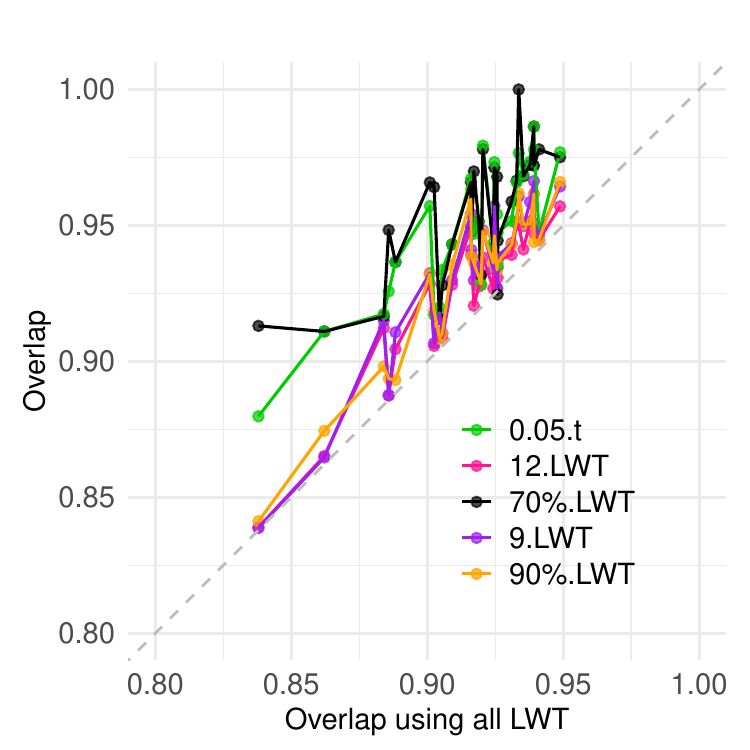}
        \includegraphics[width=0.32\linewidth]{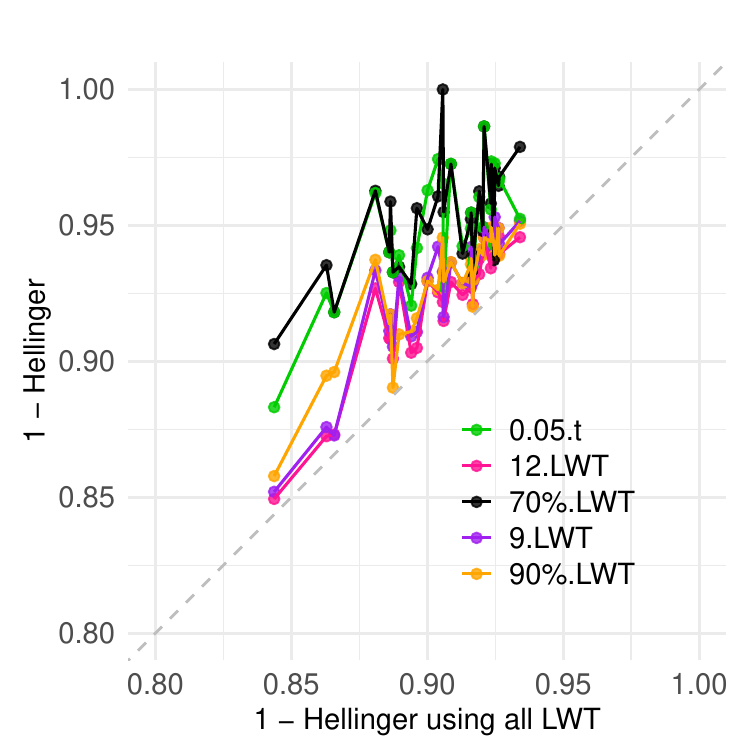}
        \caption{Conditioned on U}
    \end{subfigure}
    \caption{\textbf{Left column:} Pairwise relationships between the overlap, Bhattacharyya coefficient, and $1-\text{Hellinger}$ distance computed from the \textbf{cond rf} of ERA5 and the GCM trajectory \texttt{ec\_earth3\_aerchem} across grid points in the Iberian Peninsula. \textbf{Middle column:} Overlap values obtained under different WT-selection strategies, plotted against the default approach including all WT categories. \textbf{Right column:} Corresponding $1-\text{Hellinger}$ distances for the same WT-selection strategies, also compared to the default method.}
    \label{fig:box.cond}
\end{figure}
\FloatBarrier

\subsection{Quantifying the distance to the diagonal}

To quantify how similar the different WT selection options are to using all WT, and to enable comparison between overlap- and Hellinger-based distances, we compute a standardized distance $D_{\text{opt}}$. For each option, this metric is obtained by subtracting the overlap (or Hellinger) obtained using all WT from that of each alternative option, summing across all points, and dividing by the standard deviation of the reference (All WT) case.

The distance is defined as:

$$
D_{\text{opt}} = \frac{\sum_{S} \left( y_{\text{opt}}(s) - x(s) \right)}{\sigma(x(s))}
$$

where $y_{\text{opt}}(s)$ denotes the overlap (or Hellinger similarity) at point $s$ for a given option, $x(s)$ denotes the corresponding value for the All WT reference case, and $\sigma(x(s))$ is the standard deviation of the reference distribution across points. This formulation provides a measure of deviation from the full WT case, allowing direct comparison across similarity measures.

\begin{table}[h!]
\centering
\caption{Standardized differences ($D_{\text{opt}}$) between alternative WT options and the reference configuration (All WT) across daily and conditonal distributions. Values are shown separately for overlap-based and Hellinger-based distance metrics.}
\begin{tabular}{lrrrrrrrrrr}
\toprule
& \multicolumn{5}{c}{Overlap} & \multicolumn{5}{c}{Hellinger} \\ 
\cmidrule(lr){2-6} \cmidrule(lr){7-11}
WT & 12 & 9  & 90\% & 70\% & 0.05 t & 12 & 9 & 90\% & 70\% & 0.05 t \\ 
\midrule
Daily & 10.30 & 14.38 & 13.01 & 28.83 & 24.92 & 16.31 & 21.28 & 20.55 & 42.74 & 37.12 \\ 
  PA & 13.03 & 17.71 & 16.67 & 31.38 & 31.72 & 24.98 & 33.01 & 33.03 & 57.41 & 58.24 \\ 
  PC & 10.36 & 14.56 & 8.15 & 18.49 & 21.79 & 18.78 & 23.75 & 15.55 & 33.73 & 43.86 \\ 
  PDNE & 7.13 & 15.05 & 15.88 & 23.46 & 22.22 & 17.16 & 31.04 & 34.62 & 47.66 & 48.49 \\ 
  U & 14.25 & 19.78 & 18.57 & 47.39 & 40.93 & 22.67 & 30.27 & 30.57 & 66.22 & 60.77 \\ 
 \bottomrule
\end{tabular}
\subcaption{\texttt{ec\_earth3\_aerchem}}
\begin{tabular}{lrrrrrrrrrr} 
\midrule
Daily & 31.99 & 41.50 & 35.50 & 63.48 & 56.69 & 51.17 & 64.42 & 55.50 & 89.22 & 83.20 \\ 
  PA & 21.65 & 27.55 & 22.93 & 38.01 & 36.58 & 59.01 & 75.20 & 66.73 & 104.16 & 102.96 \\ 
  PC & 14.45 & 18.24 & 12.05 & 20.96 & 27.83 & 25.46 & 29.98 & 21.03 & 36.76 & 50.79 \\ 
  PDNE & 6.49 & 9.54 & 11.07 & 19.59 & 17.03 & 18.02 & 25.44 & 28.46 & 43.17 & 45.42 \\ 
  U & 31.30 & 40.36 & 44.96 & 85.66 & 73.95 & 62.91 & 78.86 & 81.40 & 144.06 & 126.57 \\ 
 \bottomrule
\end{tabular}
\subcaption{\texttt{cnrm\_cm6\_1\_hr}}
\begin{tabular}{lrrrrrrrrrr} 
\midrule
Daily & 9.75 & 15.38 & 13.84 & 36.50 & 30.48 & 12.89 & 20.48 & 18.32 & 45.93 & 39.51 \\ 
  PA & 12.38 & 18.53 & 15.58 & 32.26 & 36.77 & 25.69 & 34.67 & 32.14 & 58.75 & 67.58 \\ 
  PC & 10.37 & 15.28 & 8.48 & 20.94 & 20.67 & 27.05 & 34.60 & 20.39 & 46.48 & 55.67 \\ 
  PDNE & 4.88 & 11.41 & 10.89 & 20.81 & 15.09 & 16.15 & 27.14 & 29.81 & 45.97 & 41.06 \\ 
  U & 20.26 & 29.34 & 27.76 & 101.12 & 86.26 & 22.90 & 35.77 & 38.13 & 110.66 & 99.78 \\ 
 \bottomrule
\end{tabular}
\subcaption{\texttt{noresm2\_mm}}
\end{table}

\FloatBarrier

\section{Extended results for the reproduction of daily and conditional distributions across GCM trajectory}

The following analysis evaluates 36 GCM trajectories, categorized into 16 \textbf{retained models} (those passing the daily overlap filter) and 20 \textbf{discarded (``bad'') models}. Performance is assessed using overlap metrics across 30 grid points, based on the daily distribution and different conditioning WTs. Results are summarized in Tables~\ref{tab:daily:ranges}--\ref{tab:U:ranges} and visualized in Figures~\ref{fig:daily:ranges}--\ref{fig:U:ranges}. In the figures, vertical line segments represent the range between minimum and maximum overlap values, crosses ($\times$) denote the number of grid points with overlap $\leq 0.8$, and filled points ($\bullet$) denote the number of grid points with overlap $\geq 0.88$.

\subsection{Daily distribution}

\begin{table}[ht]
\centering
\caption{Number of points in each overlap range, along with the minimum and maximum overlap values, for the 16 trajectories that pass the daily overlap filter.}
\label{tab:daily:ranges}
\begin{tabular}{lrrrrrr}
  \hline
GCM & [0,0.80] & (0.80,0.88] & (0.88,0.95) & [0.95,1] & min & max \\ 
  \hline
access\_cm2 & 7 & 15 & 8 & 0 & 0.68 & 0.92 \\ 
  cmcc\_cm2\_hr4 & 8 & 14 & 8 & 0 & 0.75 & 0.93 \\ 
  cnrm\_cm6\_1 & 1 & 24 & 5 & 0 & 0.79 & 0.90 \\ 
  cnrm\_cm6\_1\_hr & 4 & 26 & 0 & 0 & 0.78 & 0.88 \\ 
  cnrm\_esm2\_1 & 4 & 25 & 1 & 0 & 0.77 & 0.90 \\ 
  ec\_earth3 & 0 & 15 & 15 & 0 & 0.81 & 0.91 \\ 
  ec\_earth3\_aerchem & 0 & 4 & 26 & 0 & 0.80 & 0.95 \\ 
  ec\_earth3\_cc & 1 & 13 & 16 & 0 & 0.80 & 0.93 \\ 
  ec\_earth3\_veg & 1 & 13 & 16 & 0 & 0.79 & 0.94 \\ 
  ec\_earth3\_veg\_lr & 0 & 18 & 12 & 0 & 0.80 & 0.95 \\ 
  fgoals\_g3 & 6 & 14 & 10 & 0 & 0.76 & 0.93 \\ 
  hadgem3\_gc31\_mm & 6 & 15 & 9 & 0 & 0.69 & 0.92 \\ 
  ipsl\_cm6a\_lr & 0 & 11 & 19 & 0 & 0.83 & 0.94 \\ 
  mpi\_esm\_1\_2\_hr & 4 & 16 & 10 & 0 & 0.76 & 0.94 \\ 
  noresm2\_mm & 5 & 17 & 8 & 0 & 0.77 & 0.93 \\ 
  taiesm1 & 7 & 21 & 2 & 0 & 0.62 & 0.93 \\ 
   \hline
\end{tabular}
\end{table}

From Table~\ref{tab:daily:ranges}, it is observed that the retained trajectories exhibit a strong concentration of grid points in the \textbf{(0.80, 0.88]} and \textbf{[0.88, 0.95)} overlap ranges. While \texttt{taiesm1} presents the lowest minimum value (0.62), some models, such as \texttt{ipsl\_cm6a\_lr}, maintain all 30 grid points above 0.80. Although overall performance is generally high across models, no trajectory exhibits overlap values above 0.95. In particular, the \texttt{ec\_earth3\_aerchem} trajectory contains 26 grid points in the \textbf{[0.88, 0.95)} range, outperforming all other models; the next highest value in this range is 16.

\begin{figure}
    \centering
    \includegraphics[width=0.7\linewidth]{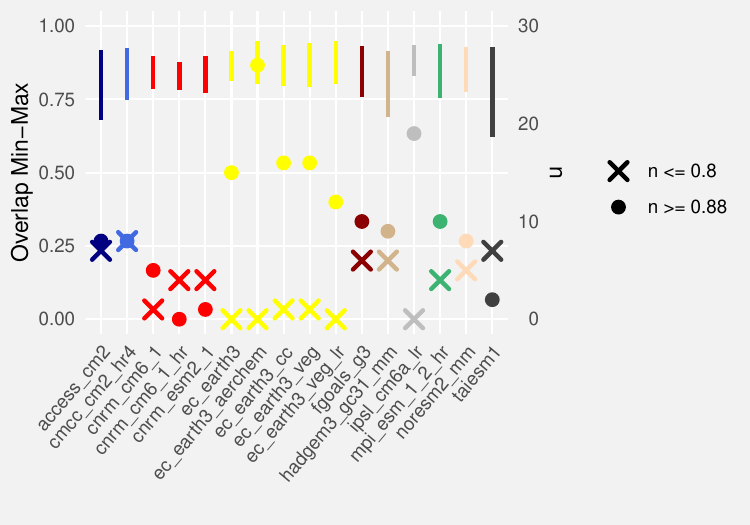}
    \includegraphics[width=0.7\linewidth]{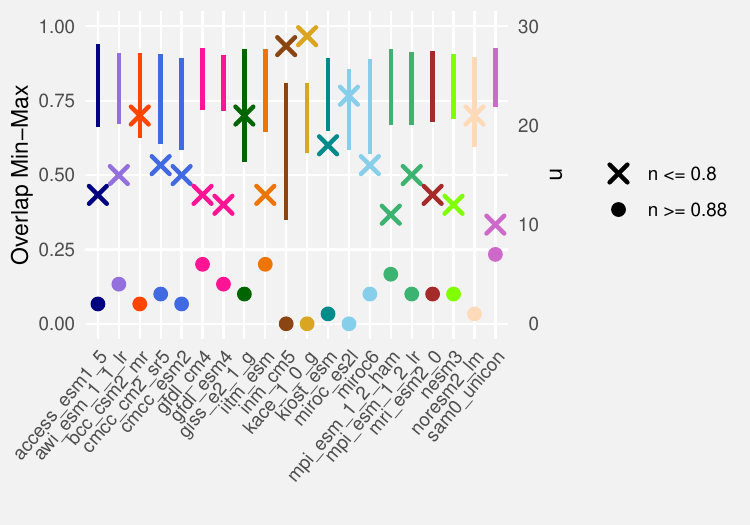}
    \caption{Summary of overlap values for 16 retained GCM trajectories (top) and 20 discarded trajectories (bottom) for the daily distribution. Vertical line segments show the range between minimum and maximum overlap values (left y-axis). Crosses indicate the number of grid points with overlap $\leq 0.8$, and filled points indicate the number of grid points with overlap $\geq 0.88$ (right y-axis). Colors denote the institute associated with each GCM trajectory.}
    \label{fig:daily:ranges}
\end{figure}

From Figure~\ref{fig:daily:ranges} (top), it is evident that \texttt{ec\_earth3\_aerchem} and \texttt{ipsl\_cm6a\_lr} show the best overall performance. The remaining trajectories from the \texttt{ec\_earth3} family, as well as those from the \texttt{cnrm} family, also perform well.

In contrast, Figure~\ref{fig:daily:ranges} (bottom) shows that discarded trajectories exhibit a much wider spread in overlap values. Models such as \texttt{kace\_1\_0\_g} and \texttt{inm\_cm5} have most grid points below 0.88 and none above this threshold.

\FloatBarrier
\subsection{Conditioned on PA}

\begin{table}[ht]
\centering
\caption{Number of points in each overlap range, along with the minimum and maximum overlap values, for the 16 trajectories that pass the daily overlap filter.}
\label{tab:PA:ranges}
\begin{tabular}{lrrrrrr}
  \hline
GCM & [0,0.80] & (0.80,0.88] & (0.88,0.95) & [0.95,1] & min & max \\ 
  \hline
access\_cm2 & 0 & 9 & 20 & 1 & 0.82 & 0.95 \\ 
  cmcc\_cm2\_hr4 & 2 & 12 & 13 & 3 & 0.78 & 0.96 \\ 
  cnrm\_cm6\_1 & 5 & 18 & 7 & 0 & 0.74 & 0.92 \\ 
  cnrm\_cm6\_1\_hr & 0 & 15 & 15 & 0 & 0.81 & 0.94 \\ 
  cnrm\_esm2\_1 & 6 & 16 & 8 & 0 & 0.76 & 0.94 \\ 
  ec\_earth3 & 1 & 4 & 23 & 2 & 0.79 & 0.97 \\ 
  ec\_earth3\_aerchem & 0 & 3 & 22 & 5 & 0.82 & 0.97 \\ 
  ec\_earth3\_cc & 1 & 2 & 22 & 5 & 0.75 & 0.96 \\ 
  ec\_earth3\_veg & 1 & 5 & 21 & 3 & 0.79 & 0.97 \\ 
  ec\_earth3\_veg\_lr & 1 & 5 & 24 & 0 & 0.79 & 0.95 \\ 
  fgoals\_g3 & 8 & 4 & 18 & 0 & 0.75 & 0.95 \\ 
  hadgem3\_gc31\_mm & 2 & 10 & 18 & 0 & 0.78 & 0.95 \\ 
  ipsl\_cm6a\_lr & 1 & 14 & 12 & 3 & 0.78 & 0.96 \\ 
  mpi\_esm\_1\_2\_hr & 0 & 12 & 17 & 1 & 0.83 & 0.95 \\ 
  noresm2\_mm & 0 & 12 & 18 & 0 & 0.81 & 0.93 \\ 
  taiesm1 & 8 & 13 & 9 & 0 & 0.68 & 0.95 \\ 
   \hline
\end{tabular}
\end{table}

When conditioning on PA, performance is slightly improved compared to the daily distribution. As shown in Table~\ref{tab:PA:ranges}, several trajectories now exhibit grid points in the \textbf{[0.95, 1]} range. In particular, the \texttt{ec\_earth3} family reaches maximum overlap values of 0.97. 

\begin{figure}
    \centering
    \includegraphics[width=0.7\linewidth]{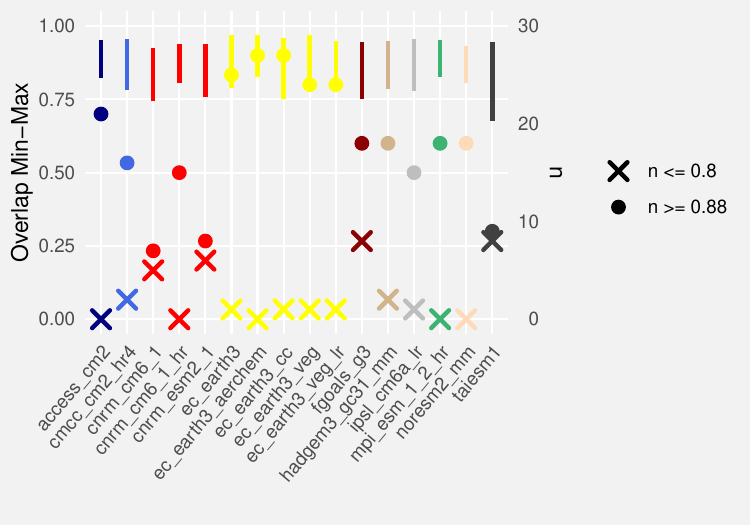}
    \includegraphics[width=0.7\linewidth]{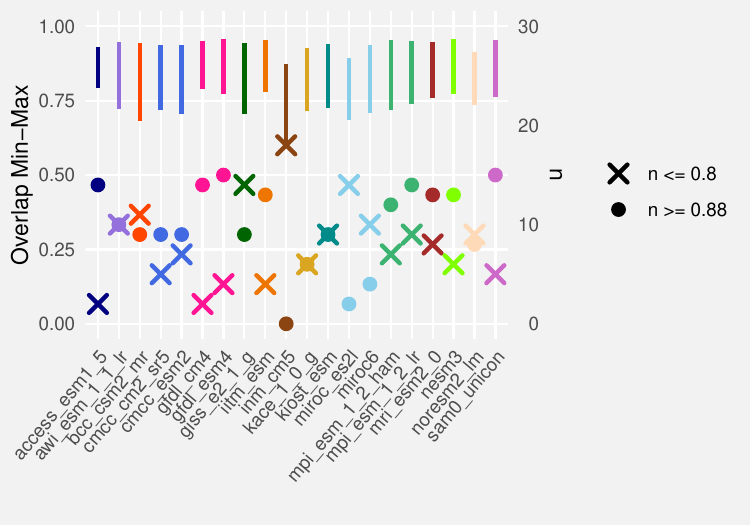}
    \caption{Summary of overlap values for 16 retained GCM trajectories (top) and 20 discarded trajectories (bottom) for the distribution conditioned on PA. Vertical line segments show the range between minimum and maximum overlap values (left y-axis). Crosses indicate the number of grid points with overlap $\leq 0.8$, and filled points indicate the number of grid points with overlap $\geq 0.88$ (right y-axis). Colors denote the institute associated with each GCM trajectory.}
    \label{fig:PA:ranges}
\end{figure}

Figure~\ref{fig:PA:ranges} (top) further shows that many trajectories concentrate a large fraction of grid points with overlap $\geq 0.88$, while only a small fraction fall below $\leq 0.8$, with the exception of \texttt{taiesm1} and \texttt{fgoals\_g3}, which each have 8 grid points in this lower range.

The discarded models also show some improvement relative to the daily distribution; however, compared to the retained trajectories, they still exhibit fewer grid points with overlap $\geq 0.88$ and a larger number of grid points with overlap $\leq 0.8$.

\FloatBarrier
\subsection{Conditioned on PDNE}

\begin{table}[ht]
\centering
\caption{Number of points in each overlap range, along with the minimum and maximum overlap values, for the 16 trajectories that pass the daily overlap filter.}
\label{tab:PDNE:ranges}
\begin{tabular}{lrrrrrr}
  \hline
GCM & [0,0.80] & (0.80,0.88] & (0.88,0.95) & [0.95,1] & min & max \\ 
  \hline
access\_cm2 & 3 & 12 & 14 & 1 & 0.74 & 0.97 \\ 
  cmcc\_cm2\_hr4 & 3 & 15 & 12 & 0 & 0.79 & 0.94 \\ 
  cnrm\_cm6\_1 & 11 & 9 & 10 & 0 & 0.70 & 0.94 \\ 
  cnrm\_cm6\_1\_hr & 8 & 16 & 6 & 0 & 0.70 & 0.92 \\ 
  cnrm\_esm2\_1 & 7 & 18 & 5 & 0 & 0.73 & 0.92 \\ 
  ec\_earth3 & 0 & 5 & 22 & 3 & 0.81 & 0.95 \\ 
  ec\_earth3\_aerchem & 0 & 5 & 21 & 4 & 0.80 & 0.96 \\ 
  ec\_earth3\_cc & 1 & 6 & 22 & 1 & 0.78 & 0.96 \\ 
  ec\_earth3\_veg & 0 & 4 & 26 & 0 & 0.81 & 0.95 \\ 
  ec\_earth3\_veg\_lr & 1 & 9 & 20 & 0 & 0.74 & 0.95 \\ 
  fgoals\_g3 & 8 & 15 & 7 & 0 & 0.68 & 0.93 \\ 
  hadgem3\_gc31\_mm & 5 & 10 & 14 & 1 & 0.76 & 0.96 \\ 
  ipsl\_cm6a\_lr & 7 & 18 & 4 & 1 & 0.74 & 0.95 \\ 
  mpi\_esm\_1\_2\_hr & 2 & 16 & 11 & 1 & 0.76 & 0.96 \\ 
  noresm2\_mm & 2 & 16 & 11 & 1 & 0.75 & 0.96 \\ 
  taiesm1 & 2 & 11 & 16 & 1 & 0.71 & 0.95 \\ 
   \hline
\end{tabular}
\end{table}

\begin{figure}
    \centering
    \includegraphics[width=0.7\linewidth]{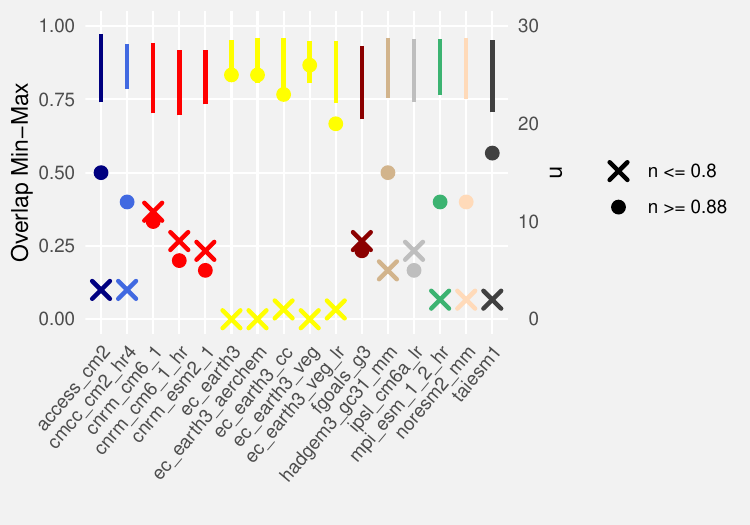}
    \includegraphics[width=0.7\linewidth]{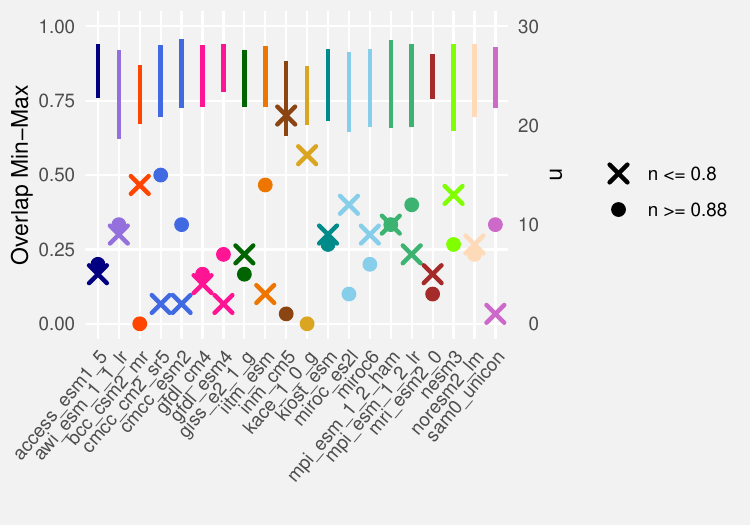}
    \caption{Summary of overlap values for 16 retained GCM trajectories (top) and 20 discarded trajectories (bottom) for the distribution conditioned on PDNE. Vertical line segments show the range between minimum and maximum overlap values (left y-axis). Crosses indicate the number of grid points with overlap $\leq 0.8$, and filled points indicate the number of grid points with overlap $\geq 0.88$ (right y-axis). Colors denote the institute associated with each GCM trajectory.}
    \label{fig:PDNE:ranges}
\end{figure}

When conditioning on PDNE, the performance of the trajectories becomes more variable. As shown in Table~\ref{tab:PDNE:ranges} and Figure~\ref{fig:PDNE:ranges} (top), the \texttt{ec\_earth3} family exhibits consistently high performance, whereas the \texttt{cnrm} family shows lower performance overall, with the \texttt{cnrm\_cm6\_1} trajectory containing 11 grid points in the \textbf{[0,0.8]} range.

In the discarded set (Figure~\ref{fig:PDNE:ranges} (bottom)), most trajectories exhibit poor performance. However, the \texttt{cmcc} family performs comparatively better, with very few grid points falling below $\leq 0.8$.

\FloatBarrier
\subsection{Conditioned on PC}

\begin{table}[ht]
\centering
\caption{Number of points in each overlap range, along with the minimum and maximum overlap values, for the 16 trajectories that pass the daily overlap filter.}
\label{tab:PC:ranges}
\begin{tabular}{lrrrrrr}
  \hline
GCM & [0,0.80] & (0.80,0.88] & (0.88,0.95) & [0.95,1] & min & max \\ 
  \hline
access\_cm2 & 12 & 11 & 7 & 0 & 0.63 & 0.94 \\ 
  cmcc\_cm2\_hr4 & 14 & 15 & 1 & 0 & 0.68 & 0.94 \\ 
  cnrm\_cm6\_1 & 7 & 20 & 3 & 0 & 0.59 & 0.92 \\ 
  cnrm\_cm6\_1\_hr & 5 & 20 & 5 & 0 & 0.73 & 0.91 \\ 
  cnrm\_esm2\_1 & 6 & 23 & 1 & 0 & 0.74 & 0.90 \\ 
  ec\_earth3 & 7 & 9 & 14 & 0 & 0.73 & 0.94 \\ 
  ec\_earth3\_aerchem & 4 & 10 & 16 & 0 & 0.74 & 0.93 \\ 
  ec\_earth3\_cc & 6 & 12 & 12 & 0 & 0.66 & 0.94 \\ 
  ec\_earth3\_veg & 7 & 10 & 13 & 0 & 0.71 & 0.94 \\ 
  ec\_earth3\_veg\_lr & 8 & 12 & 10 & 0 & 0.65 & 0.92 \\ 
  fgoals\_g3 & 9 & 16 & 5 & 0 & 0.70 & 0.91 \\ 
  hadgem3\_gc31\_mm & 7 & 13 & 10 & 0 & 0.69 & 0.92 \\ 
  ipsl\_cm6a\_lr & 13 & 16 & 1 & 0 & 0.73 & 0.89 \\ 
  mpi\_esm\_1\_2\_hr & 6 & 14 & 10 & 0 & 0.74 & 0.94 \\ 
  noresm2\_mm & 6 & 14 & 10 & 0 & 0.72 & 0.92 \\ 
  taiesm1 & 9 & 14 & 7 & 0 & 0.68 & 0.92 \\ 
   \hline
\end{tabular}
\end{table}

PC is the \textbf{most challenging conditioning factor}. There is a marked increase in the number of grid points with overlap $\leq 0.8$ among the selected trajectories (Table~\ref{tab:PC:ranges}), and no trajectories exhibit overlap values above 0.95. In addition, most trajectories are dominated by grid points in the \textbf{(0.8, 0.88]} range, whereas in previous cases the distribution was more balanced between the \textbf{(0.8, 0.88]} and \textbf{(0.88, 0.95)} ranges.

\begin{figure}
    \centering
    \includegraphics[width=0.7\linewidth]{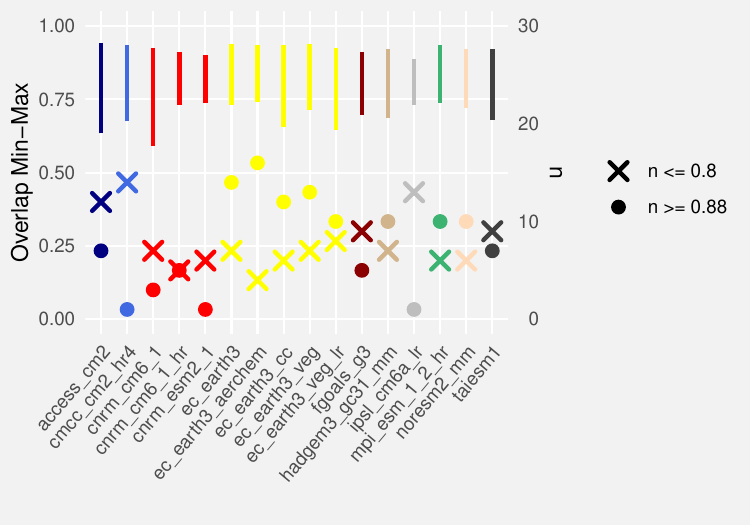}
    \includegraphics[width=0.7\linewidth]{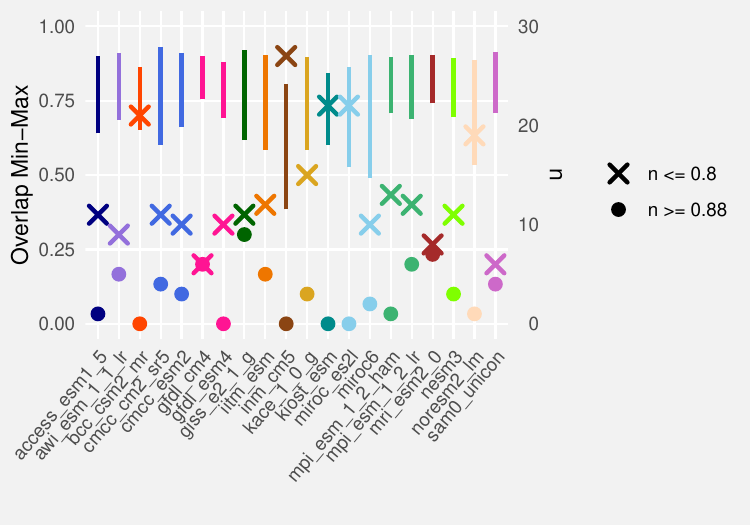}
    \caption{Summary of overlap values for 16 retained GCM trajectories (top) and 20 discarded trajectories (bottom) for the distribution conditioned on PC. Vertical line segments show the range between minimum and maximum overlap values (left y-axis). Crosses indicate the number of grid points with overlap $\leq 0.8$, and filled points indicate the number of grid points with overlap $\geq 0.88$ (right y-axis). Colors denote the institute associated with each GCM trajectory.}
    \label{fig:PC:ranges}
\end{figure}

Figure~\ref{fig:PC:ranges} (top) confirms this behaviour, showing an increase in low-overlap counts (crosses) and a decrease in high-overlap counts (filled points). In the discarded set (Figure~\ref{fig:PC:ranges} (bottom)), many trajectories contain more than 10 grid points with overlap $\leq 0.8$, and several trajectories show no grid points with overlap $\geq 0.88$.

\FloatBarrier
\subsection{Conditioned on U}

\begin{table}[ht]
\centering
\caption{Number of points in each overlap range, along with the minimum and maximum overlap values, for the 16 trajectories that pass the daily overlap filter.}
\label{tab:U:ranges}
\begin{tabular}{lrrrrrr}
  \hline
GCM & [0,0.80] & (0.80,0.88] & (0.88,0.95) & [0.95,1] & min & max \\ 
  \hline
access\_cm2 & 1 & 11 & 18 & 0 & 0.80 & 0.94 \\ 
  cmcc\_cm2\_hr4 & 1 & 17 & 12 & 0 & 0.79 & 0.93 \\ 
  cnrm\_cm6\_1 & 0 & 7 & 22 & 1 & 0.85 & 0.95 \\ 
  cnrm\_cm6\_1\_hr & 0 & 15 & 15 & 0 & 0.82 & 0.92 \\ 
  cnrm\_esm2\_1 & 0 & 12 & 18 & 0 & 0.81 & 0.94 \\ 
  ec\_earth3 & 0 & 6 & 24 & 0 & 0.84 & 0.94 \\ 
  ec\_earth3\_aerchem & 0 & 2 & 28 & 0 & 0.84 & 0.95 \\ 
  ec\_earth3\_cc & 0 & 3 & 27 & 0 & 0.84 & 0.95 \\ 
  ec\_earth3\_veg & 0 & 5 & 25 & 0 & 0.84 & 0.94 \\ 
  ec\_earth3\_veg\_lr & 0 & 7 & 23 & 0 & 0.83 & 0.92 \\ 
  fgoals\_g3 & 1 & 10 & 19 & 0 & 0.80 & 0.94 \\ 
  hadgem3\_gc31\_mm & 0 & 11 & 19 & 0 & 0.82 & 0.95 \\ 
  ipsl\_cm6a\_lr & 0 & 11 & 18 & 1 & 0.83 & 0.95 \\ 
  mpi\_esm\_1\_2\_hr & 0 & 8 & 21 & 1 & 0.84 & 0.95 \\ 
  noresm2\_mm & 0 & 14 & 16 & 0 & 0.85 & 0.93 \\ 
  taiesm1 & 3 & 13 & 14 & 0 & 0.77 & 0.92 \\ 
   \hline
\end{tabular}
\end{table}

Finally, U represents the peak performance across all distributions. Although no trajectory has more than one grid point with overlap $\geq 0.95$, most trajectories are dominated by values in the \textbf{(0.88, 0.95)} range. Very few trajectories contain grid points with overlap $\leq 0.8$, with \texttt{taiesm1} being the weakest performer in this regard, with 3 grid points in this range.

\begin{figure}
    \centering
    \includegraphics[width=0.7\linewidth]{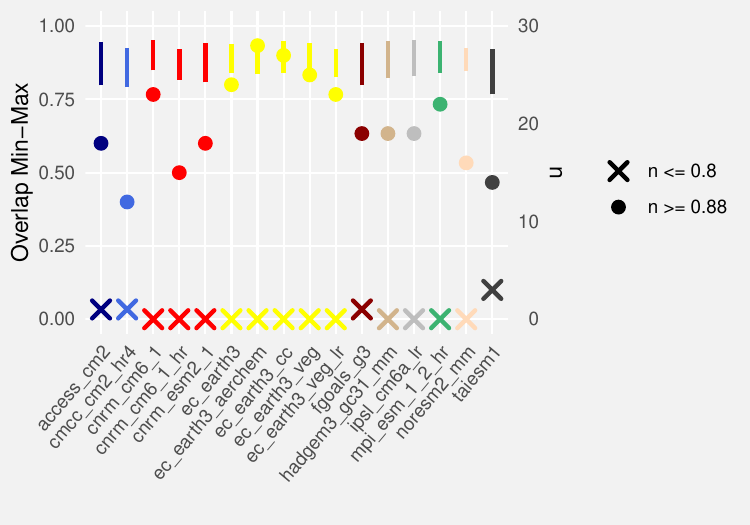}
    \includegraphics[width=0.7\linewidth]{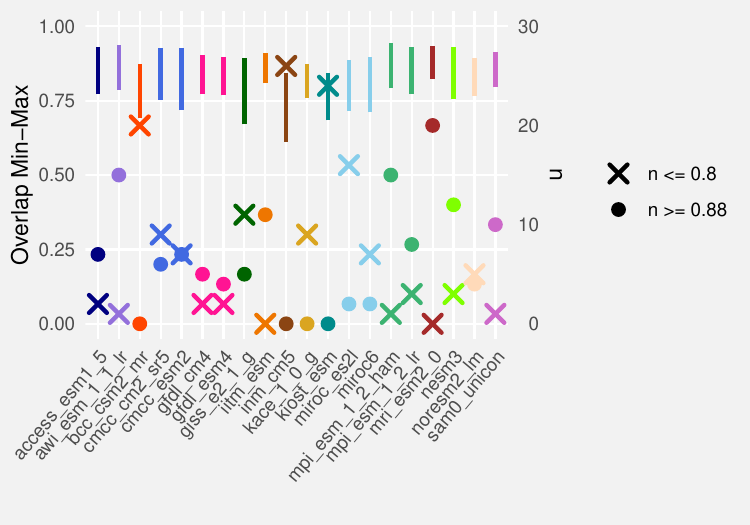}
    \caption{Summary of overlap values for 16 retained GCM trajectories (top) and 20 discarded trajectories (bottom) for the distribution conditioned on U. Vertical line segments show the range between minimum and maximum overlap values (left y-axis). Crosses indicate the number of grid points with overlap $\leq 0.8$, and filled points indicate the number of grid points with overlap $\geq 0.88$ (right y-axis). Colors denote the institute associated with each GCM trajectory.}
    \label{fig:U:ranges}
\end{figure}

Figure~\ref{fig:U:ranges} (top) confirms the strong overall performance of the trajectories, with most models showing zero grid points with overlap $\leq 0.8$, a large number of grid points with overlap $\geq 0.88$, and generally narrow min--max ranges, indicating consistent performance.

In the discarded set (Figure~\ref{fig:U:ranges} (bottom)), a clear improvement is also observed compared to previous distributions. In particular, \texttt{awi\_esm\_1\_1\_lr}, \texttt{mri\_esm2\_0}, and \texttt{mpi\_esm\_1\_2\_ham} show comparatively good performance.

\FloatBarrier
\section{Filtering summary table}

\begin{table}[ht]
\centering
\caption{
Sequential filtering of GCM trajectories based on the fraction of grid points with overlap $\leq 0.8$. The procedure starts with the Daily distribution and is progressively applied to distributions conditioned on PC and PDNE. At each stage, trajectories with at least one third of grid points below the threshold are excluded and do not propagate to subsequent columns. The final column reports the PerR score for trajectories that pass all filtering stages.
}
\begin{tabular}{lrrrr}
  \hline
GCM & Daily & $\mid$PC & $\mid$PDNE & PerR \\ 
  \hline
access\_cm2 & 7 & 12 &  &  \\ 
  access\_esm1\_5 & 13 &  &  &  \\ 
  awi\_esm\_1\_1\_lr & 15 &  &  &  \\ 
  bcc\_csm2\_mr & 21 &  &  &  \\ 
  cmcc\_cm2\_hr4 & 8 & 14 &  &  \\ 
  cmcc\_cm2\_sr5 & 16 &  &  &  \\ 
  cmcc\_esm2 & 15 &  &  &  \\ 
  cnrm\_cm6\_1 & 1 & 7 & 11 &  \\ 
  cnrm\_cm6\_1\_hr & 4 & 5 & 8 & 47.02 \\ 
  cnrm\_esm2\_1 & 4 & 6 & 7 & 49.66 \\ 
  ec\_earth3 & 0 & 7 & 0 & 40.03 \\ 
  ec\_earth3\_aerchem & 0 & 4 & 0 & 38.84 \\ 
  ec\_earth3\_cc & 1 & 6 & 1 & 41.29 \\ 
  ec\_earth3\_veg & 1 & 7 & 0 & 35.59 \\ 
  ec\_earth3\_veg\_lr & 0 & 8 & 1 & 43.27 \\ 
  fgoals\_g3 & 6 & 9 & 8 & 52.13 \\ 
  gfdl\_cm4 & 13 &  &  &  \\ 
  gfdl\_esm4 & 12 &  &  &  \\ 
  giss\_e2\_1\_g & 21 &  &  &  \\ 
  hadgem3\_gc31\_mm & 6 & 7 & 5 & 41.66 \\ 
  iitm\_esm & 13 &  &  &  \\ 
  inm\_cm5 & 28 &  &  &  \\ 
  ipsl\_cm6a\_lr & 0 & 13 &  &  \\ 
  kace\_1\_0\_g & 29 &  &  &  \\ 
  kiost\_esm & 18 &  &  &  \\ 
  miroc6 & 16 &  &  &  \\ 
  miroc\_es2l & 23 &  &  &  \\ 
  mpi\_esm\_1\_2\_ham & 11 &  &  &  \\ 
  mpi\_esm\_1\_2\_hr & 4 & 6 & 2 & 41.70 \\ 
  mpi\_esm\_1\_2\_lr & 15 &  &  &  \\ 
  mri\_esm2\_0 & 13 &  &  &  \\ 
  nesm3 & 12 &  &  &  \\ 
  noresm2\_lm & 21 &  &  &  \\ 
  noresm2\_mm & 5 & 6 & 2 & 43.10 \\ 
  sam0\_unicon & 10 & 6 & 1 &  \\ 
  taiesm1 & 7 & 9 & 2 & 45.17 \\ 
   \hline
\end{tabular}
\end{table}

\end{appendices}


\end{document}